\documentclass[12pt]{iopart}

\usepackage[utf8]{inputenc}
\usepackage{acro}
\usepackage{units}
\usepackage{url}
\usepackage{xcolor}
\usepackage{graphicx}
\usepackage{float}
\usepackage{hyperref}
\usepackage[normalem]{ulem}
\usepackage{amssymb}
\usepackage{subcaption}
\usepackage{multirow}
\usepackage{cite}
\usepackage{adjustbox}
\usepackage{orcidlink}

\newcommand{\detchar}{DetChar~}

\DeclareAcronym{ADC}{
short = {ADC},
long = {analog-to-digital converter}
}

\DeclareAcronym{AHU}{
short = {AHU},
long = {air handling unit}
}

\DeclareAcronym{ALS}{
short = {ALS},
long = {arm length stabilization}
}

\DeclareAcronym{ASD}{
short = {ASD},
long = {Amplitude Spectral Density},
long-plural-form = {amplitude spectral densities}
}

\DeclareAcronym{BBS}{
short = {BBS},
long = {Bigger Beam Splitter}
}

\DeclareAcronym{BNS}{
short = {BNS},
long = {binary neutron star}
}

\DeclareAcronym{BS}{
short = {BS},
long = {beamsplitter}
}

\DeclareAcronym{BSC}{
short = {BSC},
long = {Beam Splitter Chamber}
}

\DeclareAcronym{CAT1}{
short = {CAT1},
long = {Category 1}
}

\DeclareAcronym{CAT2}{
short = {CAT2},
long = {Category 2}
}

\DeclareAcronym{CBC}{
short = {CBC},
long = {compact binary coalescences}
}

\DeclareAcronym{CER}{
short = {CER},
long = {Corner Station Electronics Room}
}

\DeclareAcronym{CS}{
short = {CS},
long = {corner station}
}

\DeclareAcronym{CW}{
short = {CW},
long = {continuous wave}
}

\DeclareAcronym{DARM}{
short = {DARM},
long = {differential arm length}
}

\DeclareAcronym{DCPD}{
short = {DCPD},
long = {DC Photodiode}
}

\DeclareAcronym{DQ}{
short = {DQ},
long = {data quality}
}

\DeclareAcronym{DQR}{
short = {DQR},
long = {data quality report}
}

\DeclareAcronym{DTT}{
short = {DTT},
long = {Diagnostic Test Tool}
}

\DeclareAcronym{ESD}{
short = {ESD},
long = {electrostatic drive}
}

\DeclareAcronym{ETMX}{
short = {ETMX},
long = {X-arm end test mass}
}

\DeclareAcronym{ETMY}{
short = {ETMY},
long = {Y-arm end test mass}
}

\DeclareAcronym{EV}{
short = {EV},
long = {event validation}
}

\DeclareAcronym{EX}{
short = {EX},
long = {End-X}
}

\DeclareAcronym{EY}{
short = {EY},
long = {End-Y}
}

\DeclareAcronym{FAP}{
short = {FAP},
long = {false-alarm probability}
}

\DeclareAcronym{FMC}{
short = {FMC},
long = {facilities management and control}
}

\DeclareAcronym{FSS}{
short = {FSS},
long = {frequency stabilization servo}
}

\DeclareAcronym{GV}{
short = {GV},
long = {gate valve}
}

\DeclareAcronym{GW}{
short = {GW},
long = {gravitational wave}
}

\DeclareAcronym{H1}{
short = {H1},
long = {Hanford 4K Interferometer}
}

\DeclareAcronym{HAM}{
short = {HAM},
long = {Horizontal Access Module}
}

\DeclareAcronym{HEPI}{
short = {HEPI},
long = {hydraulic external pre-isolator system}
}

\DeclareAcronym{HVAC}{
short = {HVAC},
long = {heating, ventilation, and air conditioning}
}

\DeclareAcronym{IR}{
short = {IR1},
long = {Intermediate Run 1}
}

\DeclareAcronym{ISC}{
short = {ISC},
long = {Interferometer Sensing and Control}
}

\DeclareAcronym{ISI}{
short = {ISI},
long = {internal seismic isolation}
}

\DeclareAcronym{ITM}{
short = {ITM},
long = {Input Test Mass}
}

\DeclareAcronym{ITMX}{
short = {ITMX},
long = {X-arm input test mass}
}

\DeclareAcronym{ITMY}{
short = {ITMY},
long = {Y-arm input test mass}
}

\DeclareAcronym{JAC}{
short = {JAC},
long = {Jitter Attenuation Cavity}
}

\DeclareAcronym{L1}{
short = {L1},
long = {Livingston 4K Interferometer}
}

\DeclareAcronym{LHO}{
short = {LHO},
long = {LIGO Hanford Observatory}
}

\DeclareAcronym{LIGO}{
short = {LIGO},
long = {Laser Interferometer Gravitational-wave Observatory}
}

\DeclareAcronym{LLO}{
short = {LLO},
long = {LIGO Livingston Observatory}
}

\DeclareAcronym{LVEA}{
short = {LVEA},
long = {laser and vacuum equipment area}
}

\DeclareAcronym{LVK}{
short = {LVK},
long = {LIGO-Virgo-KAGRA}
}

\DeclareAcronym{MC}{
short = {MC},
long = {mode cleaner}
}

\DeclareAcronym{NPRO}{
short = {NPRO},
long = {non-planar ring oscillator}
}

\DeclareAcronym{NTSC}{
short = {NTSC},
long = {National Television System Committee}
}

\DeclareAcronym{O1}{
short = {O1},
long = {the first observing run}
}

\DeclareAcronym{O2}{
short = {O2},
long = {second observing run}
}

\DeclareAcronym{O3}{
short = {O3},
long = {third observing run}
}

\DeclareAcronym{O4}{
short = {O4},
long = {fourth observing run}
}

\DeclareAcronym{O5}{
short = {O5},
long = {Fifth Observing Run}
}

\DeclareAcronym{OFI}{
short = {OFI},
long = {output faraday isolator}
}

\DeclareAcronym{OMC}{
short = {OMC},
long = {Output Mode Cleaner}
}

\DeclareAcronym{PEM}{
short = {PEM},
long = {physical and environmental monitoring}
}

\DeclareAcronym{PLL}{
short = {PLL},
long = {Phase-locked loop}
}

\DeclareAcronym{PSD}{
short = {PSD},
long = {power spectral density}
}

\DeclareAcronym{PSL}{
short = {PSL},
long = {pre-stabilized laser}
}

\DeclareAcronym{PUM}{
short = {PUM},
long = {Penultimate Mass}
}

\DeclareAcronym{R2}{
short = {R2},
long = {\textcolor{red}{FIXME}}
}

\DeclareAcronym{rms}{
short = {rms},
long = {root mean square}
}

\DeclareAcronym{RRT}{
short = {RRT},
long = {rapid response team}
}

\DeclareAcronym{SGWB}{
short = {SGWB},
long = {stochastic gravitational-wave background}
}

\DeclareAcronym{SNR}{
short = {SNR},
long = {signal-to-noise ratio}
}

\DeclareAcronym{UIM}{
short = {UIM},
long = {Upper Intermediate Mass}
}

\DeclareAcronym{VEA}{
short = {VEA},
long = {vacuum equipment area}
}

\DeclareAcronym{VFD}{
short = {VFD},
long = {variable frequency drive}
}

\DeclareAcronym{VPW}{
short = {VPW},
long = {vacuum preparation warehouse}
}

\graphicspath{{./figures/}}

\begin{document}

\title[LIGO Detector Characterization in O4b \& O4c]{LIGO Detector Characterization in the Second and Third Parts of the Fourth Observing Run}

\author{%
J.~Glanzer\,\orcidlink{0009-0000-0808-0795}$^{1,*}$,
A.~F.~Helmling-Cornell\,\orcidlink{0000-0002-7709-8638}$^{2,**}$,
A.~Calafat\,\orcidlink{0009-0008-7515-6305}$^{3}$,
S.~R.~Callos\,\orcidlink{0000-0003-0639-9342}$^{4}$,
E.~Capote\,\orcidlink{0009-0007-0246-713X}$^{5}$,
G.~Connolly\,\orcidlink{0009-0000-8533-0428}$^{4}$,
A.~Effler\,\orcidlink{0000-0001-8242-3944}$^{6}$,
T.~A.~Ferreira\,\orcidlink{0000-0002-1166-2005}$^{7}$,
E.~Goetz\,\orcidlink{0000-0003-2666-721X}$^{8}$,
A.~M.~Knee\,\orcidlink{0000-0003-0703-947X}$^{9}$,
J.~R.~M\'erou\,\orcidlink{0000-0002-5776-6643}$^{3}$,
D.~Malakar\,\orcidlink{0000-0003-4234-4023}$^{10}$,
B.~Mannix$^{4}$,
D.~Nandi\,\orcidlink{0009-0004-3245-9454}$^{11}$,
K.~Pham\,\orcidlink{0000-0002-7650-1034}$^{12}$,
R.~M.~S.~Schofield$^{4}$,
P.~Sharma\,\orcidlink{0009-0000-5586-9764}$^{11}$,
Z.~Yarbrough\,\orcidlink{0000-0002-9825-1136}$^{11}$,
N.~Arnaud$^{13}$,
B.~K.~Berger\,\orcidlink{0000-0002-4845-8737}$^{14}$,
K.~Burtnyk$^{5}$,
C.~M.~Compton$^{5}$,
D.~Davis\,\orcidlink{0000-0001-5620-6751}$^{15}$,
F.~Di~Renzo\,\orcidlink{0000-0002-5447-3810}$^{16,17}$,
G.~Grant\,\orcidlink{0009-0001-5047-2458}$^{18}$,
J.~C.~Martins\,\orcidlink{0000-0002-6099-4831}$^{7}$,
G.~Mo$^{19,20}$,
A.~Neunzert\,\orcidlink{0000-0003-0323-0111}$^{5}$,
L.~K.~Nuttall\,\orcidlink{0000-0001-7472-0201}$^{21}$,
J.~Oberling\,\orcidlink{0009-0001-4174-3973}$^{5}$,
J.~R.~Olson\,\orcidlink{0009-0008-8934-2459}$^{4}$,
S.~Soni\,\orcidlink{0000-0003-3856-8534}$^{22}$,
M.~Trevor\,\orcidlink{0000-0002-2728-9508}$^{23}$,
R.~Abbott$^{1}$,
I.~Abouelfettouh$^{5}$,
R.~X.~Adhikari\,\orcidlink{0000-0002-5731-5076}$^{1}$,
A.~Ahuja\,\orcidlink{0009-0006-8812-9109}$^{8}$,
S.~\'Alvarez-L\'opez\,\orcidlink{0009-0003-8040-4936}$^{24,25,26}$,
A.~Ananyeva$^{1}$,
S.~Appert$^{1}$,
S.~K.~Apple\,\orcidlink{0009-0007-4490-5804}$^{27}$,
K.~Arai\,\orcidlink{0000-0001-8916-8915}$^{1}$,
J.~Areeda\,\orcidlink{0000-0003-0266-7936}$^{28}$,
N.~Aritomi$^{5}$,
S.~M.~Aston$^{6}$,
M.~Ball$^{4}$,
S.~W.~Ballmer$^{29}$,
D.~Barker$^{5}$,
L.~Barsotti\,\orcidlink{0000-0001-9819-2562}$^{24}$,
J.~Betzwieser\,\orcidlink{0000-0003-1533-9229}$^{6}$,
Z.~S.~Bhalla\,\orcidlink{0009-0002-3488-5057}$^{30}$,
D.~Bhattacharjee\,\orcidlink{0000-0001-6623-9506}$^{10,31}$,
G.~Billingsley\,\orcidlink{0000-0002-4141-2744}$^{1}$,
S.~Biscans$^{24}$,
C.~D.~Blair$^{6,32}$,
N.~Bode\,\orcidlink{0000-0002-7101-9396}$^{33,34}$,
E.~Bonilla\,\orcidlink{0000-0002-6284-9769}$^{14}$,
V.~Bossilkov$^{6}$,
A.~Branch$^{6}$,
A.~F.~Brooks\,\orcidlink{0000-0003-4295-792X}$^{1}$,
D.~D.~Brown$^{35}$,
R.~Bruntz\,\orcidlink{0000-0002-0840-8567}$^{36}$,
J.~Bryant$^{37}$,
C.~Cahillane\,\orcidlink{0000-0002-3888-314X}$^{29}$,
H.~Cao$^{24}$,
C.~Chatterjee\,\orcidlink{0000-0001-8700-3455}$^{38}$,
N.~Christensen\,\orcidlink{0000-0002-6870-4202}$^{30}$,
F.~Clara$^{5}$,
J.~Collins$^{6}$,
R.~Cottingham$^{6}$,
D.~C.~Coyne\,\orcidlink{0000-0002-6427-3222}$^{1}$,
R.~Crouch$^{5}$,
J.~Csizmazia$^{5}$,
A.~Cumming\,\orcidlink{0000-0003-4096-7542}$^{39}$,
L.~P.~Dartez$^{6}$,
N.~Demos$^{24}$,
E.~Dohmen$^{5}$,
K.~L.~Dooley\,\orcidlink{0000-0002-1636-0233}$^{40}$,
J.~C.~Driggers\,\orcidlink{0000-0002-6134-7628}$^{5}$,
S.~E.~Dwyer$^{5}$,
A.~Ejlli\,\orcidlink{0000-0002-4149-4532}$^{40}$,
T.~Etzel$^{1}$,
M.~Evans\,\orcidlink{0000-0001-8459-4499}$^{24}$,
J.~Feicht$^{1}$,
R.~Frey\,\orcidlink{0000-0003-0341-2636}$^{4}$,
W.~Frischhertz$^{6}$,
P.~Fritschel$^{24}$,
V.~V.~Frolov$^{6}$,
M.~Fuentes-Garcia\,\orcidlink{0000-0003-3390-8712}$^{1}$,
P.~Fulda$^{41}$,
M.~Fyffe$^{6}$,
D.~Ganapathy\,\orcidlink{0000-0003-3028-4174}$^{24}$,
B.~Gateley$^{5}$,
T.~Gayer$^{29}$,
J.~A.~Giaime\,\orcidlink{0000-0002-3531-817X}$^{6,11}$,
K.~D.~Giardina$^{6}$,
R.~Goetz\,\orcidlink{0000-0002-9617-5520}$^{41}$,
G.~Gonzalez$^{11}$,
A.~W.~Goodwin-Jones\,\orcidlink{0000-0002-0395-0680}$^{1,32}$,
S.~Gras$^{24}$,
C.~Gray$^{5}$,
D.~Griffith$^{1}$,
H.~Grote\,\orcidlink{0000-0002-0797-3943}$^{40}$,
T.~Guidry$^{5}$,
J.~Gurs$^{42}$,
E.~D.~Hall\,\orcidlink{0000-0001-9018-666X}$^{24}$,
J.~Hanks$^{5}$,
J.~Hanson$^{6}$,
M.~C.~Heintze$^{6}$,
N.~A.~Holland$^{43}$,
N.-T.~Howard\,\orcidlink{0000-0002-9776-8838}$^{38}$,
D.~Hoyland$^{37}$,
H.~Y.~Huang\,\orcidlink{0000-0002-1665-2383}$^{44}$,
B.~Hughey$^{45}$,
Y.~Inoue$^{44}$,
A.~L.~James\,\orcidlink{0000-0001-9165-0807}$^{1}$,
A.~Jamies$^{1}$,
K.~Jani\,\orcidlink{0000-0003-1007-8912}$^{38}$,
R.~Jaume\,\orcidlink{0000-0001-8691-3166}$^{3}$,
A.~Jennings$^{5}$,
W.~Jia$^{24}$,
D.~H.~Jones\,\orcidlink{0000-0003-3987-068X}$^{46}$,
H.~B.~Kabagoz\,\orcidlink{0000-0002-0900-8557}$^{6}$,
S.~Kandhasamy\,\orcidlink{0000-0002-4825-6764}$^{47}$,
S.~Karat$^{1}$,
S.~Karki$^{10}$,
M.~Kasprzack\,\orcidlink{0000-0003-4618-5939}$^{1}$,
K.~Kawabe$^{5}$,
N.~Kijbunchoo\,\orcidlink{0000-0002-2874-1228}$^{35}$,
P.~J.~King$^{5}$,
J.~S.~Kissel\,\orcidlink{0000-0002-1702-9577}$^{5}$,
K.~Komori\,\orcidlink{0000-0002-4092-9602}$^{48}$,
A.~Kontos\,\orcidlink{0000-0002-1347-0680}$^{2}$,
R.~Kumar$^{5}$,
K.~Kuns\,\orcidlink{0000-0003-0630-3902}$^{24}$,
M.~Landry$^{5}$,
B.~Lantz\,\orcidlink{0000-0002-7404-4845}$^{14}$,
M.~Laxen\,\orcidlink{0000-0001-7515-9639}$^{6}$,
K.~Lee\,\orcidlink{0000-0003-0470-3718}$^{49}$,
M.~Lesovsky$^{1}$,
F.~Llamas~Villarreal$^{50}$,
E.~Lofquist-Fabris$^{8}$,
M.~Lormand$^{6}$,
B.~R.~Lott\,\orcidlink{0009-0008-6705-2100}$^{30}$,
H.~A.~Loughlin$^{24}$,
R.~Macas\,\orcidlink{0000-0002-6096-8297}$^{51}$,
M.~MacInnis$^{24}$,
C.~N.~Makarem$^{1}$,
G.~L.~Mansell\,\orcidlink{0000-0003-4736-6678}$^{29}$,
R.~M.~Martin\,\orcidlink{0000-0001-9664-2216}$^{52}$,
K.~Mason$^{24}$,
F.~Matichard$^{24}$,
N.~Mavalvala\,\orcidlink{0000-0003-0219-9706}$^{24}$,
N.~Maxwell$^{5}$,
G.~McCarrol$^{6}$,
R.~McCarthy$^{5}$,
D.~E.~McClelland\,\orcidlink{0000-0001-6210-5842}$^{46}$,
S.~McCormick$^{6}$,
J.~McIver\,\orcidlink{0000-0003-0316-1355}$^{8}$,
R.~McNeil$^{8}$,
T.~McRae$^{46}$,
F.~Mera$^{5}$,
E.~L.~Merilh$^{6}$,
F.~Meylahn\,\orcidlink{0000-0002-9556-142X}$^{33,34}$,
R.~Mittleman$^{24}$,
S.~Mohan~S\,\orcidlink{0009-0005-1202-4661}$^{53}$,
D.~Moraru$^{5}$,
G.~Moreno$^{5}$,
A.~Mullavey$^{6}$,
M.~Nakano$^{1}$,
T.~J.~N.~Nelson$^{6}$,
S.~A.~Nichols\,\orcidlink{0009-0001-1750-3531}$^{54}$,
J.~Notte$^{52}$,
T.~O'Hanlon$^{6}$,
R.~Oram$^{6}$,
C.~Osthelder$^{1}$,
D.~J.~Ottaway\,\orcidlink{0000-0001-6794-1591}$^{35}$,
H.~Overmier$^{6}$,
W.~Parker\,\orcidlink{0000-0002-7711-4423}$^{6}$,
O.~Patane\,\orcidlink{0000-0002-4850-2355}$^{5}$,
A.~Pele\,\orcidlink{0000-0002-1873-3769}$^{1}$,
S.~Perry$^{8}$,
H.~Pham$^{6}$,
M.~Pirello$^{5}$,
J.~Pullin\,\orcidlink{0000-0001-8248-603X}$^{11}$,
V.~Quetschke$^{50}$,
K.~E.~Ramirez\,\orcidlink{0000-0003-2194-7669}$^{6}$,
K.~Ransom$^{6}$,
J.~Reyes$^{52}$,
J.~W.~Richardson\,\orcidlink{0000-0002-1472-4806}$^{22}$,
K.~Riles\,\orcidlink{0000-0002-6418-5812}$^{9}$,
M.~Robinson$^{5}$,
J.~G.~Rollins\,\orcidlink{0000-0002-9388-2799}$^{1}$,
C.~L.~Romel$^{5}$,
J.~H.~Romie$^{6}$,
M.~P.~Ross\,\orcidlink{0000-0002-8955-5269}$^{27}$,
B.~I.~Rotimi$^{29}$,
K.~Ryan$^{5}$,
T.~Sadecki$^{5}$,
A.~Sanchez$^{5}$,
E.~J.~Sanchez$^{1}$,
L.~E.~Sanchez$^{1}$,
R.~L.~Savage\,\orcidlink{0000-0003-3317-1036}$^{5}$,
D.~Schaetzl$^{1}$,
M.~G.~Schiworski\,\orcidlink{0000-0001-9298-004X}$^{29}$,
R.~Schnabel\,\orcidlink{0000-0003-2896-4218}$^{42}$,
E.~Schwartz\,\orcidlink{0000-0001-8922-7794}$^{14}$,
D.~Sellers$^{6}$,
T.~Shaffer$^{5}$,
R.~W.~Short$^{5}$,
D.~Sigg\,\orcidlink{0000-0003-4606-6526}$^{5}$,
B.~J.~J.~Slagmolen\,\orcidlink{0000-0002-2471-3828}$^{46}$,
J.~R.~Smith\,\orcidlink{0000-0003-0638-9670}$^{28}$,
C.~Soike$^{5}$,
V.~Srivastava$^{29}$,
T.~Starkman\,\orcidlink{0000-0003-3583-3742}$^{8}$,
L.~Sun\,\orcidlink{0000-0001-7959-892X}$^{46}$,
D.~B.~Tanner$^{41}$,
J.~Tasson$^{30}$,
M.~Thomas$^{6}$,
P.~Thomas$^{5}$,
K.~A.~Thorne$^{6}$,
E.~M.~Todd\,\orcidlink{0009-0006-1555-9474}$^{39}$,
M.~R.~Todd$^{29}$,
C.~I.~Torrie$^{1}$,
G.~Traylor$^{6}$,
A.~S.~Ubhi\,\orcidlink{0000-0002-3240-6000}$^{37}$,
R.~P.~Udall\,\orcidlink{0000-0001-6877-3278}$^{8}$,
G.~Vajente\,\orcidlink{0000-0002-7656-6882}$^{1}$,
J.~Vanosky$^{5}$,
A.~Vecchio\,\orcidlink{0000-0002-6254-1617}$^{37}$,
P.~J.~Veitch\,\orcidlink{0000-0002-2597-435X}$^{35}$,
A.~M.~Vibhute\,\orcidlink{0000-0003-1501-6972}$^{5}$,
E.~R.~G.~von.~Reis$^{5}$,
J.~Warner$^{5}$,
B.~Weaver$^{5}$,
R.~Weiss\footnote{Deceased, August 2025.}$^{24}$,
C.~Whittle\,\orcidlink{0000-0002-8833-7438}$^{1}$,
P.~Wilcox$^{30}$,
B.~Willke\,\orcidlink{0000-0003-0524-2925}$^{33,34}$,
C.~C.~Wipf$^{1}$,
J.~L.~Wright$^{46}$,
V.~A.~Xu\,\orcidlink{0000-0002-3020-3293}$^{55}$,
H.~Yamamoto\,\orcidlink{0000-0001-6919-9570}$^{1}$,
L.~Zhang$^{1}$,
Z.~Zhang\,\orcidlink{0009-0003-3860-0335}$^{30}$,
M.~E.~Zucker$^{1,24}$}

\address{%
${}^{1}$LIGO Laboratory, California Institute of Technology, Pasadena, CA 91125, USA \\
${}^{2}$Bard College, Annandale-On-Hudson, NY 12504, USA \\
${}^{3}$IAC3--IEEC, Universitat de les Illes Balears, E-07122 Palma de Mallorca, Spain \\
${}^{4}$University of Oregon, Eugene, OR 97403, USA \\
${}^{5}$LIGO Hanford Observatory, Richland, WA 99352, USA \\
${}^{6}$LIGO Livingston Observatory, Livingston, LA 70754, USA \\
${}^{7}$Instituto Nacional de Pesquisas Espaciais, 12227-010 S\~ao Jos\'e dos Campos, SP, Brazil \\
${}^{8}$University of British Columbia, Vancouver, BC V6T 1Z4, Canada \\
${}^{9}$University of Michigan, Ann Arbor, MI 48109, USA \\
${}^{10}$Missouri University of Science and Technology, Rolla, MO 65409, USA \\
${}^{11}$Louisiana State University, Baton Rouge, LA 70803, USA \\
${}^{12}$School of Physics and Astronomy, University of Minnesota, MN 55455, USA \\
${}^{13}$Universit\'e Claude Bernard Lyon 1, CNRS, IP21 LYON / IN2P3, UMR 5822, F-69622 Villeurbanne, France \\
${}^{14}$Stanford University, Stanford, CA 94305, USA \\
${}^{15}$University of Rhode Island, Kingston, RI 02881, USA \\
${}^{16}$Universit\`a di Firenze, Sesto Fiorentino I-50019, Italy \\
${}^{17}$INFN, Sezione di Firenze, I-50019 Sesto Fiorentino, Firenze, Italy \\
${}^{18}$Department of Physics and Astronomy, Minnesota State University Moorhead, Moorhead, MN 56560, USA \\
${}^{19}$Department of Astronomy, California Institute of Technology, Pasadena, CA 91125, USA \\
${}^{20}$Carnegie Science Observatories Pasadena, CA 91101, USA \\
${}^{21}$Institute of Cosmology and Gravitation, University of Portsmouth, Portsmouth, PO1 3FX, UK \\
${}^{22}$Department of Physics, University of California, Riverside, Riverside, CA 92521, USA \\
${}^{23}$University of Maryland, College Park, MD 20742, USA \\
${}^{24}$LIGO Laboratory, Massachusetts Institute of Technology, Cambridge, MA 02139, USA \\
${}^{25}$Kavli Institute for Astrophysics and Space Research, Massachusetts Institute of Technology, Cambridge, MA 02139, USA \\
${}^{26}$Department of Physics, Massachusetts Institute of Technology, Cambridge, MA 02139, USA \\
${}^{27}$University of Washington, Seattle, WA 98195, USA \\
${}^{28}$California State University Fullerton, Fullerton, CA 92831, USA \\
${}^{29}$Syracuse University, Syracuse, NY 13244, USA \\
${}^{30}$Physics and Astronomy Department, Carleton College, Northfield, MN 55057, USA \\
${}^{31}$Kenyon College, Gambier, OH 43022, USA \\
${}^{32}$OzGrav, University of Western Australia, Crawley, Western Australia 6009, Australia \\
${}^{33}$Max Planck Institute for Gravitational Physics (Albert Einstein Institute), D-30167 Hannover, Germany \\
${}^{34}$Leibniz Universit\"{a}t Hannover, D-30167 Hannover, Germany \\
${}^{35}$OzGrav, University of Adelaide, Adelaide, South Australia 5005, Australia \\
${}^{36}$Christopher Newport University, Newport News, VA 23606, USA \\
${}^{37}$University of Birmingham, Birmingham B15 2TT, United Kingdom \\
${}^{38}$Department of Physics and Astronomy, Vanderbilt University, Nashville, TN, USA \\
${}^{39}$SUPA, University of Glasgow, Glasgow G12 8QQ, United Kingdom \\
${}^{40}$Cardiff University, Cardiff CF24 3AA, United Kingdom \\
${}^{41}$University of Florida, Gainesville, FL 32611, USA \\
${}^{42}$Universit\"{a}t Hamburg, D-22761 Hamburg, Germany \\
${}^{43}$Vrije Universiteit Amsterdam, 1081 HV Amsterdam, Netherlands \\
${}^{44}$National Central University, Taoyuan City 320317, Taiwan \\
${}^{45}$Embry-Riddle Aeronautical University, Prescott, AZ 86301, USA \\
${}^{46}$OzGrav, Australian National University, Canberra, Australian Capital Territory 0200, Australia \\
${}^{47}$Inter-University Centre for Astronomy and Astrophysics, Pune 411007, India \\
${}^{48}$University of Tokyo, Tokyo, 113-0033, Japan. \\
${}^{49}$Sungkyunkwan University, Seoul 03063, Republic of Korea \\
${}^{50}$The University of Texas Rio Grande Valley, Brownsville, TX 78520, USA \\
${}^{51}$University of Portsmouth, Portsmouth, PO1 3FX, United Kingdom \\
${}^{52}$Montclair State University, Montclair, NJ 07043, USA \\
${}^{53}$National Institute of Technology Calicut, Kozhikode, Kerala, 673601, India \\
${}^{54}$SETI Institute, Mountain View, CA 94043, USA \\
${}^{55}$University of California, Berkeley, CA 94720, USA}
\eads{$^*$\mailto{jane.glanzer@ligo.org}, $^{**}$\mailto{adrian.helmling-cornell@ligo.org}}

\begin{abstract}
LIGO detector characterization efforts enabled the confident detection of gravitational waves from hundreds of compact binary coalescences during the fourth observing run.
Reliable production of high quality detector data and rapid noise mitigation efforts allow the extraction of the most in-depth knowledge of gravitational wave sources and their progenitors.
In this paper we describe LIGO detector characterization activities during the second and third parts of O4---O4b and O4c.
We summarize changes in detector configuration and performance at the LIGO Hanford and LIGO Livingston Observatories between the end of the first part of O4a and the end of O4c, including upgrades made during the commissioning break preceding O4b and during repairs performed in O4c.
We describe instrumental investigations carried out at both sites designed to understand and subsequently mitigate the effect on detector sensitivity of transient glitches, narrowband spectral lines, and vibration-driven noise, among other data quality concerns.
We then review the tools and procedures used to validate gravitational wave candidates and the data quality products thus supplied to searches for gravitational waves from compact binary coalescences and unmodeled transients, continuous gravitational waves, and the stochastic gravitational wave background.
The efforts of the detector characterization group are essential for maintaining and improving the sensitivity and reliability of the LIGO detectors especially as observing runs lengthen and more events are detected.
We conclude with prospects for LIGO detector characterization activities in future observing runs.


\maketitle
\end{abstract}

\section{Introduction}
\label{sec:intro}
The \ac{O4} of the \ac{LVK} collaboration was the longest and most sensitive ground-based gravitational wave observing campaign to date.
Since the first direct detection of \acp{GW} from the merger of two black holes on September 14th, 2015 by the \ac{LHO} and \ac{LLO}~\cite{2016PhRvD..93l2003A}, the catalog of compact binary coalescences has grown with each successive observing run of the \ac{LIGO}~\cite{2019PhRvX...9c1040A,2021PhRvX..11b1053A,2023PhRvX..13d1039A,2025arXiv250818082T,2026arXiv260527225T}.

During \ac{O4}, the two \ac{LIGO} detectors operated as part of the \ac{LVK} network alongside the Virgo detector~\cite{2015CQGra..32b4001A} and the KAGRA detector~\cite{2019NatAs...3...35K}. The \ac{O4} observing run was divided into three periods.
The first period, \ac{O4}a, spanned from May 24, 2023, 15:00:00~UTC to January 16, 2024, 16:00:00~UTC.
\ac{LIGO} detector characterization efforts during O4a are described in~\cite{2025CQGra..42h5016S}. The corresponding efforts for Virgo~\cite{2023CQGra..40r5006A} and KAGRA~\cite{2021PTEP.2021eA102A} are reported separately.
Following a commissioning break of approximately three months for the \ac{LIGO} detectors, the second part of the fourth observing run, O4b, began on April 10, 2024 15:00:00~UTC.
By the end of O4b, a total of 390 significant detection candidates had been identified across all observing runs~\cite{2026arXiv260527225T}.
The end of O4b on January 28, 2025 17:00:00~UTC also marked the start of the third part of \ac{O4}, O4c.
During O4c, the \ac{LIGO} detectors underwent a period of commissioning during which no astronomical data was taken from April 1, 2025 15:00:00~UTC to June 11, 2025 15:00:00~UTC, to make hardware changes to the vacuum system.
O4c concluded on November 18, 2025 16:00:00~UTC.

\begin{figure}
\includegraphics[width=\textwidth]{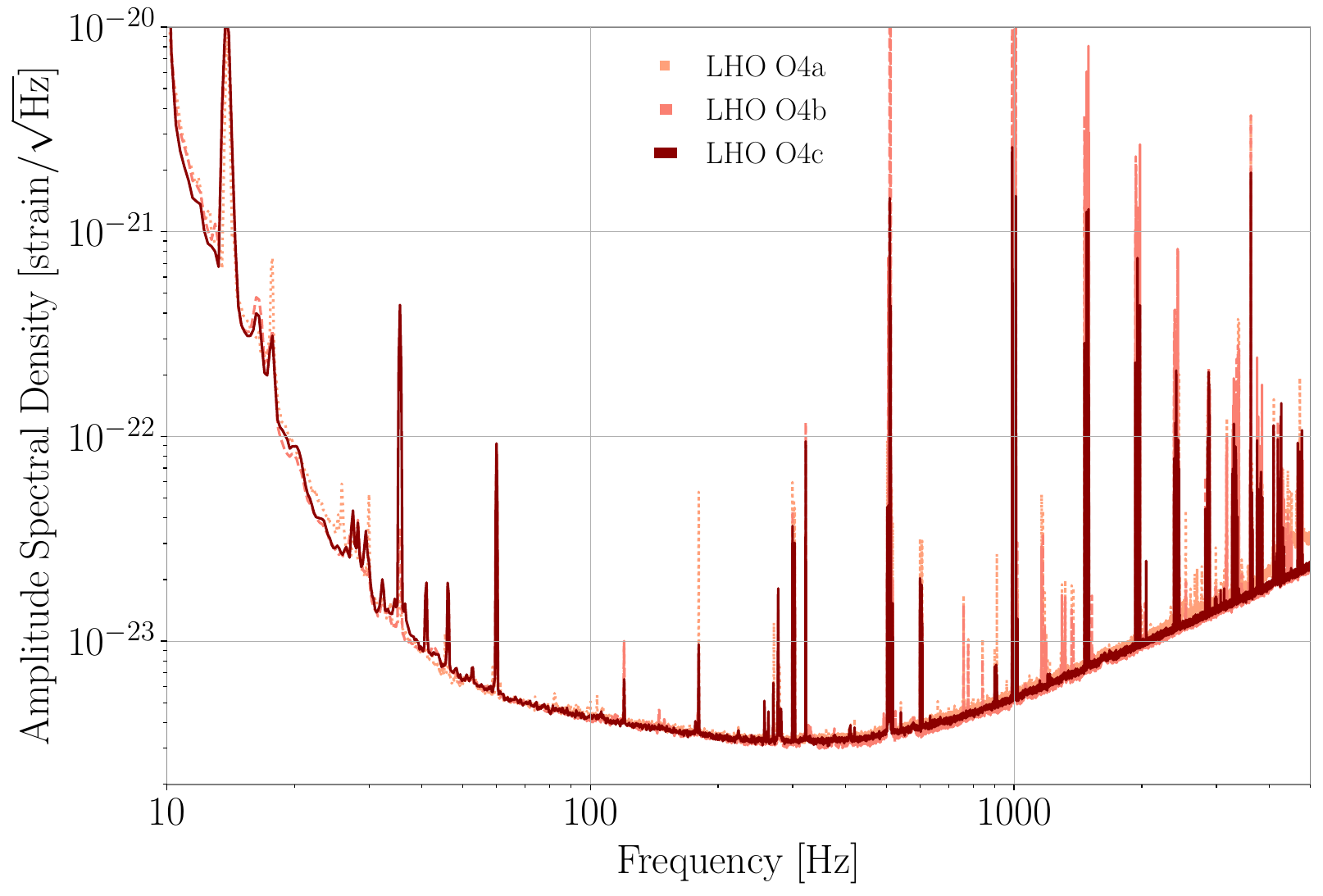}
\caption{\label{fig:H1_O4_asd} Representative \acp{ASD} of the \ac{GW} strain noise at \ac{LHO} during \ac{O4}a, \ac{O4}b, and \ac{O4}c. The broadband sensitivity is comparable across the three epochs, with slight improvements in \ac{O4}b and \ac{O4}c relative to \ac{O4}a in the $20$--$30~\mathrm{Hz}$ region. The population of narrowband spectral lines and combs varies noticeably between periods, particularly above $500~\mathrm{Hz}$. These features and their variations are discussed further in Section~\ref{sec:lines_and_combs} and in~\cite{2026arXiv260605959G}. Each \ac{ASD} is estimated from a single representative $1$-hour period of observing-mode data using the median-averaged Welch method with $8~\mathrm{s}$, $50\%$-overlapping segments, giving a frequency resolution of $0.125~\mathrm{Hz}$ and $899$ averages. The measurements start at 2023-10-24 03:13:42~UTC (\ac{O4}a), 2024-11-16 08:23:42~UTC (\ac{O4}b), and 2025-01-29 09:00:42~UTC (\ac{O4}c).}
\end{figure}

\begin{figure}
\includegraphics[width=\textwidth]{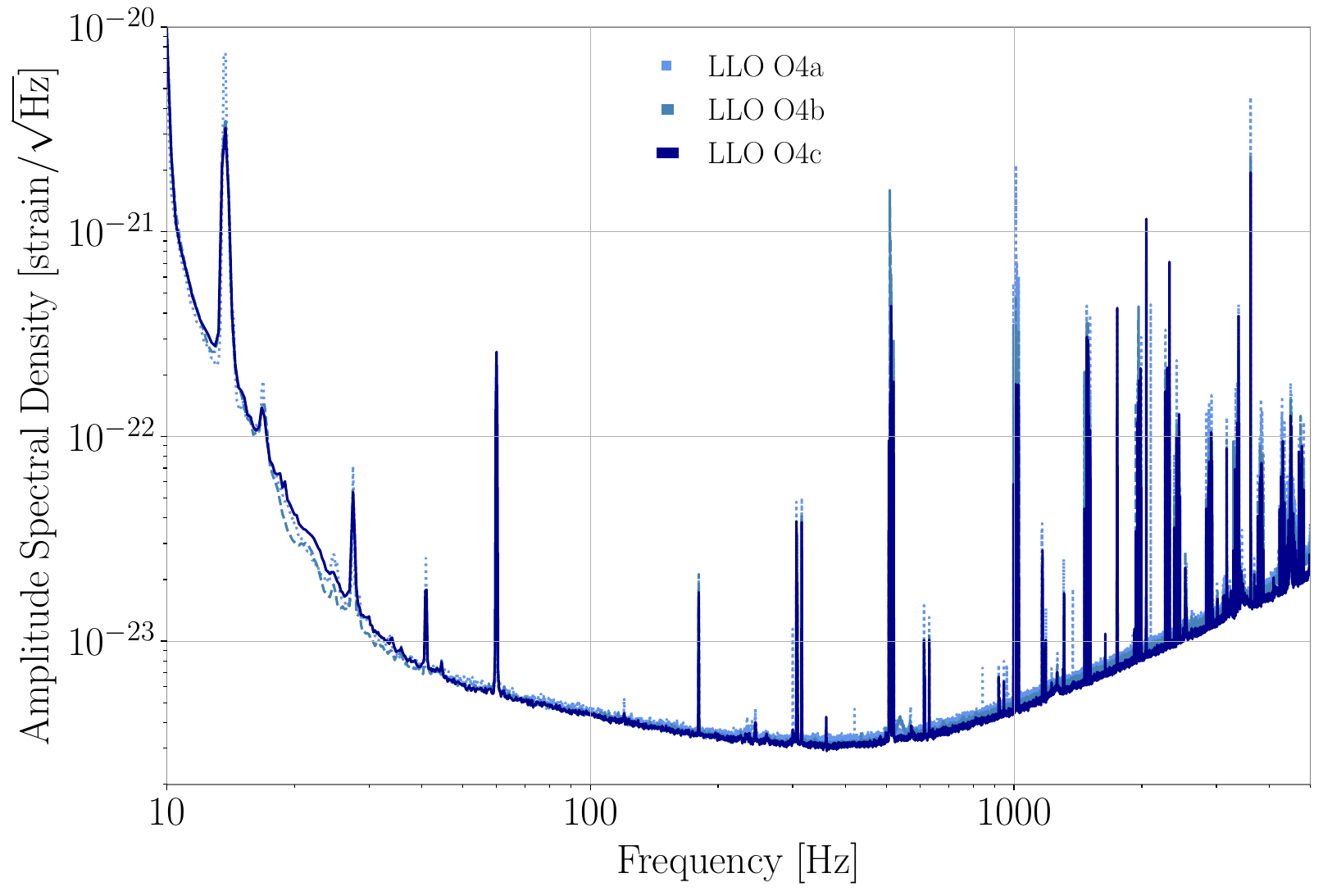}
\caption{\label{fig:L1_O4_asd} Representative \ac{ASD}s of the strain noise at \ac{LLO} during \ac{O4}a, \ac{O4}b, and \ac{O4}c. The sensitivity improved slightly between \ac{O4}a and \ac{O4}b, while \ac{O4}c shows a mild degradation relative to \ac{O4}b in the $20$--$40~\mathrm{Hz}$ band. The \acp{ASD} are estimated as described in Fig.~\ref{fig:H1_O4_asd}, from $1$-hour periods starting at 2023-11-25 07:27:42~UTC (\ac{O4}a), 2024-08-19 08:42:42~UTC (\ac{O4}b), and 2025-03-22 05:23:42~UTC (\ac{O4}c).}
\end{figure}

Figures~\ref{fig:H1_O4_asd} and~\ref{fig:L1_O4_asd} show representative \acp{ASD} of the \ac{GW} strain noise at \ac{LHO} and \ac{LLO}, respectively, across the three parts of \ac{O4}.
A comparison between the strain noise observed in \ac{O4}a at each \ac{LIGO} detector and the sum of noises predicted from physical principles or quantified via direct measurement—the \textit{noise budget}—may be found in~\cite{2025PhRvD.111f2002C}.
The \ac{GW} strain, $h = \Delta L / L$, is defined as the change in the \ac{DARM}, $\Delta L$, relative to the nominal arm length $L$, and is the primary data product of each detector from which \ac{GW} signals are extracted.
Along with \acp{ASD}, a common measure of detector sensitivity is the \ac{BNS} range, defined as the sky and orientation averaged distance at which a $1.4\,M_\odot$--$1.4\,M_\odot$ compact binary merger with a \ac{SNR} of 8 could be detected~\cite{1993PhRvD..47.2198F,2025PhRvD.111l2005D}. The \ac{BNS} range provides a convenient measure for quickly assessing detector performance.
A detailed comparison of detector performance across O4b and O4c, including the pre- and post-commissioning period of O4c is presented in Section 2.

The \ac{LIGO} detector characterization (DetChar) group is tasked with monitoring the state of the detectors, identifying and mitigating sources of noise, and providing data quality information to the astrophysical analyses that search for \ac{GW} signals.
This work includes on-site investigations of instrumental and environmental couplings, development and maintenance of software tools for characterizing the data, validation of candidate events, and the production of data quality products consumed by downstream search pipelines.

In general, the data from the \ac{LIGO} detectors is typically Gaussian and stationary.
However, noise which couples to the detector through numerous mechanisms introduces non-Gaussian features that need to be characterized and, where possible, mitigated~\cite{2021CQGra..38m5014D,2018RSPTA.37670286N,2016CQGra..33m4001A}.
These features are commonly divided into three categories: short-duration transient artifacts (often called \textit{glitches}), narrowband spectral lines and combs, and persistent broadband artifacts. 
Glitches are typically categorized by their morphology in time-frequency spectrograms~\cite{2017CQGra..34f4003Z,2023CQGra..40f5004G}.
Narrowband spectral lines are persistent artifacts that appear as peaks in the \ac{ASD} of the strain data that can often be traced back to mechanical resonances or electronics noise~\cite{2026arXiv260605959G}. A comb is a set of narrowband lines that are uniformly spaced in frequency~\cite{2026arXiv260605959G}.
Persistent broadband artifacts elevate the noise floor across a range of frequencies.
Each of these different types of noise impacts the \ac{GW} searches in distinct ways.
Glitches can impact searches for short-duration signals, including \ac{CBC} and unmodeled burst searches, where they can produce false triggers or reduce the significance of a genuine astrophysical candidate. Beyond affecting detection, a glitch that overlaps a signal can bias the estimation of source properties and, in turn, the downstream astrophysical inferences that rely on them~\cite{2022PhRvD.106j4017P,2026arXiv260407668L,2026PhRvD.113d2005U}, motivating the event validation and noise mitigation procedures described in Section 4.
Narrowband spectral lines and combs limit the sensitivity of searches for \ac{CW} signals from spinning neutron stars and may introduce cross-correlated noise, affecting searches for the stochastic \ac{GW} background.
The presence of persistent broadband artifacts affects \ac{LIGO}'s searches for both \ac{GW} transients and \ac{CW} and stochastic \ac{GW} sources by degrading detector sensitivity across large swathes of frequency space.

The scope of this paper encompasses the \ac{LIGO} commissioning and \detchar activities carried out during O4b and O4c.
Section~2 summarizes changes to the detector configurations between the end of O4a and the end of O4c and the resulting effects on metrics such as detector duty cycle.
The instrumental investigations carried out at both observatories to identify and mitigate sources of environmental noise during O4b and O4c are described in Section~3.
Section~4 details the performance tools used to validate \ac{GW} event candidates based on data quality surrounding the event time.
The use of data quality products in searches for \acp{GW} of different kinds (compact binary coalescences, unmodeled bursts, continuous waves, and stochastic signals) is discussed in Section~5.
We conclude in Section~6 with a summary of our results and outlooks for future \detchar work.

\section{Detector Configuration \& Performance in O4b \& O4c}
\label{sec:config}
\begin{figure}
\includegraphics[width=\textwidth]{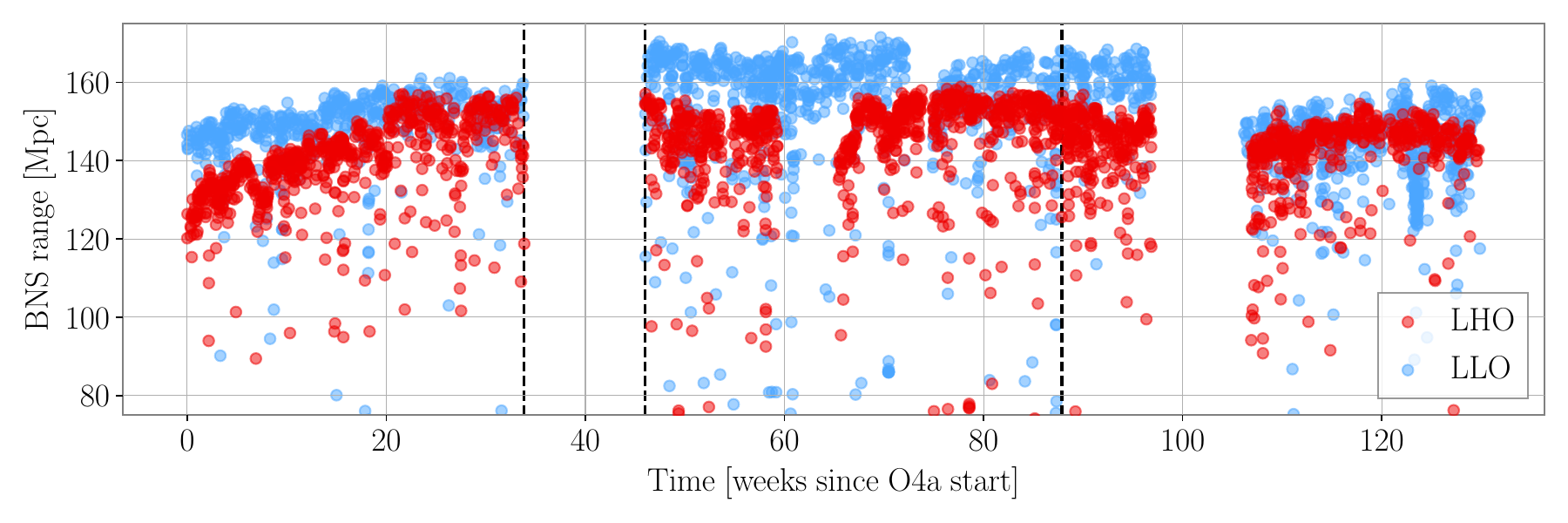}
\caption{\label{fig:bnsrange} A comparison of the \ac{BNS} range of the \ac{LIGO} detectors in each part of \ac{O4}. Dashed vertical lines indicate divisions between parts of the observing run. The large gaps correspond to commissioning periods where maintenance and upgrades were performed on the detectors. No astrophysical data were taken during these commissioning periods.}
\end{figure}

\begin{figure}
\includegraphics[width=\textwidth]{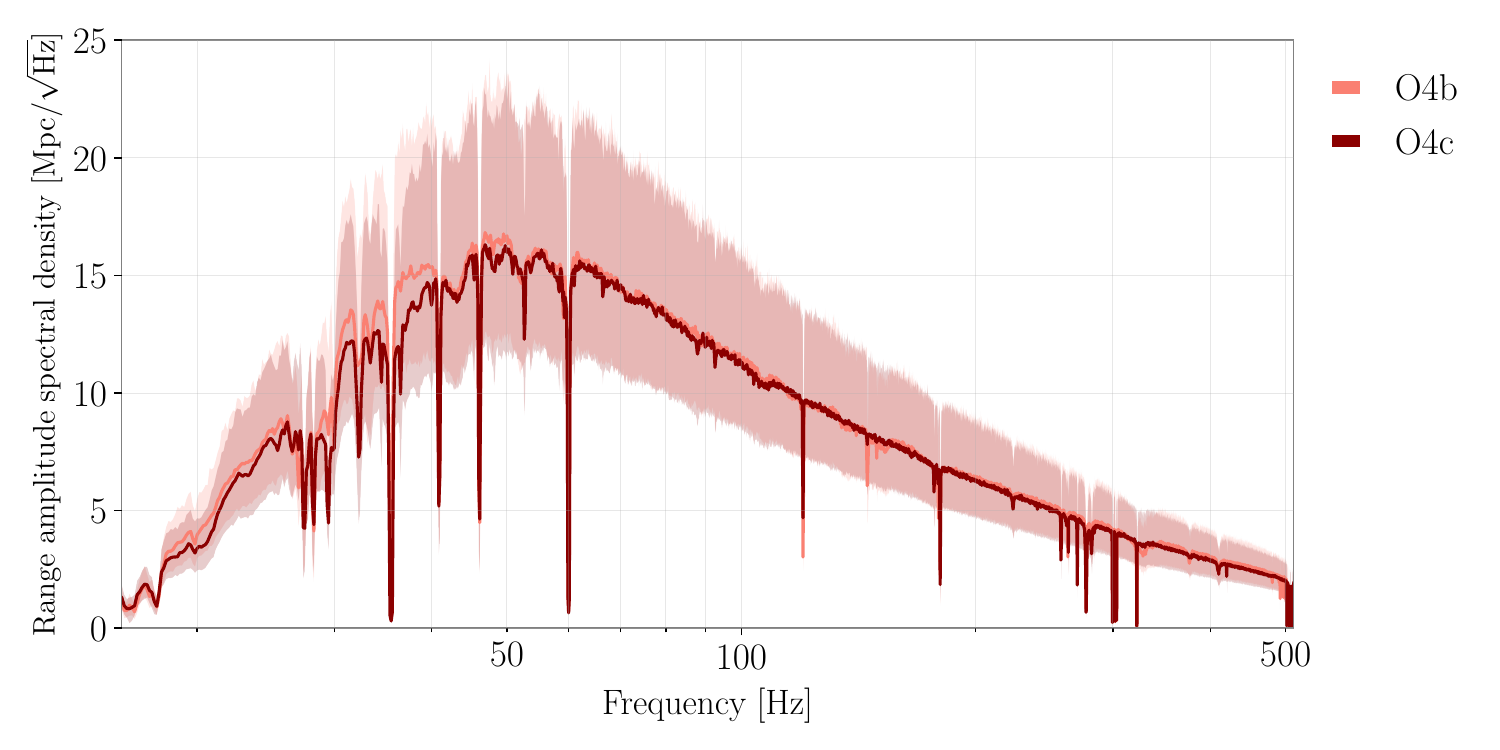}
\caption{\label{fig:h1rangeasd} \ac{BNS} range \ac{ASD} for \ac{LHO} during O4b (salmon) and O4c (maroon). Solid lines show the median over a 24-hour period of observing time; shaded bands show the spread across short segments over the day (5th--95th percentile). Narrow dips correspond to instrumental and environmental spectral lines. The presence of these noise artifacts limit the astrophysical reach of the detector in that frequency regime.}
\end{figure}

\begin{figure}
\includegraphics[width=\textwidth]{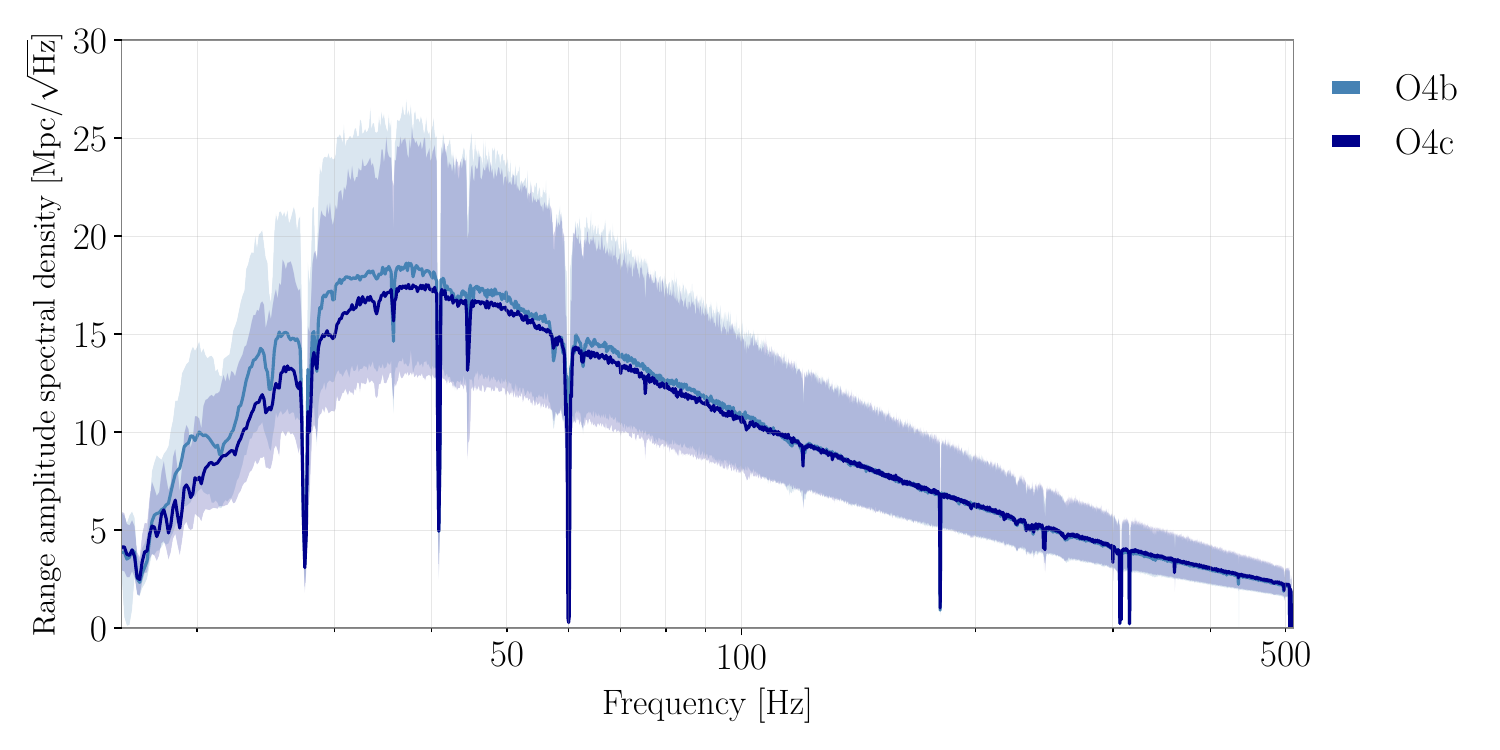}
\caption{\label{fig:l1rangeasd} \ac{BNS} range \ac{ASD} for \ac{LLO} during O4b (slate blue) and O4c (navy blue), following the plotting style of figure~\ref{fig:h1rangeasd}.}
\end{figure}

Between the end of \ac{O4}a and the start of \ac{O4}b, both \ac{LIGO} detectors underwent a commissioning break during which a number of instrumental upgrades were made.
Additional changes were implemented during the commissioning break in the middle of \ac{O4}c (April 1--June 11 2025).
This section describes those configuration changes and their impact on detector performance and sensitivity.
The \ac{BNS} range over the full duration of \ac{O4} is shown in Figure~\ref{fig:bnsrange}, and figures~\ref{fig:h1rangeasd} and~\ref{fig:l1rangeasd} show the contribution of each frequency bin to the \ac{BNS} range \ac{ASD} at both detectors across O4b and O4c.
Overall, the \ac{BNS} range was slightly greater in the second part of \ac{O4} compared to the other parts of the observing run.
Both detectors failed to return to their prior sensitivity following the mid-\ac{O4}c commissioning break; \ac{LLO}'s typical \ac{BNS} range in \ac{O4}c was less than its range in \ac{O4}a.
A comparison of the detector \ac{BNS} range by observing run part is given in Table~\ref{tab:bnsoft}.
The maximum \ac{BNS} range achieved by any \ac{LIGO} detector was 171.5~Mpc, recorded by \ac{LLO} during \ac{O4}b on September 23, 2024.
\ac{LHO}'s maximum \ac{BNS} range, also achieved during \ac{O4}b, was 158.9~Mpc, which was realized on November 18, 2024.

\begin{table}
    \begin{tabular}{c c c c}
        \br
        & O4a & O4b & O4c \\
        \mr
        LHO & 139.4 & 149.5 & 145.2 \\
        LLO & 149.4 & 159.4 & 146.8 \\
        \br
    \end{tabular}
    \caption{\label{tab:bnsoft} Comparison of the median \ac{BNS} range (in~Mpc) for the two \ac{LIGO} detectors in each part of \ac{O4}.}
\end{table}

\subsection{Hardware injections}
A critical component of \ac{GW} candidate validation is ensuring that detected events are not caused by environmental or instrumental noise sources recorded in auxiliary channels.
Several tasks described in \sref{sec:val} rely on auxiliary channel information to determine whether or not a \ac{GW} candidate is astrophysical in origin.
Since the transfer functions between the \ac{GW} strain channel and thousands of auxiliary channels are not known, there is a possibility that real astrophysical signals can sometimes appear in certain auxiliary channels due to unknown couplings between auxiliary channels and elements of the signal chain used to produce calibrated strain data.
Such auxiliary channels are called ``unsafe" and should not be used in the development of the vetoes described in section~\ref{sec:dq} since they can effectively allow a real \ac{GW} event signal to be vetoed.
To determine whether a channel responds to change in power in $h(t)$ and thus determine its ``safety," the detector characterization group performs \detchar safety injections.
These injections are done using photon calibrator to inject sine-Gaussian waveforms into the $h(t)$ channels at each detector~\cite{2016RScI...87k4503K}.
A channel is considered safe only if it shows no response to these injected signals. 
We then perform a statistical analysis of approximately 5,000 auxiliary channels sampled above 16~Hz using the pointy-poisson tool~\cite{2021PhRvD.103d2003E}, classifying each channel as either ``safe" (suitable for vetoing candidate \ac{GW} events) or ``unsafe."
The final list of safe channels is subsequently used in downstream data-quality analyses.
\detchar hardware injections and the subsequent analysis were performed routinely during O4b at both detectors.
No significant changes were observed in the safety of auxiliary channels during this time.

\subsection{Commissioning changes at both detectors}

\begin{figure}
\includegraphics[width=\textwidth]{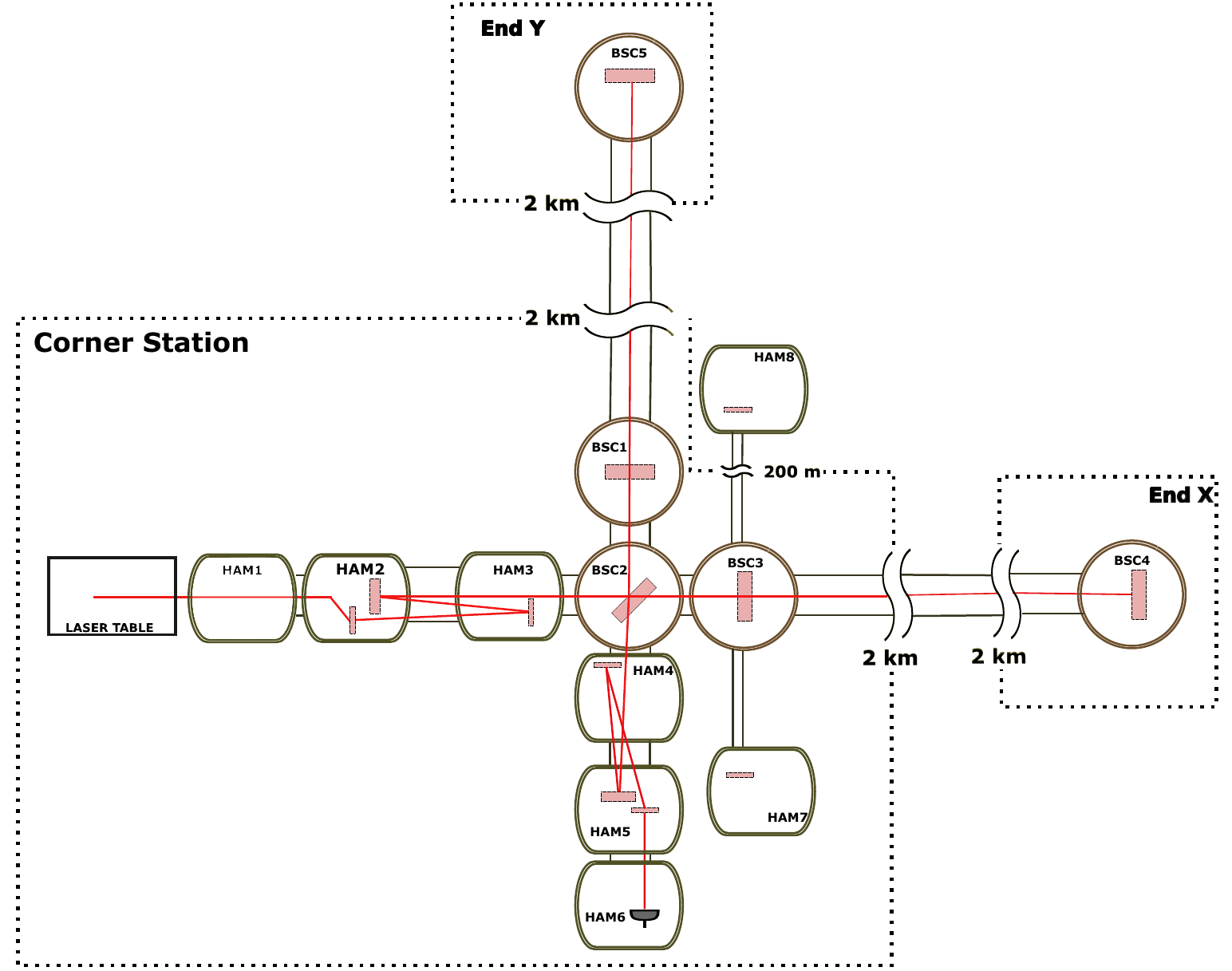}
\caption{\label{fig:map} Schematic of \ac{LLO} vacuum chambers adapted from~\cite{PEMpage}. The path of the main laser (red) off core optics (pink) is also shown. Except for the laser table, the entire system is kept under vacuum. Also shown is the filter cavity used to realize frequency-dependent squeezing~\cite{2023PhRvX..13d1021G}. Each vacuum chamber is labeled with the name used throughout this paper, though not all are discussed in this work. Dashed lines indicate which buildings house which vacuum chamber. Not depicted here are the midstations, located halfway down the length of each \ac{LIGO} arm, as no vacuum chambers besides a segment of the beamtube are housed in these outbuildings, nor the filter cavity end station, as the chamber it contains was not the subject of any noise investigations discussed in this work.}
\end{figure}

As \ac{O4} was the longest observing run---to date---for the \ac{LIGO} detectors, many changes, repairs, and improvements were made to both \ac{LLO} and \ac{LHO} during the course of the observing run.
These adjustments frequently affected the duty cycle and environmental coupling of detectors. In this section, we catalog these changes, beginning with those made to both interferometers.
Figure~\ref{fig:map} depicts \ac{LLO} with individual vacuum chambers labeled and shows their corresponding buildings.

\subsubsection{EQ mode}
Earthquakes and periods of heightened microseismic noise affect the duty cycle and sensitivity of the \ac{LIGO} detectors.
An automated global control method called ``EQ mode" was implemented in the \ac{O3} to enhance the ability of the detector to maintain lock during earthquakes which occur more than 1000~km away from the sites~\cite{2020CQGra..37w5007S}.
In \ac{O4}c, a new feature was added to EQ mode which switched arm alignment sensing and control loops to a high-bandwidth state during earthquakes~\cite{alog:85740}.
This change in EQ mode controls increased the fraction of earthquakes for which the detectors were able to maintain lock, even at higher ground motion~\cite{alog:78473}.
Table~\ref{tab:EQ_table} shows a slight improvement in the fraction of earthquakes for which \ac{LLO} maintained lock in \ac{O4}c compared with \ac{O4}b.
Further study ahead of Intermediate Run 1, which will begin in late 2026, is ongoing to determine if the new EQ mode can be further optimized to improve lock survival during earthquakes.
Maintaining lock---even though observation-quality data is not taken during EQ mode times---during periods of high ground motion increases the total duty cycle of the detector, as the relocking procedure can take tens of minutes to hours depending on environmental conditions.

\begin{table}
    \begin{adjustbox}{width=\columnwidth,center}
    \begin{tabular}{c c c c c c}
        \br
        \textbf{Detector} & \textbf{Observation Run} & \textbf{Total Earthquakes} & \textbf{Survive Lock} & \textbf{Broke Lock} & \textbf{Lock Probability (\%)} \\
        \mr
        
        \multirow{2}{*}{LLO} 
            & O4b & 176 & 53  & 123 & 30.1\% \\ \cline{2-6}
            & O4c & 234 & 71  & 163 & 30.3\% \\ 
        \mr
        \multirow{2}{*}{LHO} 
            & O4b & 193 & 102 & 91  & 52.8\% \\ \cline{2-6}
            & O4c & 203 & 112 & 91  & 55.2\% \\ 
        \br
    \end{tabular}
    \end{adjustbox}
    \caption{Performance of \ac{LIGO} detectors during earthquakes in the second and third parts of \ac{O4}. Total Earthquakes refer to all earthquakes for which the detector was locked before the earthquake's arrival. Earthquakes were identified, following the approach in~\cite{2020CQGra..37w5007S}, by coincident peaks in vertical-direction ground motion in the 30--100~mHz band in two or more ground motion witnesses in different buildings at \ac{LHO} or \ac{LLO}.}
        \label{tab:EQ_table}
\end{table}

\subsubsection{HAM-1 ISI install}\label{sssec:ham1isi}

\begin{figure}
\includegraphics[width=\textwidth]{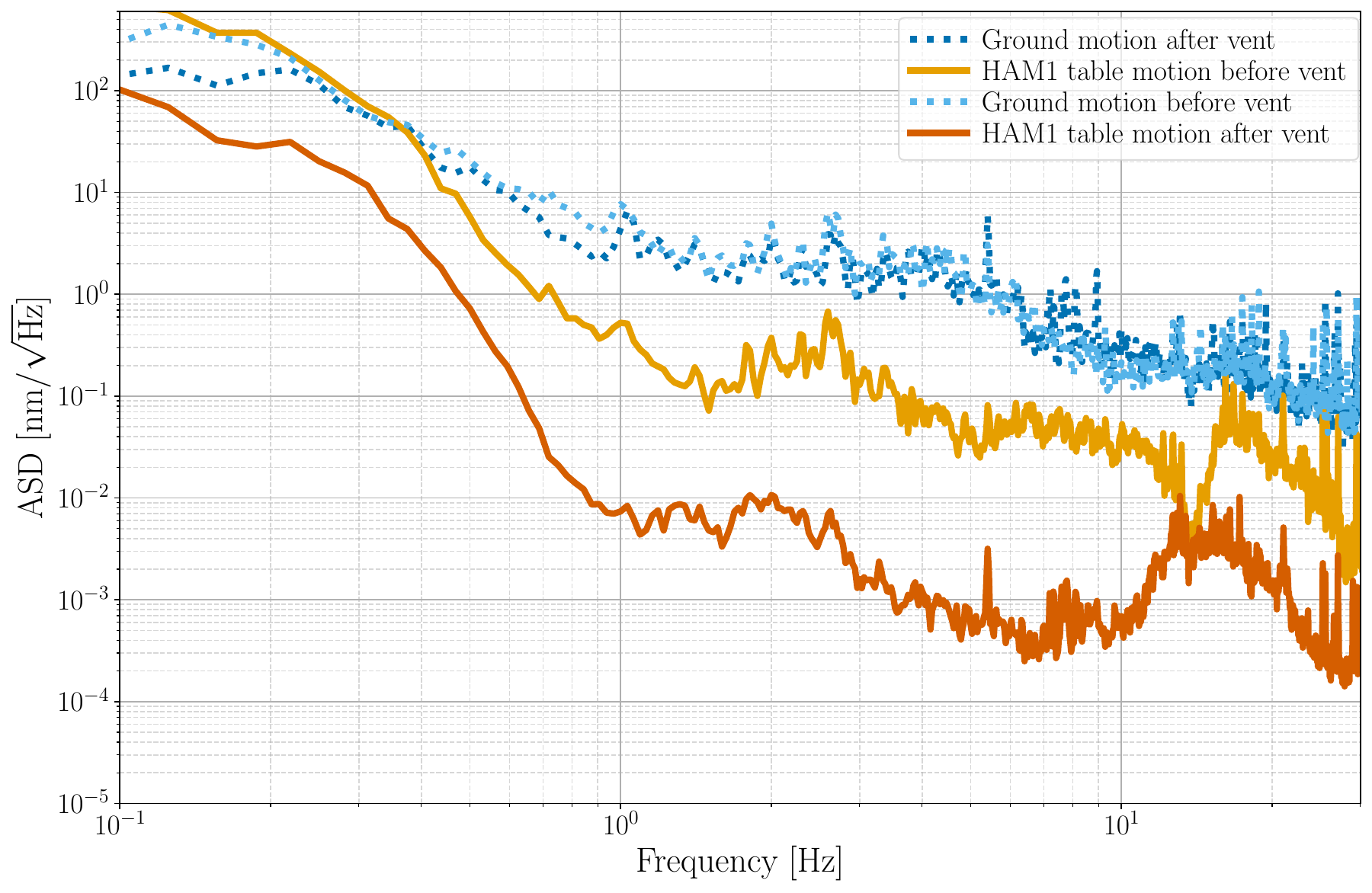}
\caption{\label{fig:Ham1improvment}  Vibration of the \ac{HAM}~1 table-top before (yellow) and after (orange) the \ac{ISI} was installed at \ac{LHO} at comperable levels of ground motion.}
\end{figure}

As of the start of \ac{O4}, both \ac{LLO} and \ac{LHO} contained one passively-damped in-vacuum table which was susceptible to excess noise contamination, especially in the 10--30 Hz band~\cite{2025PhRvD.111f2002C}.
This table, located in \ac{HAM}~1, houses interferometric sensors and single-suspension optics used to read out length and alignment degrees of freedom from the reflection and power-recycling cavity pick-off ports of the interferometer.
Noise from excess motion of this table can couple to the gravitational-wave readout through the alignment controls coupling~\cite{2025PhRvD.111f2002C}.
For the first two parts of O4, both detectors employed feedforward subtraction using on-table seismic sensors to suppress the noise from this table~\cite{2025PhRvD.111f2002C}.
Starting April 1, 2025, this passively-damped table was replaced at both sites with a \ac{HAM} \ac{ISI} table, which uses active seismic isolation, and has much better isolation performance than passive isolation~\cite{2015CQGra..32r5003M}.
The improved seismic isolation reduced the table motion by more than a factor of 10 above 1~Hz relative to the performance of the passively-damped table.
Figure~\ref{fig:Ham1improvment} shows the reduction in ground motion coupling of the \ac{HAM}~1 optics table improvement after active seismic isonlation was realized.

As a part of this install, an alignment sensor that was previously installed in-air was moved onto \ac{HAM}~1 at both sites.
This installion included a new single-stage suspension for beam steering onto the sensor.
The alignment control loops that control the power-recycling cavity alignment degree of freedom were recommissioned to use this in-vacuum sensor, which improves stability of the alignment controls during high power operation.
Moving this sensor onto an isolated platform yields less risk of noise contamination from backscattering or acoustic noise.

\subsubsection{Recalibration of Vault Magnetometers}

At each site, two LEMI fluxgate magnetometers are buried in vaults outside the detector buildings, with one unit oriented parallel to the X-arm and one parallel to the Y-arm~\cite{2021CQGra..38n5001N}.
They are located away from local sources of electromagnetic interference, which allows them to measure the ambient magnetic field.
This includes contributions from Schumann resonances, which enter the magnetic noise budget for \ac{SGWB} searches~\cite{2025PhRvD.112j2003J,2025PhRvD.111h2005V,2025arXiv250820721T}.

The LEMI magnetometers were initially calibrated in 2018 and 2019 prior to the start of \ac{O3}.
Recalibration campaigns were conducted at both sites during \ac{O4}, in May 2025 at \ac{LHO}~\cite{alog:84924} and October 2025 at \ac{LLO}~\cite{alog:79137}.
To calibrate, wire coils were driven at a known voltage and frequency to generate a magnetic field of known amplitude at the sensor location while a co-located pre-calibrated Bartington magnetometer served as a reference.
The LEMI calibration factor was then derived from the ratio of the signals recorded by the two sensors.
The updated calibration factors are approximately $20\%$ lower at \ac{LHO} and $70$--$75\%$ lower at \ac{LLO} compared to the pre-\ac{O3} values.
At both sites, the discrepancy is attributed to the previous calibration campaigns not having the reference magnetometer fully isolated from the LEMI unit; in the present campaigns, the two sensors were kept completely independent.

\subsection{Changes to LIGO Livingston}
\ac{LLO} generally operated with a larger \ac{BNS} range and duty cycle compared to \ac{LHO} in \ac{O4}b and \ac{O4}c.
While a commissioning break, mainly undertaken to perform preventative maintenance to critical vacuum system components resulted in the largest gap in \ac{O4} observing time, this time was also used to reduce sources of scattered light in the detector.

\subsubsection{Gate Valve 11 Repair}
The \ac{LIGO} detectors use \acp{GV} to separate portions of the evacuated detector beamtube and groups of \acp{BSC} and \acp{HAM} housing in-vacuum detector optics.
A \ac{GV} at \ac{LLO} required repair so it could be closed during periods of severe weather in the 2025 North Atlantic hurricane season.
A mid-\ac{O4}c gap spanning April 1, 2025-June 4, 2025 was used to complete this work, which restored the \ac{GV} to normal operation.
This gap in observing time can be seen in Figure~\ref{fig:bnsrange}.

\subsubsection{Cage Baffles}
\label{subsubsection:cagebaffles}
During the \ac{GV} repair undertaken in \ac{O4}c, a new set of baffles was installed at \ac{LLO} near each test mass mirror.
These \textit{cage baffles} are attached to the suspension system which holds the quadruple suspension of the test masses.
Positioned just in front of the mirrors, they are designed to block some of the light scattering from near the edge of each mirror, where the reflective coating does not extend all the way to the edge of the optic.
The beam from the opposite test mass, $4$~km away, has expanded by the time it reaches the end mirror, such that a small fraction of light falls at this diameter near the edge of the coating, making it a source of scattered light.
The cage baffles were installed primarily to address the scattered light noise that limited \ac{LLO} sensitivity during periods of elevated microseismic ground motion. Their resulting effect is described in Section~\ref{subsubsection:High SNR glitches}.

\subsubsection{L2 Split Drive}
In early October 2025, \ac{LLO} suffered repeated lock losses, reducing the duty cycle during this period.
The control system did not have enough range to hold the detector steady aginst transient low-frequency motion.
To fix this, the \ac{DARM} actuation was shared across both end test mass suspensions rather than a single one, which roughly doubled the available range and improved stability during lock aquisition~\cite{alog:78903}.
Driving the additional suspension excited one of the \ac{ETMY} mechanical resonances at 13.7 Hz, which was suppressed with a notch filter~\cite{alog:78942}.

\subsection{Changes to LIGO Hanford}
\ac{LHO} commissioning activities during the second and third parts of \ac{O4} reduced broadband quantum noise sources through improved squeezing performance.
However, degradation of critical equipment over the course of the observing run reduced the detector's duty cycle.

\subsubsection{Squeezer Performance in O4b and O4c}
\begin{figure}
\includegraphics[width=\textwidth]{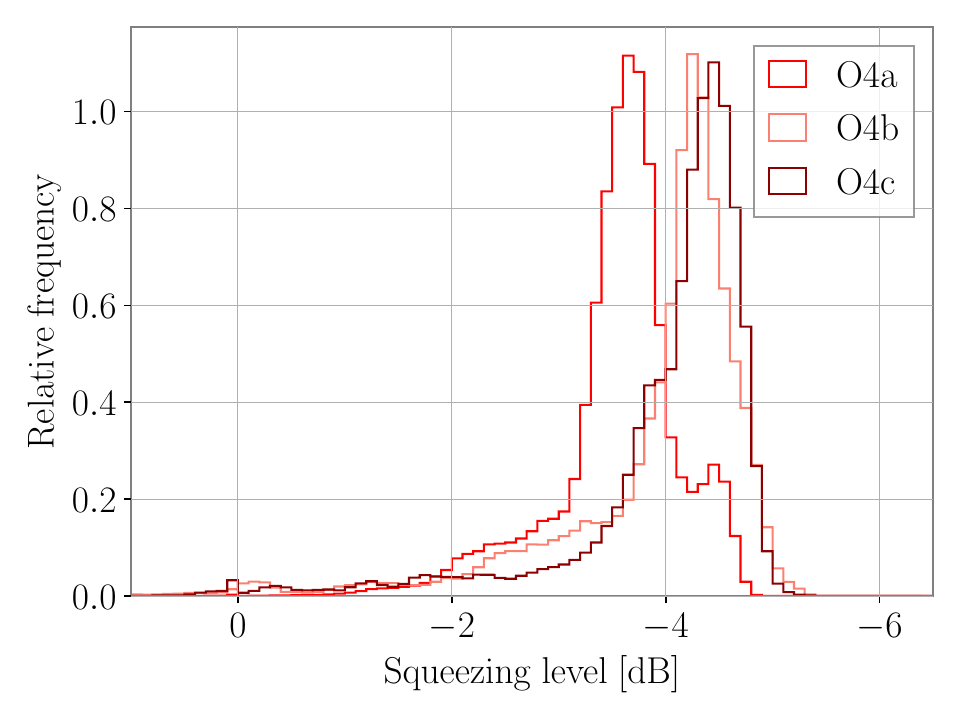}
\caption{\label{fig:lhosqzperf} Quantum noise reduction from \ac{LHO}'s squeezed light system in \ac{O4}a (red), \ac{O4}b (salmon), and \ac{O4}c (maroon). The data shown here are measured around 1.7~kHz. In general, the performance of the squeezer at high frequencies was substantially improved following \ac{O4}a, while lower frequencies saw more modest changes.}
\end{figure}

Prior to \ac{O4}a, the \ac{LIGO} detectors were reconfigured to use frequency-dependent squeezing---rather than frequency-independent squeezing, used in \ac{O3}---to simultaneously reduce radiation pressure noise (the dominant quantum noise source at low frequencies) and shot noise (the dominant quantum noise source at high frequencies)~\cite{2023PhRvX..13d1021G}.
A detailed history of detector configuration changes affecting the performance of quantum squeezing in \ac{O4}a is found in~\cite{2025PhRvD.111f2002C}.
Continued commissioning work to improve the performance of the squeezer in \ac{O4}b led to percent-level improvement in the squeezing level near 135~Hz, with improvements of half of a~dB or more at higher frequencies in \ac{O4}b and \ac{O4}c compared to \ac{O4}a.
For the 1.7~kHz band shown in Figure~\ref{fig:lhosqzperf}, the squeezer achieved 4.5~dB or more of squeezing for 27\% (20\%) of observing time in \ac{O4}c (\ac{O4}b), compared to just 4\% in \ac{O4}a.

\subsubsection{Output Faraday Isolator Failure}
On April 23, 2024, it was observed that \ac{LHO}'s output arm optical gain slightly decreased which increased scattered light noise below $\sim80$~Hz~\cite{alog:77368}.
Commissioning efforts identified a change in the performance of the \ac{OFI} as the cause of the degradation in detector sensitivity and performance~\cite{alog:77392,alog:77427}.
The \ac{OFI} rejects stray light from re-entering the interferometer arms and is located on an optics table between the \ac{BS} and the output port of the detector.
The output arm optics were realigned to reduce noise, but this change severly affected the performance of the squeezed light system~\cite{alog:77400}.

Beginning on July 12, 2024, another large change in optics alignment settings was required to keep the detector locked~\cite{alog:79082,alog:79101}.
This change prompted a decision to suspend observatory operations to enter the detector and investigate the output arm in-vacuum equipment for problems.
From the beginning of \ac{O4}b to July 17, 2024, the duty cycle of \ac{LHO} was $54.2\%$, compared to its \ac{O4}a duty cycle of $67.5\%$, in part due to instability associated with the detector in its April--July alignment.

Following replacement of failed components in the \ac{OFI} polarizer~\cite{alog:79326} and wedge~\cite{alog:79363}, astrophysical data-taking resumed on August 24, 2024 with a \ac{BNS} range of $\sim140$~Mpc, a slight decrease from its typical pre-intervention \ac{BNS} range~\cite{alog:79690}.
By September 6, 2024, however, \ac{LHO} consistently returned to observing at $\sim160$~Mpc.
The impromptu break in astronomical data-taking and subsequent recovery can be seen in Figure~\ref{fig:bnsrange} beginning about $60$ weeks from the start of \ac{O4}a.

\subsection{Input laser system glitching}
\begin{figure}
\centering
\includegraphics[width=\textwidth]{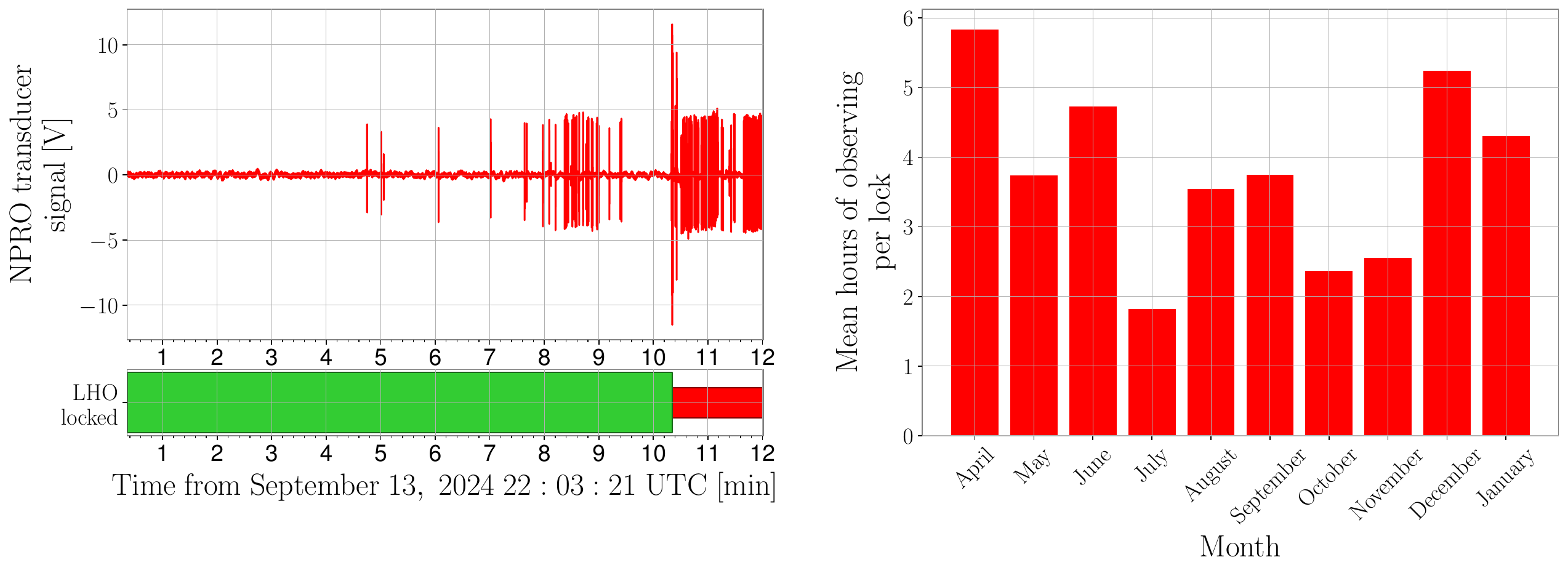}
         \caption{Left: voltage data from the \ac{FSS} subsystem preceding an \ac{NPRO} glitch which caused the detector to lose lock. Many smaller glitches are seen prior to the larger one which broke lock $\sim10$~min from the start of the depicted time. The observatory lock status is shown by the ribbon below the plot, with the thick green ribbon representing locked times, and the thinner red ribbon corresponding to unlocked times. Comparison of timeseries of many possible witness channels in different optical cavities around \ac{LHO} is what localized the glitching to the input laser system. Right: average observing-mode segment duration for each month of \ac{O4}b.}
         \label{fig:pztts}
\end{figure}

Beginning September 13, 2024, channels in the \ac{PSL}, which produces the 1064~nm light used in \ac{LIGO}'s arm cavities for astrophysical observation, started witnessing glitches.
The glitches occurred frequently, and were often sufficiently severe to cause the interferometer to lose lock~\cite{alog:81193}.
Through September and October of 2024, the duty cycle of the detector was reduced as these glitches caused lock to be broken frequently.
Analysis of locklosses with an associated glitch witnessed by \ac{FSS} monitors identified the \ac{NPRO}, the source of the 2~W in-air laser beam in the \ac{PSL}, as the ultimate origin of these glitches.
Between October 23 and 29, 2024, the \ac{NPRO} was replaced with a spare unit used in \ac{O3}~\cite{alog:80837}.
An example timeseries of one of the glitches in a channel monitoring the current sent to the \ac{NPRO} piezoelectric transducer, part of the \ac{FSS} subsystem, is shown in Figure~\ref{fig:pztts}.

The newly-installed, \ac{O3}-era \ac{NPRO} also suffered from glitches~\cite{alog:80908,alog:81386} and needed crystal temperature adjustments to prevent mode hopping~\cite{alog:81107,alog:81193}.
The glitches produced by the \ac{O3}-era \ac{NPRO} had a different morphology in \ac{FSS} data compared to the behavior shown in figure~\ref{fig:pztts}. 
Consequently, this \ac{NPRO} unit was also replaced, this time with the \ac{NPRO} used in \ac{O1} and the \ac{O2}~\cite{alog:81409}.
This unit had been refurbished by the manufacturer prior to its re-installation.
The installation of the second spare \ac{NPRO} on November 22, 2024 ended the frequent, lock-breaking glitches experienced at \ac{LHO}. 

\section{Instrumental Investigations}
\label{sec:instinv}
\subsection{Transient Noise}
The strain data recorded by the LIGO detectors is affected by short-duration disturbances of non-astrophysical origin, commonly referred to as transient noise or glitches.
These events arise from a variety of instrumental and environmental sources and exhibit a wide range of time–frequency morphologies, enabling their classification into distinct glitch groups~\cite{2017CQGra..34f4003Z}.

Glitches are commonly identified using Omicron~\cite{2020SoftX..1200620R}, which detects excess power by applying a Q-transform~\cite{1991ASAJ...89..425B} to project the data into the time-frequency domain.
This process produces trigger-based representations characterized by central time, frequency, bandwidth, and \ac{SNR}, enabling statistical studies of glitch populations and supporting their classification.

\begin{figure}
    \centering
    \begin{subfigure}{0.49\columnwidth}
        \centering
        \includegraphics[width=\linewidth]{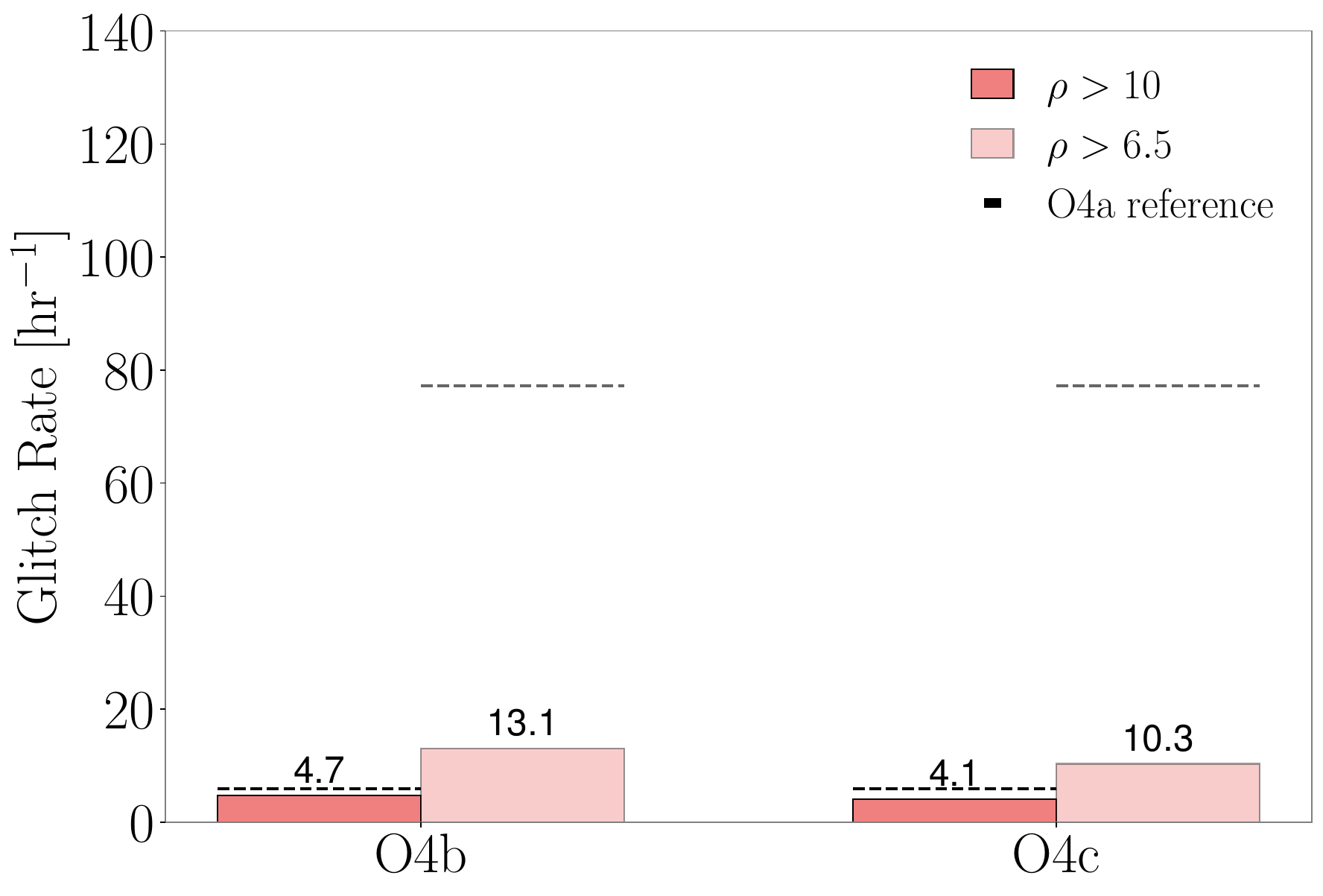}
        \caption{LHO}
        \label{fig:glitch_rate_lho}
    \end{subfigure}
    \hfill
    \begin{subfigure}{0.49\columnwidth}
        \centering
        \includegraphics[width=\linewidth]{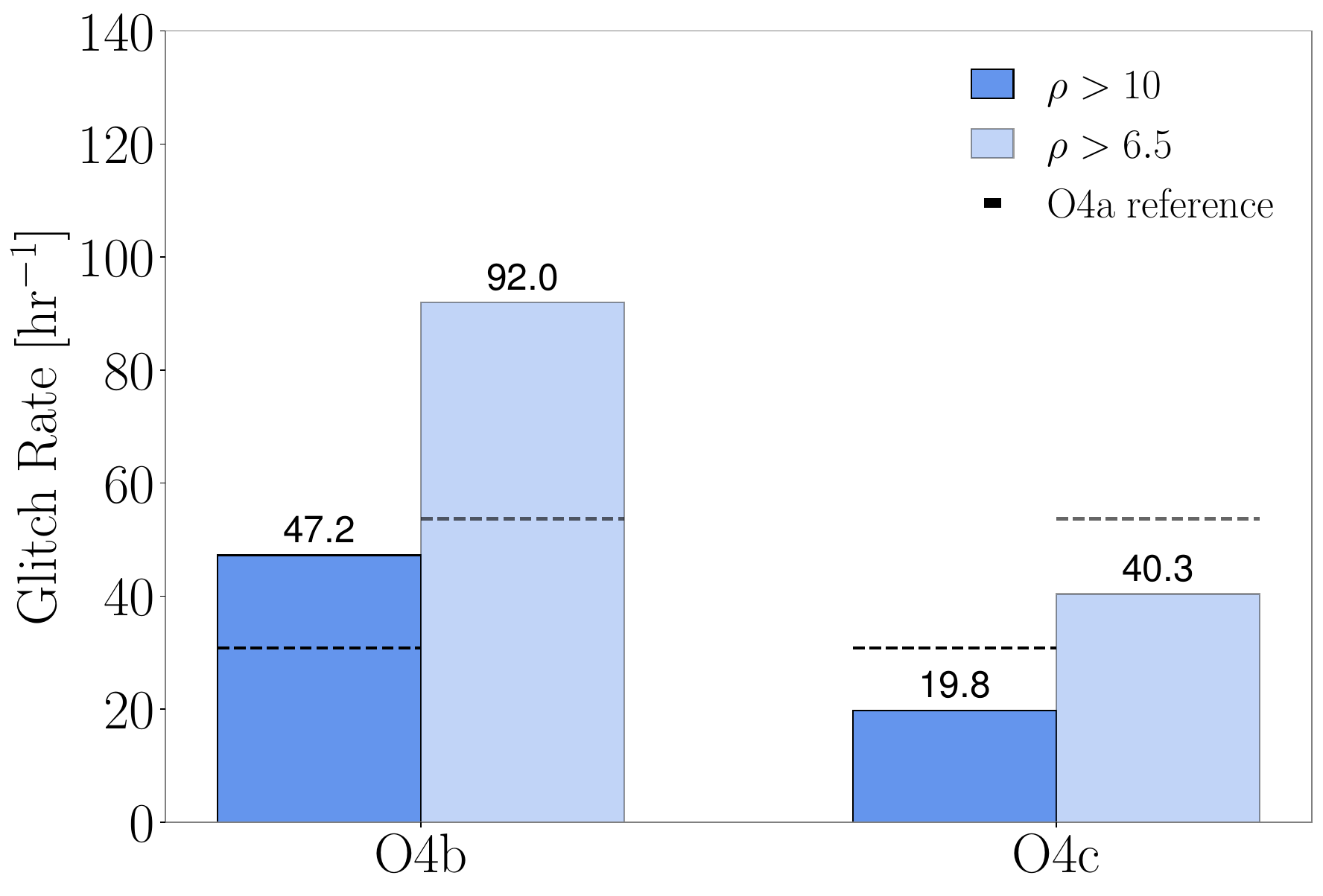}
        \caption{LLO}
        \label{fig:glitch_rate_llo}
    \end{subfigure}
    \caption{
    Glitch rate per hour during O4b and O4c for both detectors. In each panel, darker bars correspond to glitches with $\rho > 10$, while lighter bars include glitches with $\rho > 6.5$, where $\rho$ denotes the \ac{SNR}. The dashed lines indicate the corresponding \ac{O4}a rates for each \ac{SNR} threshold. Glitches were identified using the Omicron algorithm~\cite{2020SoftX..1200620R} on the ``uncleaned'' (\texttt{GDS\_CALIB\_STRAIN}) channel for consistency in comparisons with \ac{O4}a, rather than each detector's noise-subtracted (``cleaned'') strain channel~\cite{2019CQGra..36e5011D}.}
    \label{fig:glitch_rate}
\end{figure}

A comparison between total glitch rates in each part of \ac{O4} is shown in Figures~\ref{fig:glitch_rate_lho} and~\ref{fig:glitch_rate_llo}.
\ac{LHO} experienced a substantial reduction in glitch rate during both O4b and O4c following the implementation of a new feedback control configuration~\cite{alog:74939} that reduced the incidence of low-SNR broadband glitches.
The \ac{LHO} glitch rate during \ac{O4}c is the lowest observed in any detector during the advanced detector era.
Blip glitches~\cite{2019CQGra..36o5010C}, a class of broadband glitches with no known auxiliary witness which mainly degrade the sensitivity of unmodeled searches, saw a decrease in rate by 50\% at \ac{LHO} compared to \ac{O3} following this change.
Table~\ref{tab:bliptab} gives the rate of blip glitches in \ac{O3} and \ac{O4}.
At LLO, the dominant glitch population remained associated with ground motion, while the reduction observed during \ac{O4}c is related to the installation of cage baffles and an \ac{ISI} platform in the \ac{HAM}~1 chamber~\cite{2026arXiv260514143N}.
The results of these changes at \ac{LLO} are detailed in section~\ref{subsubsection:Low SNR glitches}.

\begin{table}
\begin{tabular}{cc}
\br
Observing run & Blip glitches per hour (\ac{LHO})\\
\mr
O3 & 1.68 \\
O4a & 1.24 \\
O4b & 1.13 \\
O4c & 0.84\\
\br
\end{tabular}
\caption{\label{tab:bliptab}Comparison of the \ac{LHO} blip glitch rate in \ac{O3} and \ac{O4}~\cite{2021CQGra..38m5014D,2025CQGra..42h5016S}. Blip glitches were classified using GravitySpy~\cite{2017CQGra..34f4003Z}. Improved feedback control reduced the incidence of blip glitches in \ac{O4}b and \ac{O4}c.}
\end{table}

\begin{figure}
    \centering
    \includegraphics[width=0.95\columnwidth]{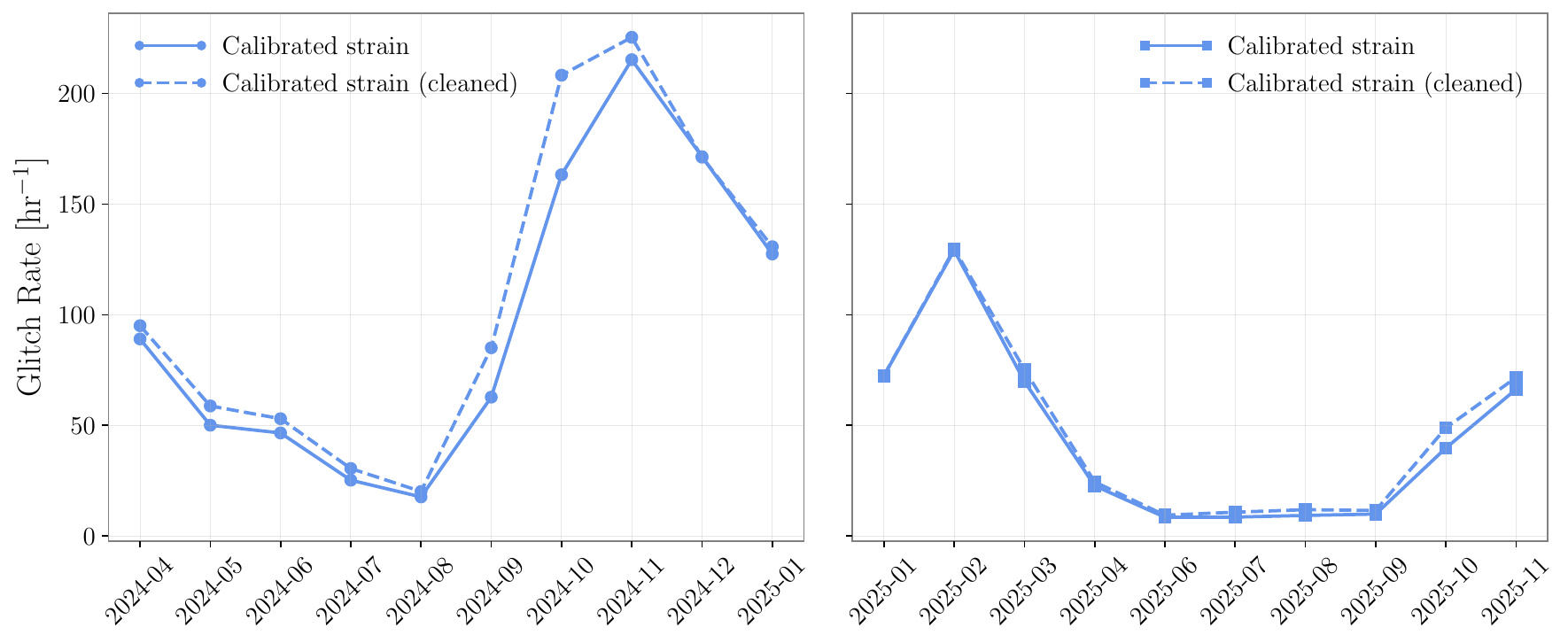}
    \caption{
    Monthly Omicron glitch rate at LLO for $\mathrm{SNR} > 6.5$ comparing results obtained using the standard calibrated \ac{GW} strain and the cleaned \ac{GW} strain data. The left plot covers \ac{O4}b and the right covers \ac{O4}c. The cleaned data show a higher number of identified glitches, particularly during \ac{O4}b, indicating that an additional population of low-amplitude glitches is recovered after the removal of narrow spectral features associated with calibration lines. The effect is less pronounced in \ac{O4}c because the ambient low-frequency ground motion which produces scattering glitches was lower in amplitude during most months~\cite{2021CQGra..38b5016S}.
    }
    \label{fig:strain_clean_comparison}
\end{figure}

A relevant change introduced in \ac{O4}b is the use of the \texttt{GDS\_CALIB\_STRAIN\_CLEAN} channel as input to Omicron, instead of the previously used \texttt{GDS\_CALIB\_STRAIN} channel. See Figure~\ref{fig:strain_clean_comparison} for a comparison of the two.
The cleaned channel remains calibrated and includes additional noise subtraction steps, including the removal of calibration lines, which reduce narrow spectral features and instrumental contributions that can otherwise mask transient signals in specific frequency bands.
This improves the sensitivity in these regions, enabling the detection of additional glitches~\cite{alog:69626,alog:69850}.

Figure~\ref{fig:strain_clean_comparison} shows a comparison of the monthly glitch rate in LLO obtained using the original calibrated strain and the cleaned strain channels for glitches with $\mathrm{SNR} > 6.5$.
An overall increase in the number of identified glitches is observed when using the cleaned data.
This difference can be ascribed to observing more, and higher-\ac{SNR}, scattering-related glitches, as these typically occur in frequency regions affected by calibration lines.

During October 2024, using the cleaned strain channel for glitch identification led to $\sim9,000$ more glitches being detected compared to the standard calibrated strain during the same month of \ac{O4}b.
The effect is more pronounced for lower SNR thresholds, as weaker glitches are more susceptible to being hidden by narrow spectral features, while higher SNR events are less affected.
This change will enable better studies of glitch populations, morphologies, and correlations with auxiliary witness sensors.

\subsection{LIGO Hanford}
\subsubsection{Vibration Coupling}
As the low-frequency sensitivity of the interferometer improved during \ac{O4}, \ac{DARM} noise peaks were revealed in the sub-$100$~Hz range. 
Many of these peaks were driven by vibration, though there was no coherence observed between \ac{DARM} and the network of \ac{PEM} accelerometers placed around \ac{LHO}~\cite{2021CQGra..38n5001N}.
Rather, the coupling was nonlinear, which suggested the coupling was likely from light scattering off optical or vacuum system components~\cite{2021CQGra..38b5016S}.
The investigations below address the drivers of these noise sources and their coupling mechanisms.
In general, the greatest reductions in environmental noise contaminating the \ac{GW} data were achieved by addressing the coupling mechanism between the detector and the noise source, rather than the source alone.
One example of this, described in~\cite{2025CQGra..42h5016S}, is the post-\ac{O4}a damping of baffles at \ac{EX}. 

\paragraph{Input Arm Baffles}
A second example of a vibrational noise source identified and mitigated in \ac{O4} was a broad peak in \ac{DARM} data centered at 13~Hz which also added noise up to $\sim$7 harmonics of its fundamental frequency.
Targeted noise injection campaigns localized the coupling site to \ac{MC} tube baffles and nozzles~\cite{alog:74175,alog:74772}.
During the \ac{O4}a-\ac{O4}b break, the \ac{MC} tube baffles were re-angled and nozzle baffles installed.
These interventions mitigated the noise above 15~Hz. Remaining noise at below 15~Hz was sufficiently mitigated by adjusting the \ac{ITMY} compensation plate yaw~\cite{alog:76969}.

\paragraph{HVAC units}
In the summer of 2024, multiple peaks in \ac{DARM} appeared in the 24-40~Hz band.
Several of these peaks were caused by \ac{HVAC} systems scattered around the detector~\cite{alog:79546,alog:80182,alog:80655,alog:82986}.
Many of these \ac{HVAC} units employ variable frequency drives which allow for a range of operating frequencies for fans in the \acp{AHU}.
Noise driven by these \ac{HVAC} units appeared in \ac{DARM} data when their fan's operating frequency range overlapped mechanical resonances of interferometer components.
In instances where the \ac{AHU} introducing the noise could be identified, the operating range of the \ac{AHU} variable frequency drives was moved away from the frequencies which coupled strongly to \ac{DARM}.

Beginning in July 2025, a large peak in \ac{DARM} at 20~Hz with a harmonic at 40~Hz appeared sporadically for several months.
It was most prominent during work hours, especially during the summer months.
The noise source was identified as an \ac{HVAC} unit located in a nearby building on \ac{LHO}'s grounds.
Despite being vibrationally isolated, it still coupled to \ac{DARM} while in operation~\cite{alog:86257}.

\subsubsection{Lines and combs investigations}
\label{sec:lines_and_combs} 
Throughout \ac{O4}, \ac{LHO} suffered more contamination from lines than \ac{LLO}, and many investigations were carried out during the observing run~\cite{2026arXiv260605959G}.
Several of the combs that were identified in \ac{LHO} data during the first part of \ac{O4} were successfully mitigated, so during the second and third parts of the run, investigations focused on two other problematic artifacts: a prolific comb with near-30~Hz spacing and various offsets seen at many harmonics; and a $\sim$100~Hz frequency band near the suspension ``violin resonances'' intermittently impacted by a large number of lines.
The near-30~Hz comb was observed in \ac{O3} \ac{LHO} data as well, though it was less prolific across the \ac{LIGO} sensitive frequency band.
The change in severity between \ac{O3} and \ac{O4} could be due to an increase in coupling of the external noise into the \ac{GW} strain channel or an increase in the strength of the noise itself.
Investigations using magnetometers strategically placed in various locations and other in-situ measurements indicated the source of the near-30~Hz comb to be video cameras located in the \ac{PSL} enclosure area~\cite{alog:87414}.
These cameras were in-use at \ac{LHO} since at least \ac{O3} (and stay powered on during observing times) and are different than the \ac{LLO} cameras (which stay powered off during observing times), which explains why \ac{LLO} data does not have the same contamination.
When these cameras were powered off, the near-30~Hz comb disappeared.
Efforts are underway at \ac{LHO} to mitigate the near-30~Hz comb.
A near-100~Hz comb in \ac{LHO} data thought to be related to the near-30~Hz comb was unaffected by powering off the cameras; further investigations are ongoing to identify this source of noise or coupling mechanism.

At times during \ac{O4}, there were a large number of additional lines observed in a $\sim$100~Hz band near the fundamental of the violin resonances~\cite{2015CQGra..32g4001L,2011CQGra..28x5001L} of \ac{LHO} data~\cite{alog:71501,alog:71800}.
This contamination was observed intermittently, i.e., not persistent throughout \ac{O4}, and was not observed at \ac{LLO}.
Normally, the presence of the high-quality-factor violin resonance lines in the spectral data at their fundamental and higher harmonics is only problematic for the narrow region in frequency where the resonances reside (a $\sim$20~Hz band at the fundamental frequency region).
The additional lines observed in a broader, $\sim$100~Hz band impacts narrowband \ac{GW} searches more significantly, warranting additional investigation.
Previously observed intermodulation between violin modes and loud calibration lines~\cite{alog:48161,alog:53439} provides an explanation for a fraction of the lines observed in this band but not all~\cite{alog:79825}.
The intermodulation and observation of noise at only \ac{LHO} motivated studies of the \ac{GW} readout differences between \ac{LHO} and \ac{LLO}~\cite{LIGO-T2100200,LIGO-T2400204,alog:71964,alog:82320,alog:79579}.
A quadratic intermodulation model was investigated, allowing calibration lines and violin-mode peaks, or pairs of violin-mode peaks within or across harmonic regions to produce sum- and difference-frequency artifacts~\cite{alog:87923}.
The effective quadratic term was estimated at two points along the readout, first from the analog photodetector current and then, after analog-to-digital conversion, in digital counts; the resulting patterns were qualitatively similar, but did not uniquely identify where the nonlinearity enters the readout path~\cite{alog:87923}.
Further effort is needed to validate the consistency of the model with detector measurements, and, if consistent, to determine which part of the readout contributes to the intermodulation.

\subsubsection{Electronics Ground Noise}
Noise from electronics ground potential fluctuations was investigated in the first part of \ac{O4} where variation of current flowing from building electronics ground to true neutral earth were observed to be correlated with noise in \ac{GW} strain data~\cite{2025CQGra..42h5016S}.
The coupling mechanism is thought to be fluctuations in the potential of the electronics ground system due to the variations in current across the finite resistance between electronics ground and neutral earth, which has been measured to be $\sim$2~$\Omega$ at \ac{LHO} across many grounding spikes connected in parallel~\cite{alog:67075}.
The same noise was observed at LLO through electrical injections onto the grounding system and variations in the \ac{ESD} bias for each test mass~\cite{alog:64609LLO,alog:67149}.
At LHO, a change in the \ac{ETMX} measured charge over the break between \ac{O4}a and \ac{O4}b suggested that there might be a change in electronics ground fluctuation coupling.
A sweep of the \ac{ETMX} \ac{ESD} voltage found that the minimum in coupling had changed from an ESD bias of $\sim$150 V in August 2023~\cite{alog:72118} to 58 V by June 2024.
Changing the bias setting to this new minimum reduced the coupling by about a factor of 10~\cite{alog:78194}.
Similar measurements of the \ac{ITMX} and \ac{ITMY} \acp{ESD} found that an \ac{ESD} bias settings of 0~V and -40~V, respectively minimized the noise observed in the coherence between \ac{DARM} and the current clamp on the electronics grounding cable~\cite{alog:78734, alog:78925}.

\subsubsection{Magnetic Noise}
Beginning in \ac{O3}, weekly magnetic field injections were made to monitor magnetic coupling to \ac{DARM} at some locations in the \ac{CS}~\cite{2021CQGra..38n5001N}.
These injections were modified to include injections at the end stations in \ac{O4}a, and were continued throughout \ac{O4}b and \ac{O4}c.
Fluctuating magnetic couplings at high frequencies were observed in O4a~\cite{2024CQGra..41n5003H}, but the cause has not been identified.
In contrast, \ac{O4}b and \ac{O4}c exhibit very little change in weekly magnetic coupling variation compared to previous runs~\cite{2021CQGra..38n5001N}.

\subsubsection{Post-O4 Environmental Noise Injections}
The coupling of the \ac{LIGO} detectors to their environment is quantified through periodic injections of acoustic, magnetic, and vibration noise.
Detailed knowledge of the environmental coupling is used both for noise investigations and event validation~\cite{2016CQGra..33m4001A,2024CQGra..41n5003H,2023ApPhL.122r4101B}.
Noise injection and coupling function computation procedures are detailed in~\cite{2015CQGra..32c5017E,2021CQGra..38n5001N}.
Prior to \ac{O4}a, environmental noise injection campaigns were performed at both observatories~\cite{alog:69745,alog:64609}.
\ac{PEM} injections were again performed at \ac{LHO} to compare the pre-run coupling to that measured at the conclusion of \ac{O4}c~\cite{alog:89929}.

\begin{figure}
\includegraphics[width=\textwidth]{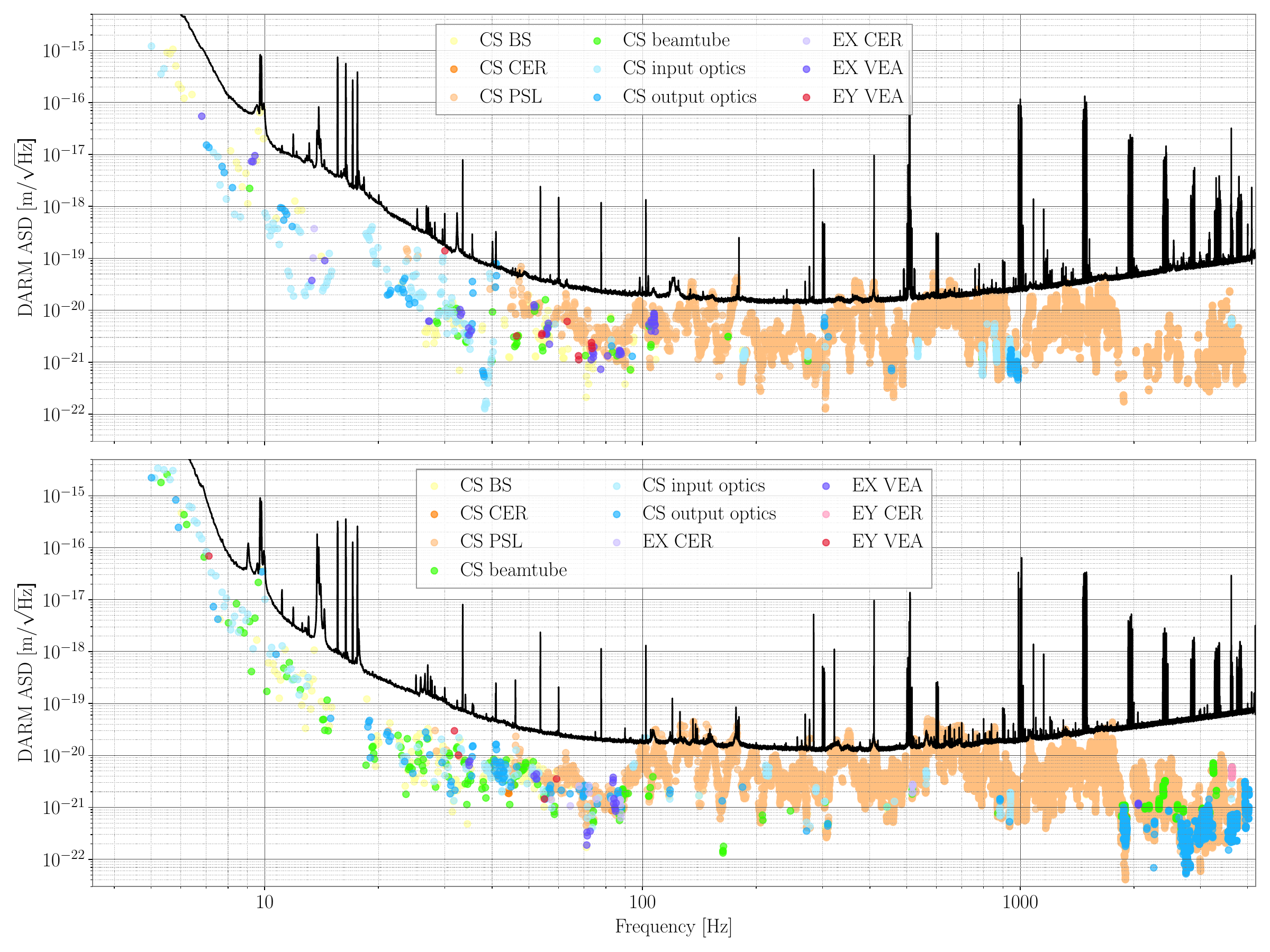}
\caption{\label{fig:H1_inj_acou} The top panel shows the results of the pre-\ac{O4} \ac{PEM} vibration noise injection campaign at \ac{LHO}. \ac{PEM} sensors, of which there are $\sim$100 at each observatory, are grouped by location, denoted by different colors. The location with the most severe coupling in each frequency bin is plotted. Sensors locations are generalized by their location in the \ac{CS} \ac{LVEA} or their location in endstation vacuum and equipment areas (VEAs) or corner or endstation controls and electronics rooms (CERs). Complete sensor coupling details may be found at~\cite{PEMpage}. Plotted points indicate frequency bins where a loud response was seen in \ac{DARM} when noise was injected following the procedure in~\cite{2021CQGra..38n5001N}. The black trace is a representative \ac{DARM} \ac{ASD} taken at a time close to when \ac{PEM} injections were performed, but when no environmental noise was added. The circles indicate the \ac{DARM} trace expected for normal levels of vibrational noise witnessed by \ac{PEM} accelerometers and microphones. Vibrational coupling is largely driven by linear and nonlinear vibrational coupling to the input laser beam witnessed by \ac{CS} \ac{PSL} sensors. Frequencies where the prediction for \ac{DARM} exceeds the observed value of \ac{DARM} are within the acceptable factor of 2 uncertainty on the true value of sensor coupling to \ac{DARM}~\cite{2021CQGra..38n5001N}. While this is deleterious for noise hunting, it does not affect the performance of event validation tasks based on \ac{PEM} sensor data in \ac{O4}~\cite{2024CQGra..41n5003H}. The bottom panel shows the results of the post-\ac{O4} acoustic noise injection campaign. Baffle adjustments overall reduced the vibrational coupling, while jitter remained troublesome. The \ac{DARM} trace was taken near the end of \ac{O4}c.}
\end{figure}

Compared to the pre-\ac{O4} coupling shown in the top panel of Figure~\ref{fig:H1_inj_acou}, the total vibrational coupling of \ac{LHO} decreased.
The post-\ac{O4} measured vibrational coupling is shown in the bottom panel of Figure~\ref{fig:H1_inj_acou}.
The most dramatic changes in \ac{EX} vibrational coupling came from the \ac{EX} cryobaffle damping performed before \ac{O4}b~\cite{alog:76969,2025CQGra..42h5016S}.
While vibrational coupling was reduced in the \ac{CS} with changes to \ac{MC} tube baffling, a worrying source of vibrational coupling, is jitter of the input laser beam~\cite{2025PhRvD.111f2002C}, which is not well-described by the current linear coupling function calculation procedure, although study of other methods, like the one described in~\cite{2021CQGra..38l5005W}, is ongoing.

\begin{figure}
\includegraphics[width=\textwidth]{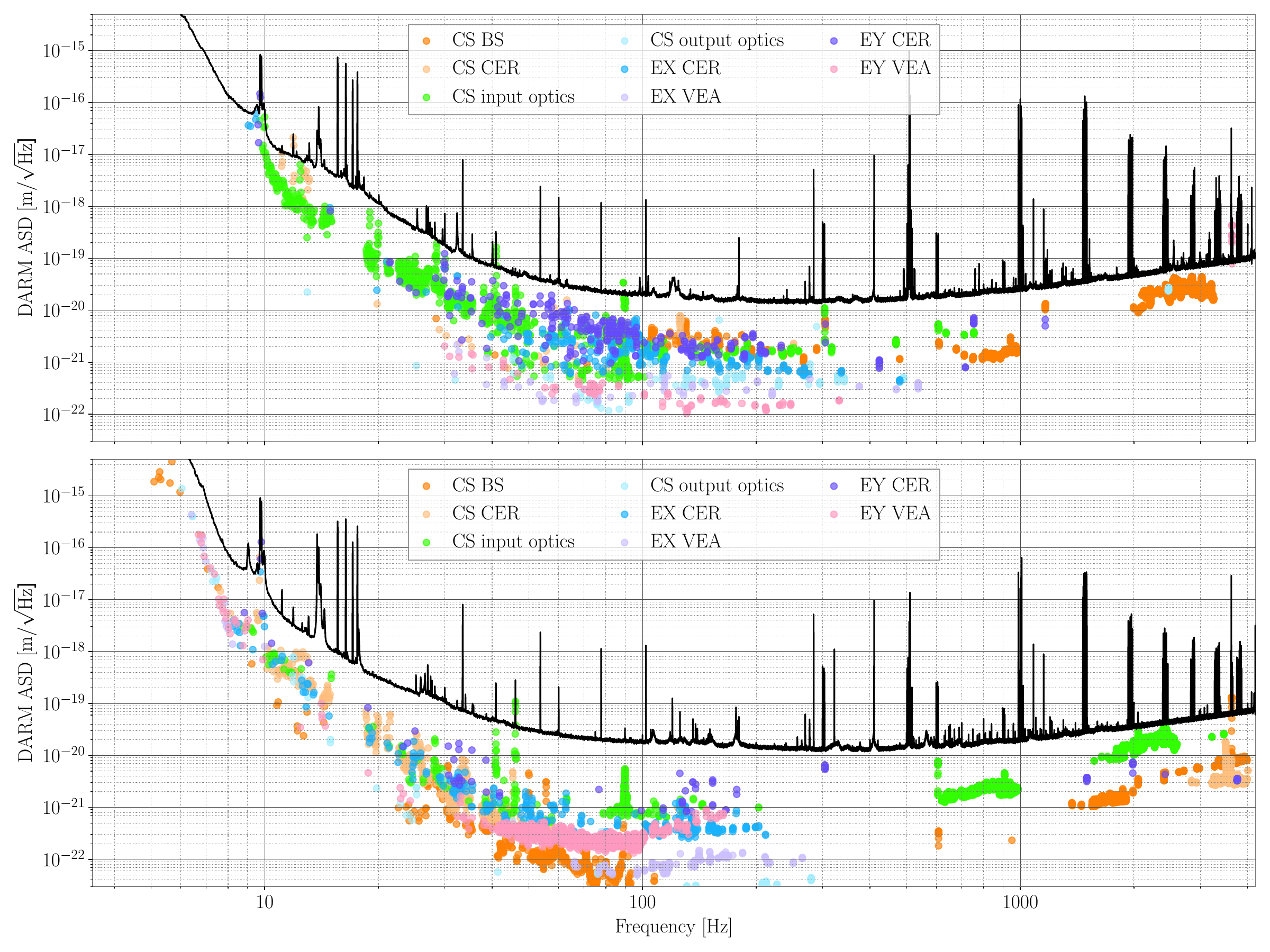}
\caption{\label{fig:H1_inj_mag} The top panel shows the results of the pre-\ac{O4} \ac{PEM} magnetic noise injection campaign. Magnetic coupling in the input arm and \ac{BS} areas in \ac{O4}a is motivated by how close the estimated ambient value is at high frequencies to the true value of \ac{DARM} and the weekly fluctuations noted in~\cite{2025CQGra..42h5016S}. The bottom panel shows the results of the post-\ac{O4} \ac{PEM} magnetic noise injection campaign. Improvements in both the measurement procedure and postprocessing of results permitted better understanding of the magnetic coupling following \ac{O4}c.}
\end{figure}

During the November 2025 post-run \ac{PEM} injection study, we approached saturation of the accelerometers and magnetometers more closely than in past injection campaigns in order to lower the upper limits.
The final post-run magnetic injections were done with more narrow bands while still covering the full range of frequencies of interest to avoid saturation.
This change to noise injection procedure is most visible when comparing the magnetic injections shown in the top panel of figure~\ref{fig:H1_inj_mag}, especially at higher frequencies, to the 2025 injections in the bottom panel of figure~\ref{fig:H1_inj_mag}.

\subsection{LIGO Livingston}
\subsubsection{Scattered-Light Noise}
Noise created by light scattering off vibrating surfaces can appear as glitches in the data, especially in the $10$-$40$~Hz regime~\cite{2021CQGra..38b5016S,2026arXiv260514143N}. Low frequency glitches  constituted about $88\%$ of all Omicron triggers during O4a and O4b and about $75\%$ in O4c. Most of these glitches are due to scattered-light noise. There were two populations of scattered-light glitches: one having high \ac{SNR} (\ac{SNR}$>$20) and another having low \ac{SNR} (SNR$<$20). The high \ac{SNR} glitches appeared anywhere between $10$-$40$ Hz depending on the magnitude of the microseismic ground motion.
The low SNR glitches appeared only around $20$ Hz and they were primarily modulated by the high frequency ($10$-$30$ Hz) vertical ground motion at the \ac{CS}~\cite{2026arXiv260514143N}. The instrumental investigations to better understand these two groups of glitches are discussed in the following subsections.  

\begin{figure}
  \centering
  \includegraphics[width=0.65\textwidth]{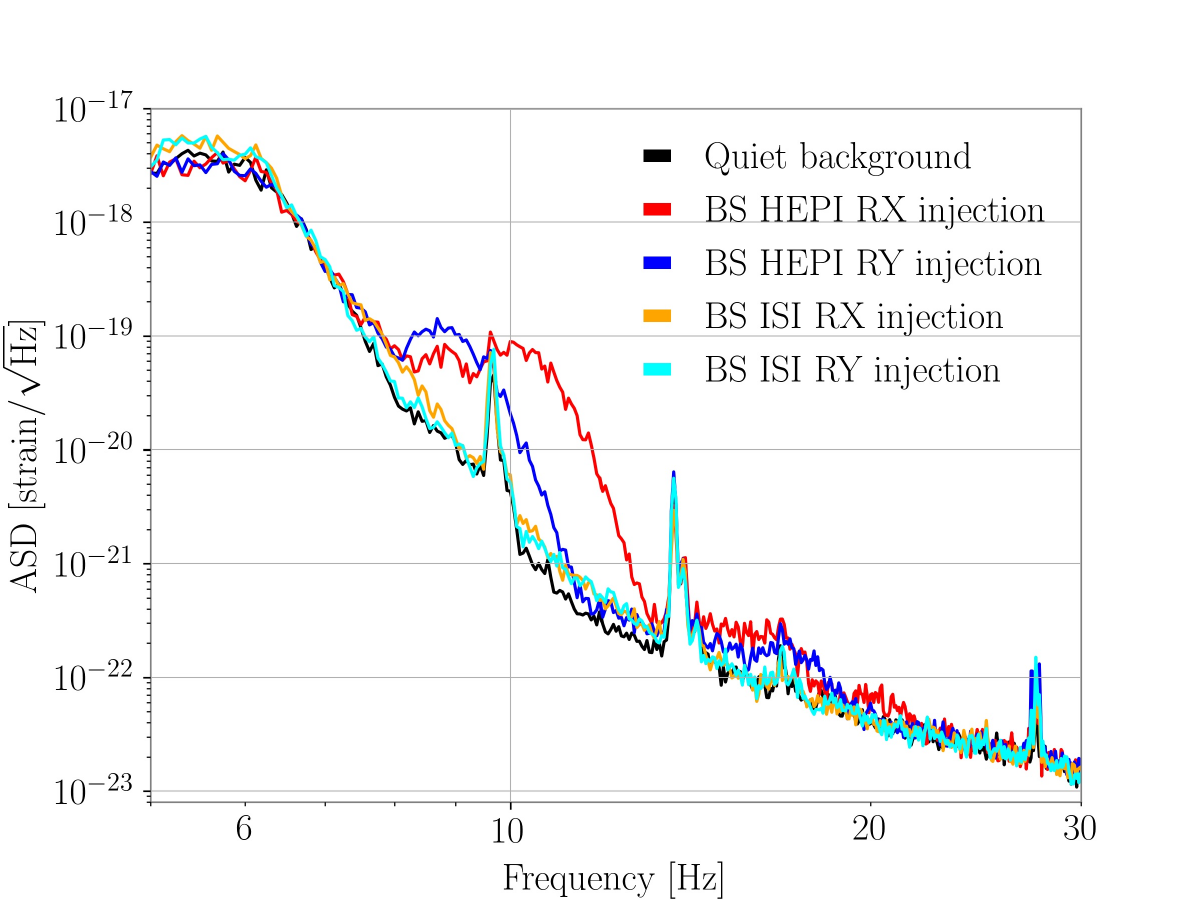}
  \caption{Scatter shelves (broadband high amplitude noise in between $8$-$15$ Hz) produced in DARM by the $0.2$ Hz BS injections compared to a reference time with no scatter shelf present. Each spectra was computed from $180$~s of data. Here, ``R'' denoates rotation about the labeled axis. Similar amount of motion was injected at longitudinal DOFs of BS \acp{HEPI} as well but no scatter shelves were observed during those injections.}
  \label{fig:bs_inj_darm}
\end{figure}

\paragraph{High SNR glitches}\label{subsubsection:High SNR glitches}
These glitches are likely caused by light scattering off vibrating surfaces.
In the first part of \ac{O4}, it was found that the scattering surface(s) causing these glitches did not have any sharp resonances at a particular frequency; instead, the rate of high \ac{SNR} scattering glitches very closely followed ground motion between $0.1$-$1.0$~Hz at \ac{LLO}~\cite{2025CQGra..42h5016S}.
The component of microseismic ground motion at the \ac{CS} in the direction along the X-arm had the highest correlation to the scattered-light noise~\cite{2026arXiv260514143N}.
To find the source of the noise, sinusoidal motion was injected at microseismic frequencies at different stages of the seismic isolation and suspension system at all test mass locations and at the beam splitter.
\textit{Scatter shelves}, broadband high amplitude noise at low frequency, were observed when noise was injected at the beam splitter \ac{HEPI}~\cite{alog:78693}, as shown in Figure~\ref{fig:bs_inj_darm}. 
However, this same behavior was not seen when injecting the same amount of motion in the second stage of the beam splitter \ac{ISI} platform.
This suggests the elliptical baffles, because they are attached to the beam splitter \ac{HEPI} but not the \ac{ISI}, may be one interferometer component which scatters light that is reintroduced as noise.
However, the height of these scatter shelves were almost $100$ times greater in amplitude than the scatter shelves observed during elevated microseismic activity in \ac{O4}.
This indicates that the elliptical baffles had much more stray light illuminating them during noise injection tests than the amount of stray light which produced scatter shelves in O4.
Even if a small fraction of this scattered beam gets reflected back by some other surface in the vicinity, then it can create noise similar to that seen in O4.
When similar injections were performed at the test mass \acp{HEPI}/\acp{ISI}, we did not see any clear scatter shelf.

Following the installation of the cage baffles at \ac{LLO}, as described in Section~\ref{subsubsection:cagebaffles}, the rate of these high SNR glitches dropped significantly.
The remaining glitches also appear at lower SNR, consistent with the baffles reducing the amount of light falling on the scattering surface~\cite{alog:78226}.

\paragraph{Low SNR glitches}
\label{subsubsection:Low SNR glitches}
The rate of these low SNR glitches is primarily modulated by higher frequency ($10$-$30$ Hz) vertical ground motion at the \ac{CS}~\cite{alog:75579}.
Motion from the \ac{HAM}~1 vacuum chamber had a strong coupling to DARM~\cite{alog:72674}.
During the mid-\ac{O4}c commissioning break, an \ac{ISI} platform was installed in the \ac{HAM}~1 chamber to provide better seismic isolation to the optics table in this chamber.
After installing the \ac{ISI}, which is described in Section~\ref{sssec:ham1isi}, the low \ac{SNR} glitches did not appear again given the same amount of ground motion~\cite{2026arXiv260514143N}. 

\subsubsection{Binary Neutron Star range noise investigations}
\begin{figure}
\includegraphics[width=\textwidth]{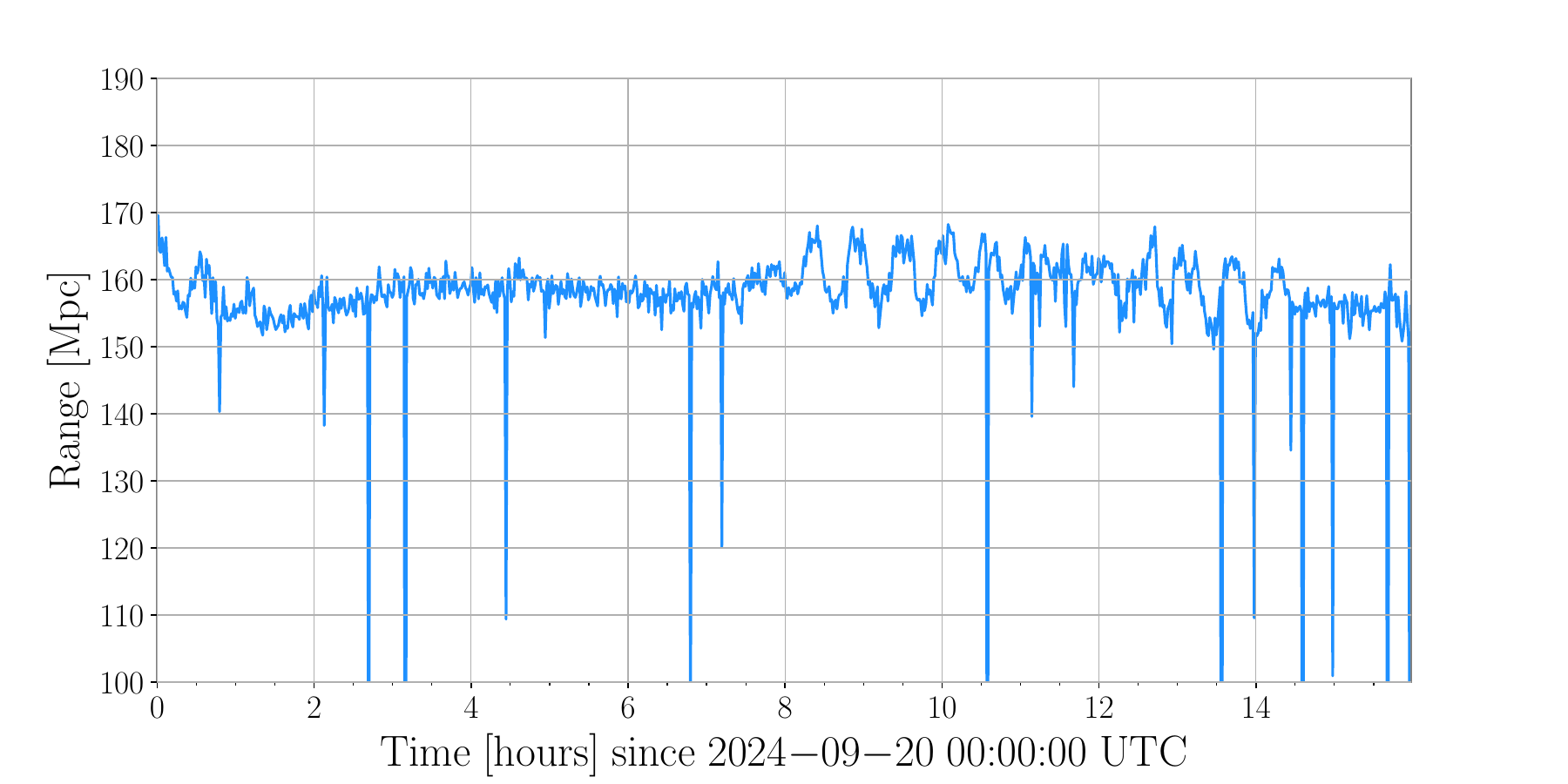}
\caption{\label{fig:30min_osc} \ac{BNS} range oscillations over a stretch of 16 hours. 30-minute oscillations are present from roughly 8 UTC to 14 UTC on this day.}
\end{figure}

\paragraph{Quasi-periodic BNS range oscillations}
The \ac{BNS} range oscillations, described in~\cite{2025CQGra..42h5016S}, continued to occur throughout \ac{O4}b and \ac{O4}c.
These oscillations occurred with a period of approximately 30 minutes, producing range variations between $5$-$15$~Mpc, and are associated with broadband excess noise in the strain data between $30$-$50$~Hz.
The oscillations were not present at all times, but when active could persist for all or part of a day, as shown in Figure~\ref{fig:30min_osc}.
Although \ac{O4}a investigations potentially implicated temperature effects at the \ac{EX}, the coupling mechanism was not identified.
Further investigations during O4b and O4c sought to identify the source more precisely.

A bakeout blanket was placed on the \ac{EX} manifold, which coincided with a reduction in the frequency and amplitude of the oscillations~\cite{alog:73481}, though this has not been formally confirmed as a causal intervention.
Low frequency tapping tests were also performed near the \ac{EX} cryopump, which produced strain noise with a character similar to the half-hour oscillations~\cite{alog:73288}.
More investigation is needed into whether or not a mechanical coupling at this location plays a role.

Periodic variation in voltage monitors which monitor the \ac{EX} \acp{ESD} was found to correlate with range oscillations, although a positive identification could not be conclusively established due to potential correlations in these channels with the ambient temperature~\cite{alog:80232,alog:80556}.

LASSO regression analyses~\cite{2018CQGra..35v5002W} applied to multiple periods of \ac{O4}b \ac{LLO} data consistently identified the \ac{EX} as the location of subsystems most strongly correlated with the strain noise.
Many auxiliary channels across several subsystems exhibited oscillatory behavior consistent with the range variations, including temperature sensors, humidity sensors, \ac{PEM} sensors, and channels associated with the \ac{ALS} system.
Among the \ac{ALS} channels flagged was \texttt{L1:ALS-X\_REFL\_CTRL\_OUTPUT}, which is electronics noise. 
However, this electronics noise does drive the \ac{ALS} voltage-controlled oscillators, which are a part of keeping the \ac{ALS} phase-locked loops locked even when the \ac{ALS} is not in use.
As a test, the phase-locked loops were unlocked at both end stations while the interferometer operated in a low-noise state beginning in February 2025.
In this configuration, the \ac{BNS} range oscillations persisted~\cite{alog:75125}.
The channels identified by LASSO spanned several subsystems and varied from day to day, reflecting the fundamental challenge that when many auxiliary channels exhibit similar oscillatory behavior, regression alone cannot determine which, if any, represent a causal driver.

To address this limitation that regression methods cannot resolve the temporal ordering of correlations, a time-lag cross-correlation method was developed to test whether changes in auxiliary channels precede or follow changes in the strain noise~\cite{jane_surf2025}.
Preliminary results suggest that channels belonging to the facilities management and control subsystem at \ac{EX}, in particular the \ac{AHU} heat switch, most frequently led the strain response, pointing toward a possible environmental driver. 
Ongoing work aims to extend this analysis to channels at the \ac{CS} and \ac{EY}.

\begin{figure}
\includegraphics[width=\textwidth]{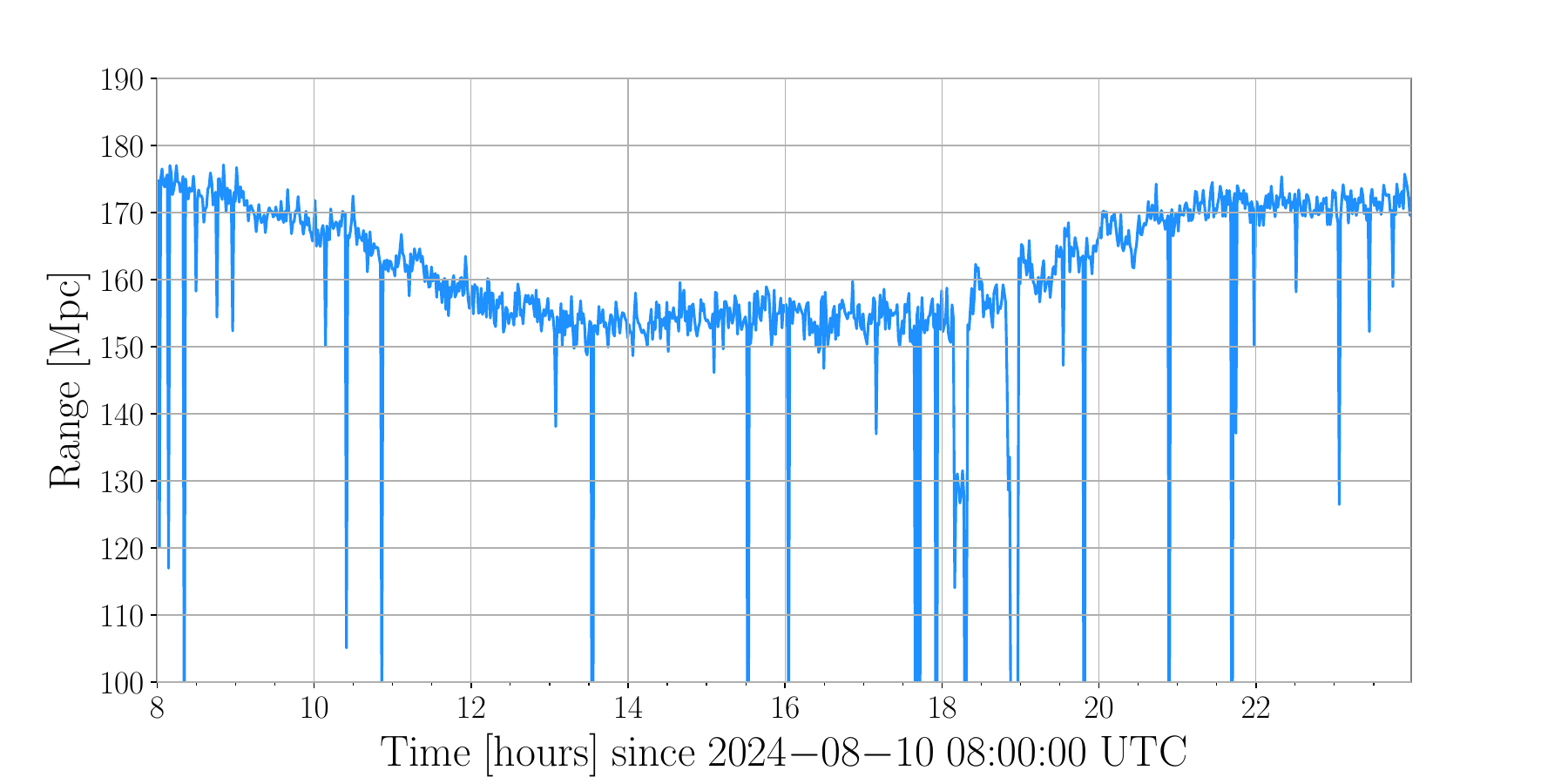}
\caption{\label{fig:long_drop} Example of the \ac{BNS} range drop associated with low frequency broadband strain noise. The range decreased by roughly 20 Mpc during this period.}
\end{figure}

\paragraph{12 hour cycle \ac{BNS} range drops}
During O4b, \ac{LLO} exhibited periodic drops in the \ac{BNS} range with a cycle of approximately twelve hours, during which the range decreased by up to $20$~Mpc relative to its nominal value before recovering.
An example of this behavior is shown in figure~\ref{fig:long_drop}.
These drops were associated with low frequency broadband strain noise.
An on/off test of the squeezer was performed, and it was concluded that the squeezer was not the source of the noise~\cite{alog:72583}.
Coherence was observed between the strain data and a photodiode monitoring transmission through the \ac{EX} test mass during these low range periods~\cite{alog:72819}.
This suggested that beam clipping or misalignment at the end test mass may have contributed to the broadband noise responsible for the drops.
After the mid-O4c commissioning break, this coherence was no longer observed~\cite{l1:alog:78194}.

Correlations between the range drops and environmental temperature sensors at the corner station were also investigated as a potential source of the noise.
Initial observations noted a potential correlation between the \ac{BNS} range and the channel \texttt{L0:FMC-CS\_LVEA\_AVTEMP}, which monitors the average temperature of the \ac{LVEA} in the \ac{CS}~\cite{alog:73009}.
This motivated an investigation into whether localized temperature fluctuations in the \ac{LVEA} were inducing thermal lenses on the input test masses that could degrade the interferometer sensitivity.
Subsequent analysis found that temperature fluctuations localized to the region around the \ac{BSC} chambers, where the input test masses are housed, were correlated with negative thermal lenses measured by the Hartmann wavefront sensor in \ac{ITMY}~\cite{alog:74834}.
Further analysis over a five-day period showed that the spherical power measured in \ac{ITMY} exhibited variations with a cycle consistent with the 12-hour range drops. Variations in the \ac{DARM} band-limited \ac{rms} channels correlated with these thermal lens fluctuations were identified as a plausible mechanism by which the thermal lensing could be lowering the range. The \ac{LVEA} temperature control was adjusted during this period~\cite{alog:78104,alog:78563,alog:78617} to make it more stable.
However, the coupling mechanism was not conclusively established, and further modeling work is ongoing.

\subsubsection{Wandering narrowband features during O4}
 \label{sec:wandering_lines_o4} 
\begin{figure}
    \centering
    \includegraphics[width=\columnwidth]{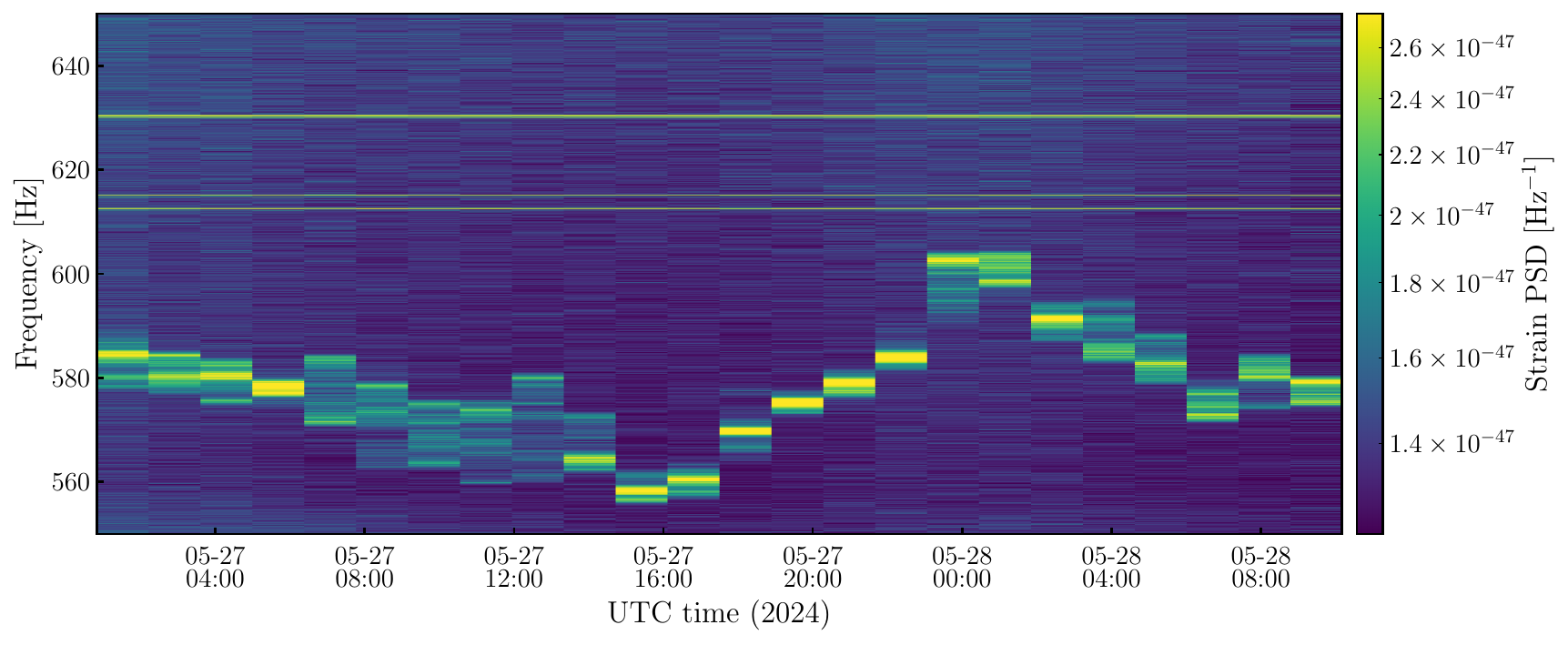}
    \caption{Spectrogram of \ac{LLO} strain data over a representative continuous observing stretch during O4, showing a wandering narrowband feature near $550$--$640\,\mathrm{Hz}$ with a frequency excursion of order $\sim 30\,\mathrm{Hz}$.}
    \label{fig:LLO_wandering_line}
\end{figure}

In addition to the stationary spectral lines in section~\ref{sec:lines_and_combs}, \ac{O4} data also contained narrowband features whose central frequencies drifted over time. One significant example observed at \ac{LLO} was a wandering feature near $\sim 550$--$640~\mathrm{Hz}$, with a frequency excursion of order $\sim 30~\mathrm{Hz}$, which was noted near the beginning of O4 and remained visible at least throughout the observing run.
This wandering behavior is shown in figure~\ref{fig:LLO_wandering_line}.
Related wandering line structures were also observed at higher frequencies, including features near $\sim 3.4~\mathrm{kHz}$, $\sim 5.0~\mathrm{kHz}$, and $\sim 7.9~\mathrm{kHz}$, with some of the higher-frequency features showing inverted frequency evolution relative to the lower-frequency line~\cite{alog:70869}.
Follow-up studies using high-sample-rate auxiliary data indicated that these structures were unlikely to be produced simply by aliasing from the $512~\mathrm{kHz}$ to $16~\mathrm{kHz}$ data stream~\cite{alog:70869}.
Auxiliary channel investigations found modest correlations with environmental and interferometer channels, including an \ac{LVEA} temperature probe and a local-oscillator monitoring channel associated with the 42~MHz squeezer system, suggesting that slow environmental or instrumental variations may contribute to the observed drift, although the detailed coupling mechanism remains unresolved~\cite{alog:70869}.

\subsection{Computational Tools}
Numerous computational tools were employed in the course of these and other commissioning and noise-hunting activities described in this section as well as section~\ref{sec:config}.
These tools include automated rapid online data access and visualization toolkits~\cite{duncan_macleod_2024_10999143,AREEDA201727,2021SoftX..1400677F}, analyses which identify correlations between auxiliary channel timeseries data~\cite{bruco,2018CQGra..35v5002W} or clustered noise artifacts~\cite{2020SoftX..1200620R} and noise/glitches in detector strain data~\cite{2011CQGra..28w5005S,2010JPhCS.243a2005I,2020arXiv200512761E,2022PhRvD.106j2006D}.
Citizen-science based strain data glitch classification results were also central to many studies~\cite{2017CQGra..34f4003Z}.
Also used were packages which specifically monitor the behavior of environmental witnesses~\cite{2021CQGra..38n5001N,ligocam} and those which search for persistent noise artifacts~\cite{goetz_2025_18776397,Meyers:2018nyo,Stochmon}.
While not all tools contributed to all investigations mentioned herein, the suite of data quality tools available to detector scientists and data analysts permits noise studies to improve the astrophysical reach of the \ac{LIGO} detectors.

\section{Event Validation}
\label{sec:val}
A central responsibility of the \ac{LIGO} \detchar group is the rapid and rigorous assessment of GW candidate events.
The candidate events are triggers identified by the online search pipelines \textsc{PyCBC}, \textsc{GstLAL}, \textsc{MBTA}, \textsc{SPIIR}, \textsc{Aframe}, \textsc{MLy} and \textsc{cWB} as possible \acp{GW}~\cite{usman2016pycbc,messick2017analysis,aubin2021mbta, luan2012towards,hooper2012summed,marx2025machine,skliris2024toward,mishra2022search}.
The \ac{EV} process determines, for each GW candidate, whether the trigger is more likely to be astrophysical or a product of instrumental or environmental noise. This process also checks if there are artifacts that could bias analysis of an astrophysical signal.
The \ac{O4}b and \ac{O4}c procedure of event validation was similar to what was done in \ac{O4}a~\cite{2025CQGra..42h5016S}, with several targeted improvements.
During \ac{O4}b and \ac{O4}c, 41 volunteers from the \ac{LVK} collaboration assessed the detector data quality around the vicinity of \ac{GW} candidates. 
 
The \ac{EV} process included verifying whether the detector was observing in a nominal configuration at the time of the candidate, recognizing noise artifacts that would bias source property estimation, checking PEM sensor outputs for any signs of environmental coupling into strain data, checking stationarity of the data, and also identifying statistical correlations with strain and auxiliary sensors.
If the data quality is found to be inadequate by the DetChar experts, additional mitigation techniques - including Bayesian noise inference and transient noise subtraction - may be applied by the data analysts~\cite{2021PhRvD.103d4006C}.  

In the following subsections, we detail the changes in detector network, \ac{DQR}, \ac{EV} infrastructure and procedure during \ac{O4}b and \ac{O4}c, and validation of events in these runs.

\subsection{Data Quality Report}

The \ac{DQR} is the principal automated tool that informs event validation.
For each candidate event, it runs a fixed suite of analyses around the time of the trigger --- including environmental noise estimates, strain--auxiliary channel correlations, glitch predictions, excess-power and stationarity tests, and detector-range monitoring --- and reports qualitative information about the state of the detectors, and for many tasks includes a label (Pass, DQ Issue, or Task Error) with a statistical estimate of noise contamination~\cite{2024arXiv240115392D,2023CQGra..40c5008V,2020CQGra..37u5014M,2024CQGra..41n5003H,2020arXiv200512761E,2023PhRvD.108f3016M,2020SoftX..1200620R,2024CQGra..41h5007A,2026arXiv260725208A}.
The architecture of the report, including its rapid (few-minute) and offline (few-hour) tiers, is unchanged from \ac{O4}a and is described in detail in~\cite{2025CQGra..42h5016S}.

In the \ac{EV} workflow, the  \ac{DQR}'s automated flags determine which candidates require human follow-up in low latency, which is on the order of 5 minutes.
Candidates for which any task reports a DQ Issue or Task Error label receive additional scrutiny from the on-shift \ac{RRT} while candidates with no flagged issues require no further human review in low latency.
All candidates are subsequently re-examined offline, regardless of the low-latency outcome.
The \ac{EV} task force comprises three teams: the event validation volunteers, the noise mitigation team, and the final reviewer. 
The event validator, assigned on a one-week shift, assesses each active detector and records a conclusion of ``Not Observing", ``No Data Quality Issues", or ``Data Quality Issues" based on the information available to the \ac{RRT} as well as \ac{DQR} tasks, the results of some of which take hours to produce a result.
The final result of \ac{EV} for a candidate which is not retracted is a recommendation, if necessary, for regions of time-frequency space that should be avoided for accurate parameter estimation.

The principal upgrade to the \ac{DQR} framework for \ac{O4}b and \ac{O4}c was the replacement of the single statistical threshold used to flag data quality issues with per-task thresholds.
Each task was assigned its own numerical threshold, calibrated using the false- and true-alarm statistics observed in \ac{O4}a; this improved the overall true-alarm rate of the report while reducing spurious flags.
In O4a, the \ac{DQR} was based entirely on the \textsc{DQRbuild}~\cite{davis2026rapid} package; in O4b, additional tools in the VirgoDQR~\cite{acernese2023virgo} package were incorporated.

\subsection{Expanded detector network support}

One major change from \ac{O4}a was the inclusion of Virgo data in \ac{O4}b and \ac{O4}c.
The \ac{DQR} infrastructure was extended to consistently handle the three detector network configuration.
\ac{DQR} tasks have been successfully run on Virgo data surrounding \ac{GW} candidates and their results have been used in GW candidate assessments.

After the Noto earthquake on January 1, 2024, KAGRA rejoined \ac{O4}c on June 11, 2025.
KAGRA data was not used during parameter estimation, and hence it was not used for validation of candidates during the remainder of \ac{O4}c.
However, KAGRA data was ingested and analyzed by \ac{DQR} tasks in O4c.

\subsection{Noise Mitigation}

The noise mitigation team analyzes the time-frequency window recommendation for parameter estimation.
In \ac{O4}b, the mitigation team searched for non-Gaussian artifacts within this recommendation window and performed noise subtraction on found glitches.
In \ac{O4}c, the noise mitigation team less frequently required noise subtraction based on the recommendations in~\cite{2025PhRvD.112h4006H}: glitches that overlap the signal in time and frequency were removed, but any other glitches with \ac{SNR} $< 50$ were kept as is.
Noise subtraction was performed by the \textsc{BayesWave} algorithm~\cite{2015CQGra..32m5012C, 2021PhRvD.103d4006C,2023PhRvX..13d1039A}.

If the noise is extended in time or frequency, then the noise mitigation team usually recommends restricting the parameter estimation analysis window.
Once the noise subtracted data frames are checked for Gaussianity, the frames and recommendations are sent to analysis groups through \textsc{CBCFlow}~\cite{Ashton2024CBCFlow} to perfrom parameter estimation and population studies.

\subsection{Validation of \ac{O4}b and \ac{O4}c events found by online search pipelines}

In \ac{O4}b, 114 significant detection candidates were shared as public alerts, out of which 9 were retracted.
The majority of these retractions were issued due to concerns from search pipelines, rather than data quality.
However, three retractions were motivated by scattered light noise overlapping the time-frequency track of the inspiral in \ac{O4}b~\cite{2025GCN.38856....1L, 2024GCN.36190....1L, 2024GCN.38070....1L} and one retraction~\cite{2024GCN.36747....1L} was informed by environmental noise transients witnessed by \ac{PEM} accelerometers coupling to strain~\cite{helmlingcornellthesis}.
In each of these cases, the retraction decision was reached by considering both data quality and search pipeline performance concerns.

\begin{figure}
	\centering
	\begin{subfigure}{\columnwidth}
		\centering
		\includegraphics[width = \linewidth]{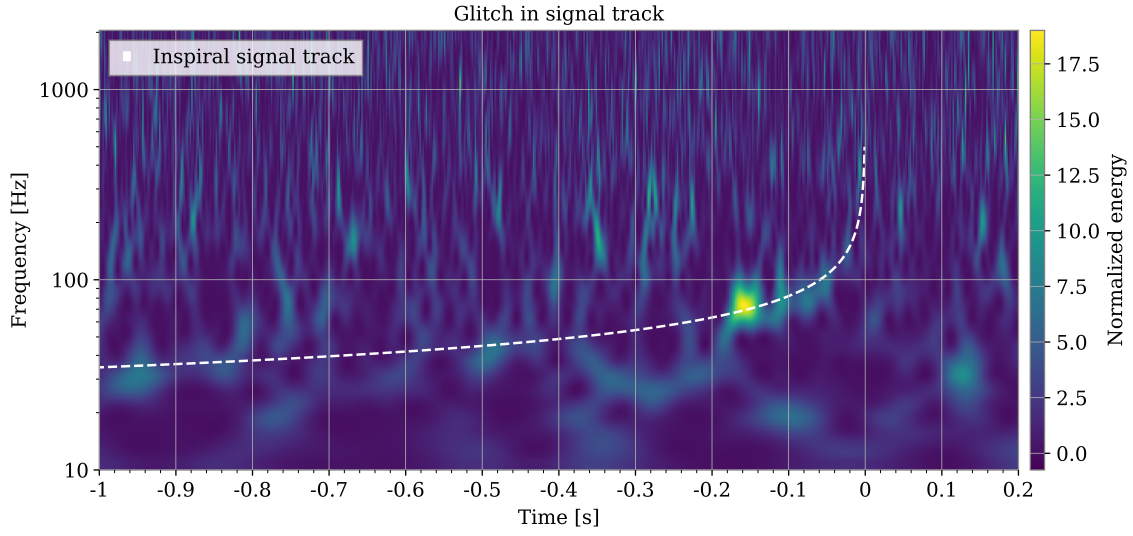}
		\caption{Frame with glitch in signal track}
		\label{fig:signal_glitch}
	\end{subfigure}
	
	\begin{subfigure}{\columnwidth}
		\centering
		\includegraphics[width = \linewidth]{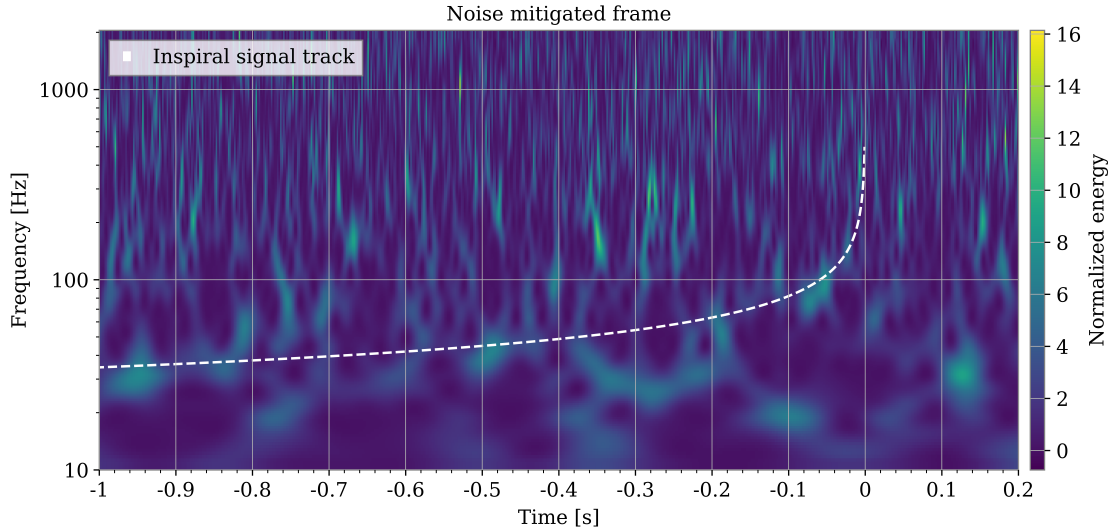}
		\caption{Noise mitigated frame}
		\label{fig:noise_mitigation}
	\end{subfigure}
	\caption{Example of glitch mitigation performed during \ac{O4}b for the event GW240520\_213616.}
	\label{fig:glitch_mitigation}
\end{figure}

Of the unretracted candidate events, 25 required glitch subtraction in \ac{O4}b~\cite{2019CQGra..36e5011D,2026arXiv260527225T}.
It was seen that not all glitches affect the parameter estimation, as recommended in~\cite{2025PhRvD.112h4006H}.
Thus the need for noise mitigation was reduced in \ac{O4}c.
An example of a glitch mitigated frame is shown in figure~\ref{fig:glitch_mitigation} for the event GW240520\_213616~\cite{2026arXiv260527225T}.

For low frequency extended noises, a low cutoff of 30 Hz was recommended for 10 event candidates in either \ac{LLO} or Virgo, and 40 Hz for one event candidate in \ac{LLO}.
There was one event candidate where the low frequency was set at 70 Hz and one at 50 Hz in Virgo due to the presence of huge numbers of glitches in the low frequency region.
In three instances where the transient noise environment at Virgo was sufficiently severe that glitch mitigation via noise subtraction was deemed intractable, the Virgo data were excluded from parameter estimation analysis for the affected event candidate.

\section{Data Quality for Astrophysical Searches}
\label{sec:dq}
\detchar plays a central role in mitigating instrumental and environmental noise sources that can affect astrophysical searches.
Although direct mitigation at the instrument level is always preferred, it is not always possible to eliminate all sources of noise.
As a result, residual non-Gaussian and non-stationary features persist in the data.
To address this, the \detchar group develops and provides a range of \ac{DQ} products that are applied across astrophysical searches to reduce the impact of these artifacts.

In this section, we discuss the data quality products produced by the \ac{LIGO} \detchar group which are shared across astrophysical analyses, as well as \ac{DQ} measures that are tailored to specific searches depending on the nature of the source and the analysis method.
Similar, but not identical \ac{DQ} products, are used in the analysis of Virgo data~\cite{2023CQGra..40r5006A,2023NIMPA104867945A}.
These search specific applications are generally organized into four main categories, which also reflect the major classes of GW sources: \acp{CBC}, burst, \ac{CW}, and \ac{SGWB}.

\subsection{Data quality products for all searches}\label{data_qual_all_search}

As in previous observing runs, periods of data considered unusable due to severe \ac{DQ} issues are removed prior to analysis through the use of \ac{DQ} flags.
These flags, which differ among search groups, are distributed via the Gravitational Wave Open Science Center~\cite{2026arXiv260527090T}.

\ac{CAT1} flags define time segments that should be excluded before any astrophysical search is performed.
They are broadly consistent across search groups, but small differences exist between search-specific \ac{CAT1} flags due to the different analysis methods. The criteria used to define \ac{CAT1} flags in O4b and O4c are consistent with those used in O4a.
Throughout \ac{O4}, \ac{CAT1} flags continued to be applied sparingly, reflecting improved robustness of search pipelines to \ac{DQ} issues.
Rather than aggressively removing data, periods with partial usability, such as those affected only in limited frequency bands, were often retained.
Nevertheless, segments with severe glitches, calibration issues, or detector configuration issues were excluded. The overall analysis time removal due to \ac{CAT1} flags remained below $\sim 0.1\%$ for each \ac{LIGO} interferometer. \ac{CAT1} flags in O4b and O4c included, but were not limited to:
\begin{itemize}
    \item Periods of incorrect line subtraction, particularly at the beginning of lock stretches.
    \item Periods of parametric instability~\cite{2015PhRvL.114p1102E} leading to large amplitude excitations before loss of lock.
    \item Issues in the squeezing control system impacting the detector sensitivity.
    \item Severe ring-ups of violin modes, affecting calibration and data stationarity.
    \item Periods of incorrect observing mode definition or missing $h(t)$ data.
\end{itemize} 

\ac{CAT2} flags target shorter duration noise transients and are typically tailored to specific analyses. As with \ac{CAT1}, the criteria defining \ac{CAT2} flags are consistent with those used in O4a.
In O4b and O4c, their use continues to diverge between search classes: \ac{CBC} searches largely avoid using \ac{CAT2} vetoes, relying instead on probabilistic \ac{DQ} metrics, while short-duration unmodelled burst searches still employ \ac{CAT2} flags to veto periods of substantial contamination of astrophysical data.
Because the latter class of searches are looking for generic excesses in strain coincident between detectors, their $\ac{DQ}$ demands are more restrictive compared to \ac{CBC} searches.

\subsection{Data quality for transient searches}

Transient searches target signals with durations ranging from sub-second to minutes, including \acp{CBC} and unmodelled burst signals. These searches must balance sensitivity with robustness to transient noise artifacts.

\subsubsection{Data quality for compact binary coalescence searches}

\ac{CBC} searches in O4b and O4c continued the O4a strategy of relying on statistical, auxiliary-channel-based data quality information rather than traditional \ac{CAT2} segment vetoes~\cite{2025CQGra..42h5016S}.
The main \ac{DQ} product generated for these searches was the iDQ timeseries~\cite{2020arXiv200512761E}, computed independently for each \ac{LIGO} interferometer from auxiliary channel activity using the Ordered Veto List (OVL) algorithm~\cite{2013CQGra..30o5010E}.
For each strain sample, iDQ quantifies the evidence that a transient noise artifact is present, expressed as a \ac{FAP}: the probability that a detector free of transient noise would produce auxiliary-channel activity at least as significant.
A small \ac{FAP} therefore corresponds to a confident identification of transient noise.
The configuration and calibration of iDQ were unchanged from O4a~\cite{2025CQGra..42h5016S}, as was its consumption by the searches: although the low-latency iDQ output was distributed to all of the online \ac{CBC} pipelines (\textsc{PyCBC Live}, \textsc{GstLAL}, \textsc{MBTA}, and \textsc{SPIIR}), the PyCBC-based analyses were the only ones to ingest it, with the remaining pipelines relying on their own internal noise mitigation.

iDQ information was used in two forms, matching the two modes in which the PyCBC search operates.
In low latency, \textsc{PyCBC Live} retained the O4a criterion of discarding candidates whose coalescence times fell within $\pm 1$~s of any sample with $\mathrm{FAP}(t) < 10^{-4}$, i.e., periods when transient noise was confidently identified~\cite{2021ApJ...923..254D, 2026arXiv260607679T}.
Offline, where longer stretches of data are available for ranking and calibrating vetoes, times satisfying $\log\mathcal{L}(t) \geq 5$, where $\mathcal{L}(t)$ is the iDQ likelihood ratio for the presence of transient noise, were expanded by $\pm 0.25$~s to form a glitch flag.
Rather than applying this flag as a hard veto, the offline PyCBC search folds it into its ranking statistic~\cite{2022PhRvD.106j2006D}: templates are grouped into bins by duration, and within each bin, candidates occurring during flagged time are down-ranked according to how strongly that bin's single-detector triggers concentrate in flagged time relative to the bin's mean trigger rate.

\begin{figure}
    \centering
    \includegraphics[width=\columnwidth]{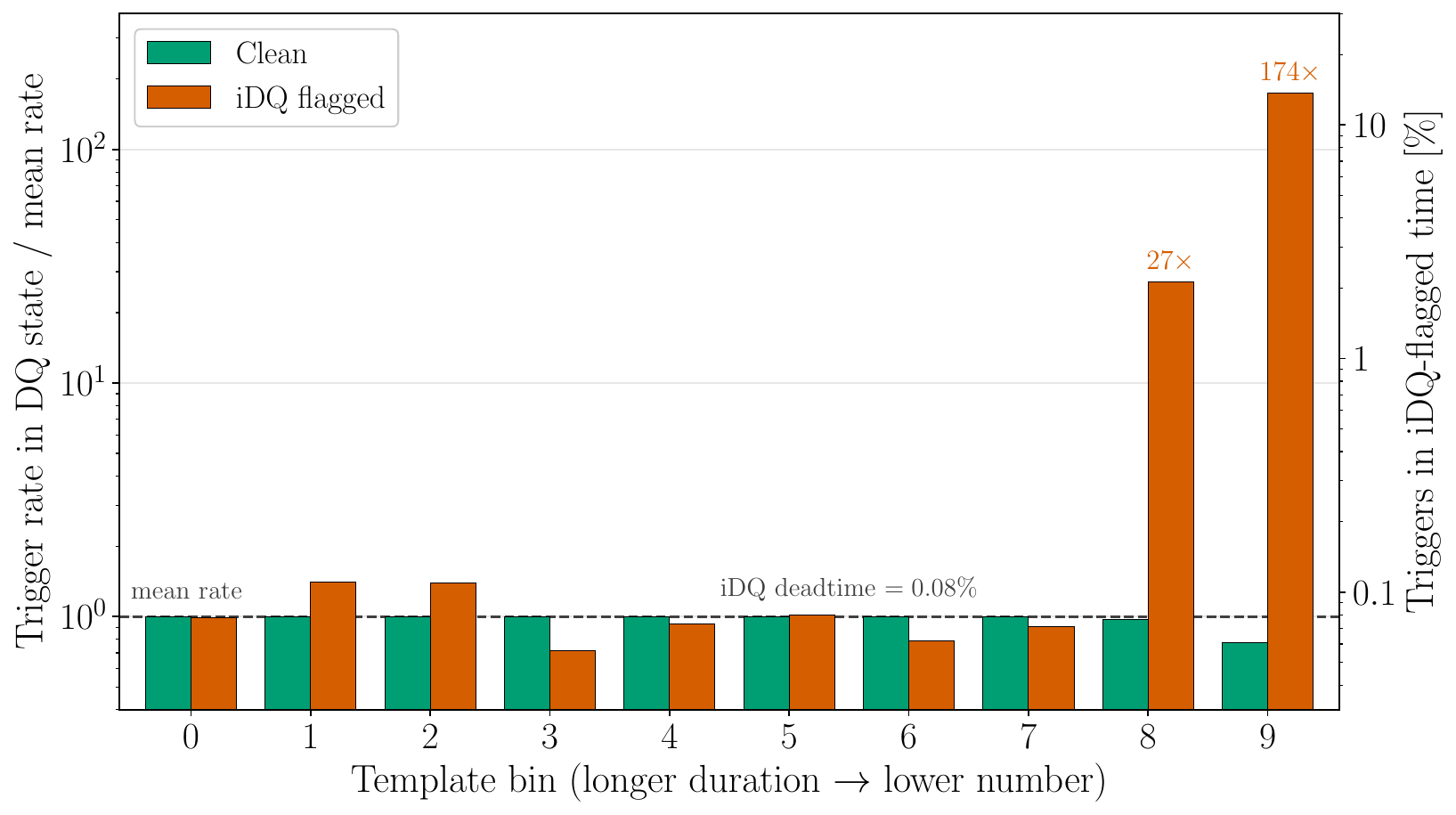}
    \caption{
    Impact of the offline iDQ glitch flag on the PyCBC search at \ac{LHO}. For each PyCBC template-duration bin, bars give the rate of single-detector triggers recorded while the \ac{DQ} is clean (green) and while the offline iDQ glitch flag is active (orange), normalised by the mean trigger rate in that bin. The dashed line marks the rate expected if triggers were distributed uniformly in time. Clean data sit on this line in every bin, whereas triggers during flagged time concentrate strongly in the shortest-duration templates, which are the most easily mimicked by transient noise. Candidates from these templates are consequently down-ranked when they occur during flagged time; candidates occurring in clean data are essentially unaffected. The figure is based on a representative two-week interval of O4c \ac{LHO} data from June 24, 2025 to July 8, 2025.
    }
    \label{fig:idq_dq_rates}
\end{figure}

Figure~\ref{fig:idq_dq_rates} shows these relative trigger rates at \ac{LHO} for a representative two-week interval of O4c, during which the glitch flag was active for $0.08\%$ of the analysed livetime.
For the shortest-duration templates, which are the most easily mimicked by transient noise, the trigger rate during flagged time exceeds the bin mean by more than two orders of magnitude; nearly 14\% of that bin’s triggers occur within the 0.08\% of the livetime when the iDQ glitch flag automatically identified periods of poor data quality.
Candidates from these templates are therefore strongly down-ranked when they occur during flagged time, while candidates in clean data---which trigger at the mean rate in every bin---are essentially unaffected, as are candidates from longer-duration templates, for which glitches are a weaker contaminant.
In this way iDQ concentrates its down-weighting where instrumental glitches are most likely to be confused with genuine \ac{CBC} signals.

No new auxiliary-channel-based \ac{CAT2} flag classes were introduced for \ac{CBC} searches during O4b and O4c; the focus instead remained on monitoring iDQ performance across the run and on the event-level validation procedures described in Section~\ref{sec:val}.

\subsubsection{Data quality for unmodelled transient searches}

Burst searches are broadly divided into two categories: short-duration and long-duration analyses.
Short-duration searches target signals with durations of $\lesssim \mathcal{O}(1\,\mathrm{s})$ and are especially susceptible to short glitches in the detector.
Many, but not all, of these searches continue to utilize both \ac{CAT1} and \ac{CAT2} data-quality flags.
Long-duration searches, which probe signals lasting from several seconds to hundreds of seconds, are performed using both \textsc{cWB} and \textsc{PySTAMPAS}~\cite{2021SoftX..1400678D,2021PhRvD.104j2005M}. Being unmodelled, these searches make minimal assumptions on signal morphology and instead identify excess power coherent across the detector network in time-frequency maps of the strain data. Each pipeline is additionally run in several configurations that differ in their time-frequency decomposition, ranking statistic, or time-frequency pixel clustering algorithm, so as to remain sensitive to this broad range of possible signal morphologies. Because these signals persist over longer timescales, long-duration analyses are correspondingly more sensitive to extended spectral features such as persistent narrowband lines.

\ac{CAT2} flags for burst searches are constructed from auxiliary channels showing strong correlations with glitches in the strain data. 
In O4b and O4c, these include flags targeting very loud glitches (e.g., $\mathrm{SNR} > 100$) as well as site-specific disturbances such as $60\,\mathrm{Hz}$-related glitches observed at LLO~\cite{64337llo}. 
The use of \ac{CAT2} flags in burst searches reduces the background rate of triggers while maintaining low deadtime (typically $\sim 0.1\%$).

Long-duration searches, particularly those targeting narrowband or quasi-monochromatic waveforms such as magnetar emission or gamma-ray burst plateau signals, are additionally sensitive to periods of narrowband line artifacts and transient line ring-ups.
These features can mimic or obscure astrophysical signals in time-frequency analyses.
In practice, analyses targeting these sources apply their own \ac{DQ} measures, including removing the known narrowbands artifacts from the time-frequency space (PySTAMPAS) or applying decision tree classifiers to filter out triggers associated with line artifacts.
The list of lines removed from the analysis is informed by the instrumental lines list compiled by \detchar investigations~\cite{2026arXiv260605959G}.

\subsection{Data quality for persistent searches}

Persistent gravitational-wave searches target signals that remain in band over long durations, including \acp{CW} and the \ac{SGWB}.
These analyses integrate data over long timescales and are therefore sensitive to non-stationarity and narrowband spectral artifacts that can accumulate over time.

Data quality efforts for persistent searches focus on mitigating transient glitches, slow variations in detector noise, and persistent or wandering spectral features of instrumental origin.
While the specific procedures differ between \ac{CW} and stochastic analyses, both rely on careful mitigation of time-domain frequency-domain contamination to ensure unbiased results.

The following sections summarize the data quality treatments applied in \ac{CW} and stochastic searches during \ac{O4}b and \ac{O4}c.
For a comprehensive discussion of \ac{CW} data quality products and methodology during O4, see~\cite{2026arXiv260605959G}.

\subsubsection{Data quality for continuous wave searches}

\paragraph{Self-gated strain}
Glitches introduce noise into the detectors that degrade the sensitivity of \ac{CW} searches.
For this reason, a procedure known as ``self-gating'' is applied to the strain data prior to ingestion by \ac{CW} analysis pipelines in which loud glitches are identified and zeroed out in the time domain.
The removal of loud glitches via gating reduces the noise floor primarily below $500$~Hz, thereby improving  sensitivity to \ac{CW} signals.
The self-gating algorithm is based on the same iterative threshold selection procedure used in \ac{O4}a, the implementation of which is described in detail in~\cite{selfgating}.

\paragraph{Lines lists}
Catalogs of line and comb artifacts (``lines lists'') observed in run-averaged spectra were produced for each detector~\cite{2026arXiv260605959G}.
These lists are an important \ac{DQ} product used by \ac{CW} searches, primarily for the purposes of vetoing non-astrophysical signal candidates.
The lines lists are assembled via inspection of run-averaged high-resolution spectra, where the spectra are generated from Hann-windowed and self-gated 7200~s $h(t)$ segments.
This results in a detailed spectrum with $\sim 0.13889$~mHz frequency resolution suitable for identifying persistent spectral contamination.
Each line and comb artifact is identified and recorded through a combination of automated line-tagging tools and manual visual inspection of the run-averaged spectra.

For each detector, we recorded artifacts in either a ``vetted'' list or an ``unvetted list''.
The vetted list includes artifacts determined to be almost certainly non-astrophysical.
This determination is usually based on correlating the appearance or disappearance of the artifact with detector hardware changes; an exception to this are combs, which are automatically added to the vetted list due to being inconsistent with \ac{CW} emission models.

By the end of the \ac{O4}, the total number of vetted line artifacts (counting individual comb teeth) in \ac{LHO} was 2399, accounting for $\sim 17.21\%$ of the $10$-$2000$~Hz observing band; in \ac{LLO}, the line count was 449, accounting for $\sim 6.37\%$ of the observing band.
Most of the vetoed band fraction for \ac{LHO} is due to the excessive line contamination flanking the violin modes; without counting these contaminated regions, the vetoed band percentage at \ac{LHO} is comparable to that of \ac{LLO}.

\subsubsection{Data quality for stochastic searches}

\paragraph{Time-domain \ac{DQ} measures: gating and non-stationarity cuts.}
Time-domain cleaning for stochastic searches begins with auto-gating within the \textsc{pygwb} workflow~\cite{2023ApJ...952...25R} to suppress short-duration transients.
Outliers in the whitened strain are identified using an amplitude threshold, and an inverse Planck-taper window is applied to smoothly suppress the data around these excursions.
This removes glitches while avoiding sharp spectral artifacts that could otherwise bias segment-level noise estimates and causing unnecessary data loss.
In O4b and \ac{O4}c, the fraction of gated data was $\lesssim 0.1\%$ for each interferometer.

After gating, the data are divided into segments, typically 192~s in duration.
For each segment $i$, the stochastic estimator uncertainty $\sigma_{Y,i}$ is used to compute $\Delta\sigma_i$, the fractional deviation from the run's median $\sigma_Y$ following
\begin{equation}
\Delta\sigma_i = \frac{\sigma_{Y,i} - \mathrm{median}(\sigma_Y)}{\mathrm{median}(\sigma_Y)}
\end{equation}
Segments with $|\Delta\sigma_i|$ above a chosen threshold are excluded.
$\sigma_Y$ depends on both the character of the noise in the detector, as well as as the shape of the \ac{SGWB}.
The nonstationarity cut could therefore be applied to different segments, dependind on a search's choice of $\alpha$---the spectral index of the \ac{SGWB}.
Recent searches for an \ac{SGWB} of the form $\Omega_{\mathrm{GW}}(f)\propto f^{\alpha}$ include $\alpha\in\{0,2/3,3\}$ in \ac{O4} data~\cite{2025arXiv250820721T,2026arXiv260823477T}.

The fraction of segments removed by non-stationarity cuts is less than $\sim10\%$ of the total coincident observing time between the 2 \ac{LIGO} detectors.
Without gating, the fraction of segments removed by non-stationarity cuts would have been $\sim27\%$, demonstrating the robustness of the gating procedure in dealing with short-duration transients and preventing unnecessary data loss.

\paragraph{Frequency-domain \ac{DQ} measures: coherence and notching.}
Frequency-domain \ac{DQ} measures target spectral artifacts that can introduce spurious correlated power between detectors and bias the cross-correlation estimator.
They are cataloged in a \ac{DQ} product known as a notch list, which defines list of frequency bins to be excluded from the analysis.
As \ac{SGWB} searches combine the information from a broad frequency range, the notch list is optimized to remove contaminated bins that could bias the estimator while preserving as much bandwidth as possible.

\begin{figure*}
    \centering
    \includegraphics[width=\textwidth]{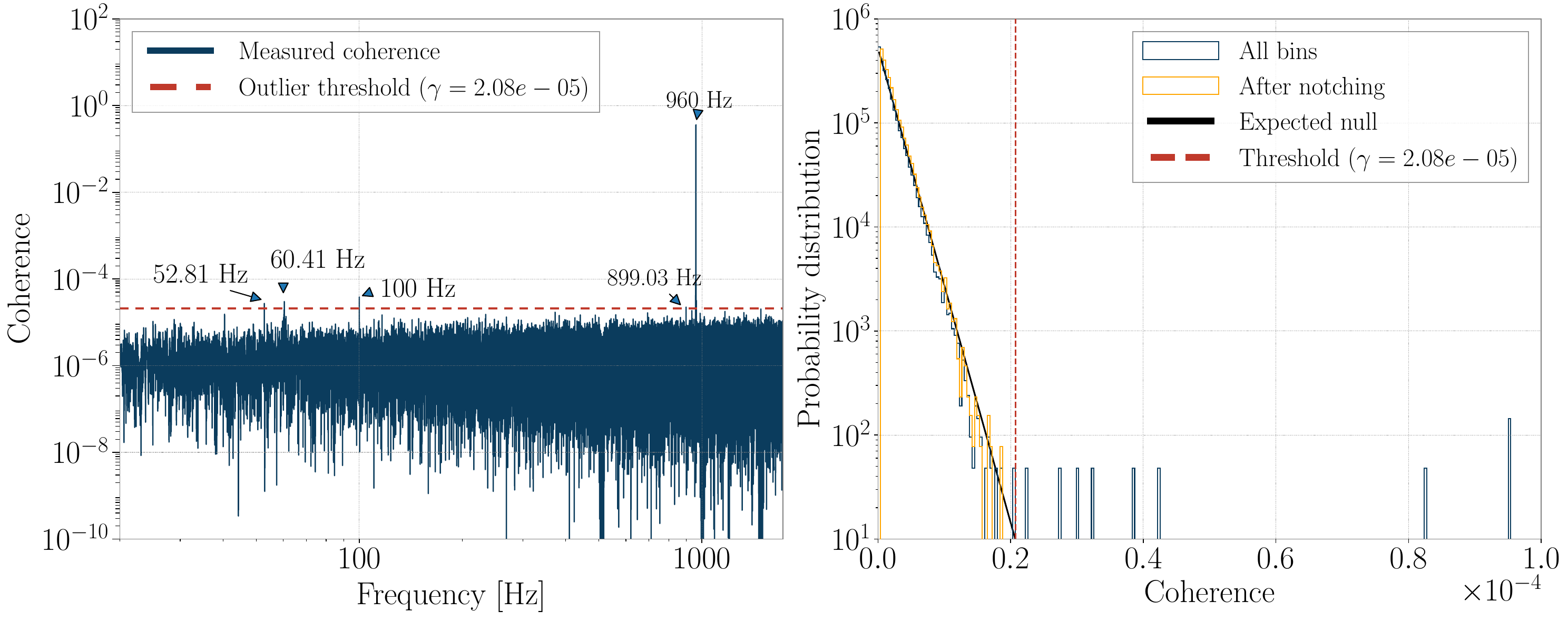}
    \caption{
    Left: coherence spectrum between \ac{LHO} and \ac{LLO} strain data over the analysis band. The dashed line indicates the statistical threshold corresponding to a fixed \ac{FAP} under the null hypothesis of uncorrelated noise. Prominent outliers correspond to known instrumental features.
    Right: distribution of coherence values across all frequency bins. The gray histogram shows all bins, while the shaded histogram shows the distribution after applying notch lists. The solid curve indicates the expected null distribution. Notching significantly suppresses the high-coherence tail.
    }
    \label{fig:coherence_notching}
\end{figure*}

The notch list combines three components. Known instrumental features -- calibration lines, simulated \ac{CW} signals, power-mains harmonics, and mechanical resonances -- are identified by direct inspection of run-averaged spectra.
Bins are also flagged as statistically significant coherence outliers, quantified relative to the expected distribution under the null hypothesis of uncorrelated noise.
Finally, bins with excess temporal variability that would not necessarily stand out in a run-averaged coherence or \ac{PSD} spectrum are added, as described below.
Figure~\ref{fig:coherence_notching} demonstrates the coherence-based component using the timeshifted data of \ac{LHO} and \ac{LLO}.

In the coherence spectra, most frequency bins lie near the statistical floor, while a small number exceed the threshold and appear as outliers.
They include a simulated \ac{CW} signal at 52.81~Hz, excess power near the 60~Hz power mains, a tooth of the 100Hz comb, one of the harmonics of the beamsplitter violin mode, as well as the cluster of the DuoTone lines at 960~Hz.
After applying notch lists, the high-coherence tail is significantly suppressed, demonstrating the removal of correlated spectral artifacts. 

\begin{figure*}
    \centering
    \includegraphics[width=\textwidth]{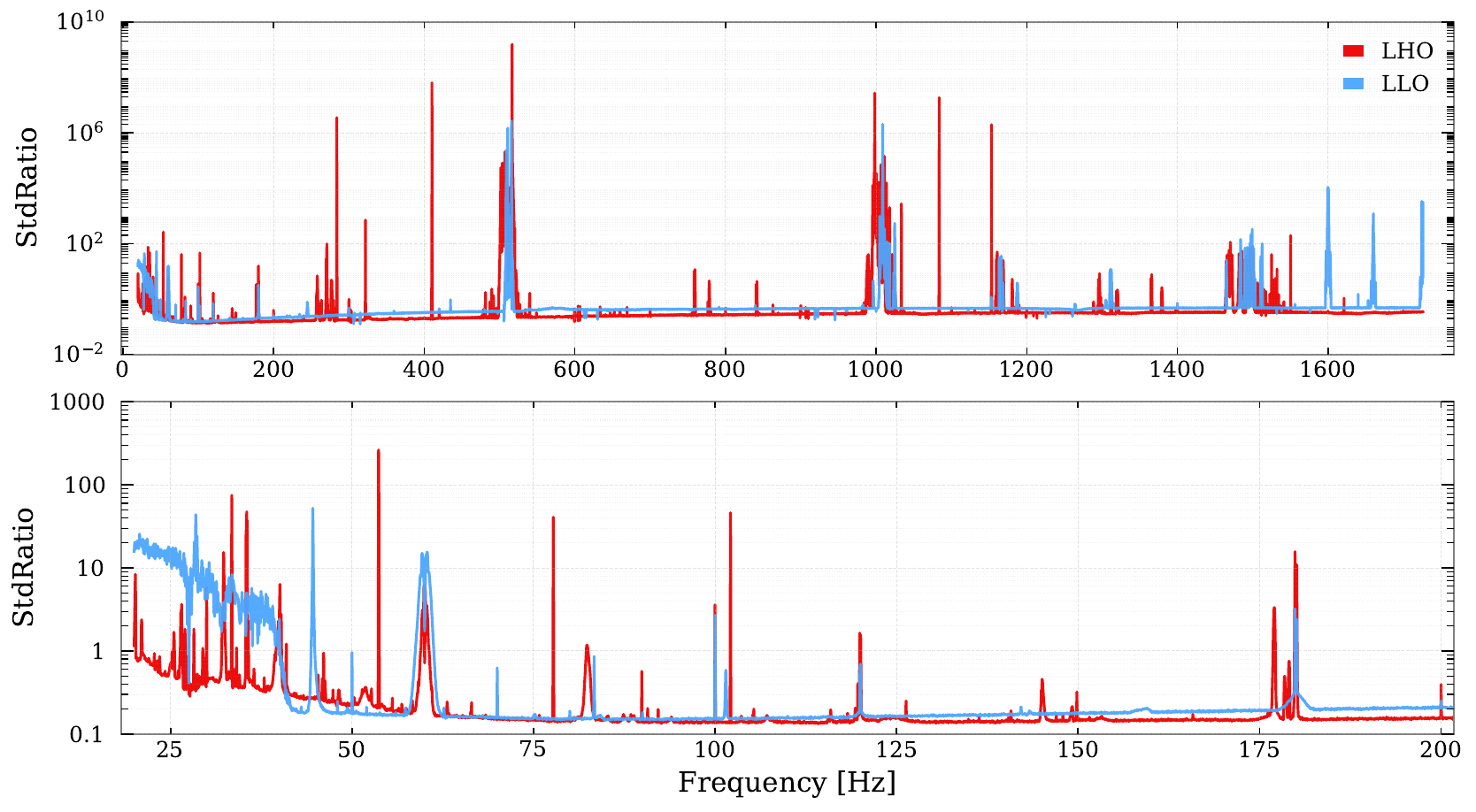}
    \caption{
    Standard deviation of normalized \acp{PSD} for \ac{LHO} (red) and \ac{LLO} over O4b and \ac{O4}c up to the mid-\ac{O4}c commissioning break when \ac{LLO}'s \ac{GV} was replaced.}
    \label{fig:variance_spectra}
\end{figure*}

Additionally, a complementary approach is to identify frequency bins exhibiting excess temporal variability in their \acp{PSD}.
Such variability can indicate non-stationary or intermittently active noise sources, which may not be fully captured in run average spectra.
This information is used to construct ``strict" notches. For each frequency bin, the \ac{PSD} is computed independently for each \textsc{pygwb} analysis job $t$ spanning the run~\cite{2023ApJ...952...25R}, $\mathrm{PSD}(f,t)$, and bins exhibiting excess temporal variability across jobs,
\begin{equation}
    \mathrm{StdRatio}(f) = \mathrm{std}_t\left[\frac{\mathrm{PSD}(f,t)}{\mathrm{median}_t\,\mathrm{PSD}(f,t)}\right],
\end{equation}
are flagged as non-stationary.
A smooth local baseline for $\mathrm{StdRatio}(f)$ is estimated by identifying peak-like structures that rise above a running median-filtered version of the spectrum, masking those regions, and interpolating linearly across the remaining bins, so that the baseline is representative of the typical variability within a local frequency band rather than a single global threshold.
Frequency bins whose $\mathrm{StdRatio}(f)$ exceeds this baseline by a factor of $1.5$ are flagged as strict-notch candidates.
Each candidate band is then manually vetted: bands that appear in only one detector and coincide with instrumental origins are retained, while marginal candidates with $\mathrm{StdRatio}(f)$ close to threshold and no corresponding structure in the run-averaged \ac{PSD} are excluded.
Figure~\ref{fig:variance_spectra} shows the resulting variability spectra for the \ac{LIGO} detectors.
Most frequencies lie near a baseline consistent with stationary noise, while peaks correspond to non-stationary spectral features.
Low-frequency scattered-light glitches at \ac{LLO}, discussed in section~\ref{subsubsection:High SNR glitches}, elevate its variability spectrum below $\sim40$~Hz, while the wandering $\sim550$--$640$~Hz narrowband feature described in Section~\ref{sec:wandering_lines_o4} dominates \ac{LLO}'s excess variability at higher frequencies and was among the strict-notch bands designed to prevent a recurrence of the spurious excess in the power-law posterior at large spectral index during O4 searches for a \ac{SGWB}~\cite{2025arXiv250820721T,2026arXiv260823477T}.
This method is particularly effective at identifying bands affected by wandering or intermittent artifacts that may not appear as strong coherence outliers but can still bias the estimator.
In total, strict notching contributes an additional $4.5\%$ of the $[20,~1726]$~Hz analysis band beyond standard notching procedures, bringing the total fraction of the band removed to $22.8\%$ for O4b and O4c data up to the end of March 2025.

\section{Future Outlook}
\label{sec:conc}
During O4, the \ac{LIGO} \detchar group monitored data from the two \ac{LIGO} detectors, characterized noise artifacts, and provided feedback to detector scientists on mitigation efforts.
Data quality products were produced and employed by \ac{GW} searches and source parameter estimation in order to improve search sensitivity, detection efficiency, and precision of astrophysical source parameter estimates.
O4 was the longest and most sensitive run to date, and the \ac{LIGO} Laboratory has, unlike during each of O1--O3, operated the two detectors through several epochs of observing and commissioning breaks during which the detectors have evolved.
During the commissioning breaks following O4a, and occasionally during required breaks in observing during O4b and O4c to repair damaged components, the \ac{LIGO} detectors adopted improved hardware and configuration changes in order to improve their overall performance.

Long observing runs offer unique opportunities to measure and investigate long-term changes of detector noise that are caused by environmental factors or detector configuration changes.
Overall, we have observed improvements in some areas but some degradations in others.
While detections of \ac{GW} signals from \ac{CBC} sources continues apace, care must be taken with each signal, as evidenced by the Data Quality Report effort made for each detection and parameter estimation follow-up.
Other classes of \ac{GW} signals have not yet been detected ranging from unmodeled, short duration ``burst'' sources to persistent sources from spinning neutron stars and \ac{SGWB} signals.
Identifying noise sources and, where possible, mitigating the sources or coupling mechanisms are crucial for detecting these new classes of signals.

We have described the changes made to the \ac{LIGO} instruments and studies carried out during O4b and O4c to investigate various noise sources impacting the \ac{LIGO} detectors.
Some of these studies were initiated because of changes in detector performance while others were initiated because of detector configuration changes.
For low-latency, online transient \ac{GW} searches, we have summarized the Data Quality Report, a standard approach for quality validation that is carried out rapidly following the appearance of a candidate signal discovered in low-latency analyses.
This procedure has proven successful to streamline follow-up investigations and rapidly determine if \ac{GW} candidate signals are impacted by noise artifacts.
Modest upgrades to the Data Quality Report since the initial deployment in O4a have further improved the utility of this framework.

Low-latency, online transient \ac{GW} searches, follow-up \ac{GW} source parameter estimates, and high-latency, offline \ac{GW} searches rely on data quality products produced as part of the \detchar group efforts.
These products are essential in order to avoid analyzing data of poor quality that could impact search results.
The Data Quality Report is an example of a data quality product developed for O4 as part of the broader \detchar group activities which automates low-latency detector data quality assessment triggered by transient \ac{GW} candidates.

Further efforts to streamline other \detchar investigation, data quality product, and \ac{GW} candidate follow-up procedures will be essential as detector sensitivity continues to improve, observing run durations increase in length, and new classes of \ac{GW} signals are detected.
While automation efforts will be essential in the years ahead, we expect that future detector improvements will reveal new noise sources and coupling pathways that may require development of new investigative tools and mitigation strategies.
Integrating new tools with existing ones, maintaining these computational workflows, and being responsive to new developments requires an agile and dedicated team of personnel involved in LIGO \detchar activities.

\ack
The authors thank Beverly Berger for helpful comments and feedback while preparing the manuscript.

This material is based upon work supported by NSF's LIGO Laboratory which is a major facility fully funded by the National Science Foundation.
LIGO was constructed by the California Institute of Technology and Massachusetts Institute of Technology with funding from the National Science Foundation and operates under Cooperative Agreement PHY-2309200.
The authors are grateful for computational resources provided by the LIGO Laboratory and supported by National Science Foundation Grants PHY-0757058 and PHY-0823459.
LIGO Data Grid computing for LIGO detector characterization was supported by National Science Foundation awards 2110594 and 2513358.

AHC is grateful for support from the NSF grant PHY-2308793.
AC and JRM are supported by the Universitat de les Illes Balears (UIB) with funds from the Programa de Foment de la Recerca i la Innovaci\'o de la UIB 2024-2026 (supported by the yearly plan of the Tourist Stay Tax ITS2023-086); the Spanish Agencia Estatal de Investigaci\'on grants PID2022-138626NB-I00, RED2024-153978-E, RED2024-153735-E, funded by MICIU/AEI/10.13039/501100011033 and the ERDF/EU; and the Comunitat Autònoma de les Illes Balears through the Conselleria d’Educaci\'o i Universitats with funds from the ERDF (SINCO2022/18146 - Plataforma HiTech-IAC3-BIO).
JRM is also supported by the Spanish Ministerio de Ciencia, Innovaci\'on y Universidades (FPU22/01187).
This work was supported by a grant from the Simons Foundation International [SFI-MPS-SSRFA-00023625, DD].

This manuscript bears reference number P-2600357.

\section*{References}
\bibliography{references}

@article{2015CQGra..32g4001L,
	title        = {{Advanced LIGO}},
	author       = {Aasi, J. and others},
	year         = 2015,
	month        = apr,
	journal      = {Classical Quantum Gravity},
	volume       = 32,
	number       = 7,
	pages        = {074001},
	doi          = {10.1088/0264-9381/32/7/074001},
	collaboration = {{LIGO Scientific Collaboration}},
	eid          = {074001},
	archiveprefix = {arXiv},
	eprint       = {1411.4547},
	primaryclass = {gr-qc},
	adsurl       = {https://ui.adsabs.harvard.edu/abs/2015CQGra..32g4001L}
}

@ARTICLE{1993PhRvD..47.2198F,
       author = {{Finn}, Lee Samuel and {Chernoff}, David F.},
        title = "{Observing binary inspiral in gravitational radiation: One interferometer}",
      journal = {Phys. Rev. D},
         year = 1993,
        month = mar,
       volume = {47},
       number = {6},
        pages = {2198-2219},
          doi = {10.1103/PhysRevD.47.2198},
archivePrefix = {arXiv},
       eprint = {gr-qc/9301003},
 primaryClass = {gr-qc},
       adsurl = {https://ui.adsabs.harvard.edu/abs/1993PhRvD..47.2198F}
}

@ARTICLE{2026arXiv260823477T,
       author = {{Abac}, A.~G. and others},
        title = "{Updated Upper Limits on the Isotropic Gravitational-Wave Background from LIGO, Virgo, and KAGRA Data through April 2025}",
      journal = {arXiv e-prints},
	  collaboration = {{LIGO-Virgo-KAGRA Collaboration}},
         year = 2026,
        month = aug,
          eid = {arXiv:2608.23477},
        pages = {arXiv:2608.23477},
          doi = {10.48550/arXiv.2608.23477},
archivePrefix = {arXiv},
       eprint = {2608.23477},
 primaryClass = {gr-qc},
       adsurl = {https://ui.adsabs.harvard.edu/abs/2026arXiv260823477T}
}

@ARTICLE{2023NIMPA104867945A,
       author = {{Arnaud}, N.},
        title = {{LIGO and Virgo detector characterization and data quality: Contributions to the O3 run and preparation for O4}},
      journal = {Nucl. Instrum. Methods Phys. Res. A},
         year = 2023,
        month = mar,
       volume = {1048},
          eid = {167945},
        pages = {167945},
          doi = {10.1016/j.nima.2022.167945},
       adsurl = {https://ui.adsabs.harvard.edu/abs/2023NIMPA104867945A}
}

@article{2021CQGra..38l5005W,
	title        = {{Method for environmental noise estimation via injection tests for ground-based gravitational wave detectors}},
	author       = {{Washimi}, T. and {Yokozawa}, T. and {Tanaka}, T. and {Itoh}, Y. and {Kume}, J. and {Yokoyama}, J.},
	year         = 2021,
	month        = jun,
	journal      = {Classical Quantum Gravity},
	volume       = 38,
	number       = 12,
	pages        = 125005,
	doi          = {10.1088/1361-6382/abf89a},
	eid          = 125005,
	archiveprefix = {arXiv},
	eprint       = {2012.09294},
	primaryclass = {gr-qc},
	adsurl       = {https://ui.adsabs.harvard.edu/abs/2021CQGra..38l5005W}
}

@article{2026arXiv260725208A,
	title        = {{GSpyNetTree-O4: an event validation tool used in the fourth LIGO-Virgo-KAGRA observing run}},
	author       = {{\'Alvarez-L\'opez}, Sofia and {Chan}, Man Leong and {Herbst}, Franz S. and {Raghunathan}, Dhatri and {Ahuja}, Airene and {Liyanage}, Annudesh and {Ding}, Julian and {Garcia-Varela}, Alejandro and {Ng}, Raymond and {McIver}, Jess},
	year         = 2026,
	month        = jul,
	archiveprefix = {arXiv},
	eprint       = {2607.25208},
	primaryclass = {gr-qc},
	adsurl       = {https://ui.adsabs.harvard.edu/abs/2026arXiv260725208A}
}

@misc{PEMpage,
	title        = {PEM Central},
	url          = {https://pem.ligo.org/}
}

@article{2023CQGra..40r5006A,
	title        = {{Virgo detector characterization and data quality: results from the O3 run}},
	author       = {{Acernese}, F. and others},
	year         = 2023,
	month        = sep,
	journal      = {Classical Quantum Gravity},
	volume       = 40,
	number       = 18,
	pages        = 185006,
	doi          = {10.1088/1361-6382/acd92d},
	eid          = 185006,
	archiveprefix = {arXiv},
	eprint       = {2210.15633},
	primaryclass = {gr-qc},
	adsurl       = {https://ui.adsabs.harvard.edu/abs/2023CQGra..40r5006A}
}

@article{AREEDA201727,
	title        = {{LigoDV-web}: Providing easy, secure and universal access to a large distributed scientific data store for the {LIGO} scientific collaboration},
	author       = {J.S. Areeda and J.R. Smith and A.P. Lundgren and E. Maros and D.M. Macleod and J. Zweizig},
	year         = 2017,
	journal      = {Astron. Comput.},
	volume       = 18,
	pages        = {27--34},
	doi          = {https://doi.org/10.1016/j.ascom.2017.01.003},
	issn         = {2213-1337},
	url          = {https://www.sciencedirect.com/science/article/pii/S2213133716301196}
}

@article{2021SoftX..1400677F,
	title        = {{DQSEGDB: A time-interval database for storing gravitational wave observatory metadata}},
	author       = {{Fisher}, Ryan P. and {Hemming}, Gary and {Bizouard}, Marie-Anne and {Brown}, Duncan A. and {Couvares}, Peter F. and {Robinet}, Florent and {Verkindt}, Didier},
	year         = 2021,
	month        = jun,
	journal      = {SoftwareX},
	volume       = 14,
	pages        = 100677,
	doi          = {10.1016/j.softx.2021.100677},
	eid          = 100677,
	archiveprefix = {arXiv},
	eprint       = {2008.11316},
	primaryclass = {astro-ph.IM},
	adsurl       = {https://ui.adsabs.harvard.edu/abs/2021SoftX..1400677F}
}

@software{bruco,
	title        = {{BruCo: Brute-force coherence noise analsysis}},
	author       = {Vajente, G.},
	year         = 2023,
	month        = may,
	url          = {https://git.ligo.org/gabriele-vajente/bruco},
	version      = {0.2.1}
}

@software{ligocam,
	title        = {{LIGO Channel Activity Monitor }},
	author       = {Talukder, D.},
	year         = 2017,
	month        = sep,
	url          = {https://github.com/dipongkar/ligocam}
}

@article{2011CQGra..28w5005S,
	title        = {{A hierarchical method for vetoing noise transients in gravitational-wave detectors}},
	author       = {{Smith}, Joshua R. and {Abbott}, Thomas and {Hirose}, Eiichi and {Leroy}, Nicolas and {MacLeod}, Duncan and {McIver}, Jessica and {Saulson}, Peter and {Shawhan}, Peter},
	year         = 2011,
	month        = dec,
	journal      = {Classical Quantum Gravity},
	volume       = 28,
	number       = 23,
	pages        = 235005,
	doi          = {10.1088/0264-9381/28/23/235005},
	eid          = 235005,
	archiveprefix = {arXiv},
	eprint       = {1107.2948},
	primaryclass = {gr-qc},
	adsurl       = {https://ui.adsabs.harvard.edu/abs/2011CQGra..28w5005S}
}

@phdthesis{Meyers:2018nyo,
	title        = {{Cross-correlation searches for persistent gravitational waves with Advanced {LIGO} and noise studies for current and future ground-based gravitational-wave detectors}},
	author       = {Meyers, Patrick Michael},
	year         = 2018,
	school       = {Minnesota U.}
}

@techreport{Stochmon,
	title        = {\textit{{Stochmon: A LIGO Data Analysis Tool}}},
	author       = {Hernandez, G and Thrane. E},
	year         = 2023,
	number       = {T1400205},
	url          = {https://dcc.ligo.org/LIGO-T1400205/public},
	institution  = {LSC}
}

@inproceedings{2010JPhCS.243a2005I,
	title        = {{Used percentage veto for LIGO and virgo binary inspiral searches}},
	author       = {{Isogai}, Tomoki and {LIGO Scientific Collaboration} and {Virgo Collaboration}},
	year         = 2010,
	month        = aug,
	booktitle    = {J. Phys. Conf. Ser.},
	publisher    = {IOP},
	series       = {J. Phys. Conf. Ser.},
	volume       = 243,
	pages        = {012005},
	doi          = {10.1088/1742-6596/243/1/012005},
	eid          = {012005},
	adsurl       = {https://ui.adsabs.harvard.edu/abs/2010JPhCS.243a2005I}
}

@article{2021PTEP.2021eA102A,
	title        = {{Overview of KAGRA: Calibration, detector characterization, physical environmental monitors, and the geophysics interferometer}},
	author       = {{Akutsu}, T. and others},
	year         = 2021,
	month        = may,
	journal      = {Progress of Theoretical and Experimental Physics},
	volume       = 2021,
	number       = 5,
	pages        = {05A102},
	doi          = {10.1093/ptep/ptab018},
	eid          = {05A102},
	archiveprefix = {arXiv},
	eprint       = {2009.09305},
	primaryclass = {gr-qc},
	adsurl       = {https://ui.adsabs.harvard.edu/abs/2021PTEP.2021eA102A}
}

@article{2019NatAs...3...35K,
	title        = {{KAGRA: 2.5 generation interferometric gravitational wave detector}},
	author       = {{Kagra Collaboration}},
	year         = 2019,
	month        = jan,
	journal      = {Nat. Astron.},
	volume       = 3,
	pages        = {35--40},
	doi          = {10.1038/s41550-018-0658-y},
	archiveprefix = {arXiv},
	eprint       = {1811.08079},
	primaryclass = {gr-qc},
	adsurl       = {https://ui.adsabs.harvard.edu/abs/2019NatAs...3...35K}
}

@article{2015CQGra..32b4001A,
	title        = {{Advanced Virgo: a second-generation interferometric gravitational wave detector}},
	author       = {{Acernese}, F. and others},
	year         = 2015,
	month        = jan,
	journal      = {Classical Quantum Gravity},
	volume       = 32,
	number       = 2,
	pages        = {024001},
	doi          = {10.1088/0264-9381/32/2/024001},
	eid          = {024001},
	archiveprefix = {arXiv},
	eprint       = {1408.3978},
	primaryclass = {gr-qc},
	adsurl       = {https://ui.adsabs.harvard.edu/abs/2015CQGra..32b4001A}
}

@article{2022PhRvD.106j4017P,
	title        = {{Curious case of GW200129: Interplay between spin-precession inference and data-quality issues}},
	author       = {{Payne}, Ethan and {Hourihane}, Sophie and {Golomb}, Jacob and {Udall}, Rhiannon and {Davis}, Derek and {Chatziioannou}, Katerina},
	year         = 2022,
	month        = nov,
	journal      = {Phys. Rev. D},
	volume       = 106,
	number       = 10,
	pages        = 104017,
	doi          = {10.1103/PhysRevD.106.104017},
	eid          = 104017,
	archiveprefix = {arXiv},
	eprint       = {2206.11932},
	primaryclass = {gr-qc},
	adsurl       = {https://ui.adsabs.harvard.edu/abs/2022PhRvD.106j4017P}
}

@article{2026arXiv260407668L,
	title        = {{Coalescing Compact Binary Parameter Estimation with Gravitational Waves in the Presence of non-Gaussian Transient Noise}},
	author       = {{Lecoeuche}, Yannick and {McIver}, Jess and {Knee}, Alan M. and {Udall}, Rhiannon and {Rink}, Katie and {Hourihane}, Sophie and {Miller}, Simona J. and {Chatziioannou}, Katerina and {Massinger}, TJ and {Davis}, Derek},
	year         = 2026,
	month        = apr,
	doi          = {10.48550/arXiv.2604.07668},
	archiveprefix = {arXiv},
	eprint       = {2604.07668},
	primaryclass = {gr-qc},
	adsurl       = {https://ui.adsabs.harvard.edu/abs/2026arXiv260407668L}
}

@article{2026PhRvD.113d2005U,
	title        = {{Inferring the spins of merging black holes in the presence of data-quality issues}},
	author       = {{Udall}, Rhiannon and {Bini}, Sophie and {Chatziioannou}, Katerina and {Davis}, Derek and {Hourihane}, Sophie and {Lecoeuche}, Yannick and {McIver}, Jess and {Miller}, Simona},
	year         = 2026,
	month        = feb,
	journal      = {Phys. Rev. D},
	volume       = 113,
	number       = 4,
	pages        = {042005},
	doi          = {10.1103/hyz2-3wxy},
	eid          = {042005},
	archiveprefix = {arXiv},
	eprint       = {2510.05029},
	primaryclass = {gr-qc},
	adsurl       = {https://ui.adsabs.harvard.edu/abs/2026PhRvD.113d2005U}
}

@article{2025arXiv250818082T,
	title        = {{GWTC-4.0: Updating the Gravitational-Wave Transient Catalog with Observations from the First Part of the Fourth LIGO-Virgo-KAGRA Observing Run}},
	author       = {Abac, A. and others},
	year         = 2025,
	month        = aug,
	doi          = {10.48550/arXiv.2508.18082},
	collaboration = {{LIGO-Virgo-KAGRA Collaboration}},
	archiveprefix = {arXiv},
	eprint       = {2508.18082},
	primaryclass = {gr-qc},
	adsurl       = {https://ui.adsabs.harvard.edu/abs/2025arXiv250818082T}
}

@article{2026arXiv260527225T,
	title        = {{GWTC-5.0: Observations from the Second Part of the Fourth LIGO-Virgo-KAGRA Observing Run and Updates to the Gravitational-Wave Transient Catalog}},
	author       = {Abac, A. and others},
	year         = 2026,
	month        = may,
	doi          = {10.48550/arXiv.2605.27225},
	collaboration = {{LIGO-Virgo-KAGRA Collaboration}},
	archiveprefix = {arXiv},
	eprint       = {2605.27225},
	primaryclass = {gr-qc},
	adsurl       = {https://ui.adsabs.harvard.edu/abs/2026arXiv260527225T}
}

@article{2026arXiv260527090T,
	title        = {{Open Data from LIGO, Virgo, and KAGRA through the Second Part of the Fourth Observing Run}},
	author       = {Abac, A. and others},
	year         = 2026,
	month        = may,
	doi          = {10.48550/arXiv.2605.27090},
	archiveprefix = {arXiv},
	eprint       = {2605.27090},
	primaryclass = {gr-qc},
	adsurl       = {https://ui.adsabs.harvard.edu/abs/2026arXiv260527090T}
}

@article{2025arXiv250820721T,
	title        = {{Upper Limits on the Isotropic Gravitational-Wave Background from the first part of LIGO, Virgo, and KAGRA's fourth Observing Run}},
	author       = {{Abac}, A.~G. and others},
	year         = 2025,
	month        = aug,
	doi          = {10.48550/arXiv.2508.20721},
	collaboration = {{LIGO-Virgo-KAGRA Collaboration}},
	archiveprefix = {arXiv},
	eprint       = {2508.20721},
	primaryclass = {gr-qc},
	adsurl       = {https://ui.adsabs.harvard.edu/abs/2025arXiv250820721T}
}

@article{2016CQGra..33m4001A,
	title        = {{Characterization of transient noise in Advanced LIGO relevant to gravitational wave signal GW150914}},
	author       = {{Abbott}, B.~P. and other},
	year         = 2016,
	month        = jul,
	journal      = {Classical Quantum Gravity},
	volume       = 33,
	number       = 13,
	pages        = 134001,
	doi          = {10.1088/0264-9381/33/13/134001},
	eid          = 134001,
	archiveprefix = {arXiv},
	eprint       = {1602.03844},
	primaryclass = {gr-qc},
	adsurl       = {https://ui.adsabs.harvard.edu/abs/2016CQGra..33m4001A}
}

@article{2016PhRvD..93l2003A,
	title        = {{GW150914: First results from the search for binary black hole coalescence with Advanced LIGO}},
	author       = {{Abbott}, B.~P. and others},
	year         = 2016,
	month        = jun,
	journal      = {Phys. Rev. D},
	volume       = 93,
	number       = 12,
	pages        = 122003,
	doi          = {10.1103/PhysRevD.93.122003},
	eid          = 122003,
	archiveprefix = {arXiv},
	eprint       = {1602.03839},
	primaryclass = {gr-qc},
	adsurl       = {https://ui.adsabs.harvard.edu/abs/2016PhRvD..93l2003A}
}

@article{2019PhRvX...9c1040A,
	title        = {{GWTC-1: A Gravitational-Wave Transient Catalog of Compact Binary Mergers Observed by LIGO and Virgo during the First and Second Observing Runs}},
	author       = {{Abbott}, B.~P. and others},
	year         = 2019,
	month        = jul,
	journal      = {Phys. Rev. X},
	volume       = 9,
	number       = 3,
	pages        = {031040},
	doi          = {10.1103/PhysRevX.9.031040},
	collaboration = {{LIGO-Virgo Collaboration}},
	eid          = {031040},
	archiveprefix = {arXiv},
	eprint       = {1811.12907},
	primaryclass = {astro-ph.HE},
	adsurl       = {https://ui.adsabs.harvard.edu/abs/2019PhRvX...9c1040A}
}

@article{2021PhRvX..11b1053A,
	title        = {{GWTC-2: Compact Binary Coalescences Observed by LIGO and Virgo during the First Half of the Third Observing Run}},
	author       = {{Abbott}, R. and others},
	year         = 2021,
	month        = apr,
	journal      = {Phys. Rev. X},
	volume       = 11,
	number       = 2,
	pages        = {021053},
	doi          = {10.1103/PhysRevX.11.021053},
	collaboration = {{LIGO-Virgo Collaboration}},
	eid          = {021053},
	archiveprefix = {arXiv},
	eprint       = {2010.14527},
	primaryclass = {gr-qc},
	adsurl       = {https://ui.adsabs.harvard.edu/abs/2021PhRvX..11b1053A}
}

@article{2023PhRvX..13d1039A,
	title        = {{GWTC-3: Compact Binary Coalescences Observed by LIGO and Virgo during the Second Part of the Third Observing Run}},
	author       = {{Abbott}, R. and others},
	year         = 2023,
	month        = oct,
	journal      = {Phys. Rev. X},
	volume       = 13,
	number       = 4,
	pages        = {041039},
	doi          = {10.1103/PhysRevX.13.041039},
	eid          = {041039},
	archiveprefix = {arXiv},
	eprint       = {2111.03606},
	collaboration = {{LIGO-Virgo-KAGRA Collaboration}},
	primaryclass = {gr-qc},
	adsurl       = {https://ui.adsabs.harvard.edu/abs/2023PhRvX..13d1039A}
}

@article{aubin2021mbta,
	title        = {The MBTA pipeline for detecting compact binary coalescences in the third LIGO--Virgo observing run},
	author       = {Aubin, Florian and Brighenti, Francesco and Chierici, Roberto and Estevez, Dimitri and Greco, Giuseppe and Guidi, Gianluca Maria and Juste, Vincent and Marion, Fr{\'e}d{\'e}rique and Mours, Benoit and Nitoglia, Elisa and others},
	year         = 2021,
	journal      = {Classical Quantum Gravity},
	publisher    = {IOP Publishing},
	volume       = 38,
	number       = 9,
	pages        = {095004}
}

@article{2023ApPhL.122r4101B,
	title        = {{Searching for the causes of anomalous Advanced LIGO noise}},
	author       = {{Berger}, B.~K. and others},
	year         = 2023,
	month        = may,
	journal      = {Appl. Phys. Lett.},
	volume       = 122,
	number       = 18,
	pages        = 184101,
	doi          = {10.1063/5.0140766},
	eid          = 184101,
	adsurl       = {https://ui.adsabs.harvard.edu/abs/2023ApPhL.122r4101B}
}

@article{1991ASAJ...89..425B,
	title        = {{Calculation of a constant Q spectral transform}},
	author       = {{Brown}, Judith C.},
	year         = 1991,
	month        = jan,
	journal      = {J. Acoust. Soc. Am.},
	volume       = 89,
	number       = 1,
	pages        = {425--434},
	doi          = {10.1121/1.400476},
	adsurl       = {https://ui.adsabs.harvard.edu/abs/1991ASAJ...89..425B}
}

@article{2019CQGra..36o5010C,
	title        = {{Blip glitches in Advanced LIGO data}},
	author       = {{Cabero}, M. and others},
	year         = 2019,
	month        = aug,
	journal      = {Classical Quantum Gravity},
	volume       = 36,
	number       = 15,
	pages        = 155010,
	doi          = {10.1088/1361-6382/ab2e14},
	eid          = 155010,
	archiveprefix = {arXiv},
	eprint       = {1901.05093},
	primaryclass = {physics.ins-det},
	adsurl       = {https://ui.adsabs.harvard.edu/abs/2019CQGra..36o5010C}
}

@article{2025PhRvD.111f2002C,
	title        = {{Advanced LIGO detector performance in the fourth observing run}},
	author       = {{Capote}, E. and others},
	year         = 2025,
	month        = mar,
	journal      = {Phys. Rev. D},
	volume       = 111,
	number       = 6,
	pages        = {062002},
	doi          = {10.1103/PhysRevD.111.062002},
	eid          = {062002},
	archiveprefix = {arXiv},
	eprint       = {2411.14607},
	primaryclass = {gr-qc},
	adsurl       = {https://ui.adsabs.harvard.edu/abs/2025PhRvD.111f2002C}
}

@article{2015CQGra..32m5012C,
	title        = {{Bayeswave: Bayesian inference for gravitational wave bursts and instrument glitches}},
	author       = {{Cornish}, Neil J. and {Littenberg}, Tyson B.},
	year         = 2015,
	month        = jul,
	journal      = {Classical Quantum Gravity},
	volume       = 32,
	number       = 13,
	pages        = 135012,
	doi          = {10.1088/0264-9381/32/13/135012},
	eid          = 135012,
	archiveprefix = {arXiv},
	eprint       = {1410.3835},
	primaryclass = {gr-qc},
	adsurl       = {https://ui.adsabs.harvard.edu/abs/2015CQGra..32m5012C}
}

@article{2021PhRvD.103d4006C,
	title        = {{BayesWave analysis pipeline in the era of gravitational wave observations}},
	author       = {{Cornish}, Neil J. and {Littenberg}, Tyson B. and {B{\'e}csy}, Bence and {Chatziioannou}, Katerina and {Clark}, James A. and {Ghonge}, Sudarshan and {Millhouse}, Margaret},
	year         = 2021,
	month        = feb,
	journal      = {Phys. Rev. D},
	volume       = 103,
	number       = 4,
	pages        = {044006},
	doi          = {10.1103/PhysRevD.103.044006},
	eid          = {044006},
	archiveprefix = {arXiv},
	eprint       = {2011.09494},
	primaryclass = {gr-qc},
	adsurl       = {https://ui.adsabs.harvard.edu/abs/2021PhRvD.103d4006C}
}

@article{2021ApJ...923..254D,
	title        = {{Real-time Search for Compact Binary Mergers in Advanced LIGO and Virgo's Third Observing Run Using PyCBC Live}},
	author       = {{Dal Canton}, Tito and {Nitz}, Alexander H. and {Gadre}, Bhooshan and {Cabourn Davies}, Gareth S. and {Villa-Ortega}, Ver{\'o}nica and {Dent}, Thomas and {Harry}, Ian and {Xiao}, Liting},
	year         = 2021,
	month        = dec,
	journal      = {Astrophys. J.},
	volume       = 923,
	number       = 2,
	pages        = 254,
	doi          = {10.3847/1538-4357/ac2f9a},
	eid          = 254,
	archiveprefix = {arXiv},
	eprint       = {2008.07494},
	primaryclass = {astro-ph.HE},
	adsurl       = {https://ui.adsabs.harvard.edu/abs/2021ApJ...923..254D}
}

@article{2026arXiv260607679T,
	title        = {{PyCBC Live Search for Compact Binary Mergers in Advanced LIGO and Virgo's Fourth Observing Run}},
	author       = {{Trevor}, Max and {Cabourn Davies}, Gareth S. and {Dal Canton}, Tito and {Dent}, Thomas and {Harry}, Ian and {Hoang}, Stephanie and {Tolley}, Arthur},
	year         = 2026,
	month        = jun,
	doi          = {10.48550/arXiv.2606.07679},
	archiveprefix = {arXiv},
	eprint       = {2606.07679},
	primaryclass = {astro-ph.IM},
	adsurl       = {https://ui.adsabs.harvard.edu/abs/2026arXiv260607679T}
}

@article{2018RSPTA.37670286N,
	title        = {{Characterizing transient noise in the LIGO detectors}},
	author       = {{Nuttall}, L.~K.},
	year         = 2018,
	month        = may,
	journal      = {Philos. Trans. R. Soc. London, Ser. A},
	volume       = 376,
	number       = 2120,
	pages        = 20170286,
	doi          = {10.1098/rsta.2017.0286},
	eid          = 20170286,
	archiveprefix = {arXiv},
	eprint       = {1804.07592},
	primaryclass = {astro-ph.IM},
	adsurl       = {https://ui.adsabs.harvard.edu/abs/2018RSPTA.37670286N}
}

@article{2021CQGra..38m5014D,
	title        = {{LIGO detector characterization in the second and third observing runs}},
	author       = {{Davis}, D. and others},
	year         = 2021,
	month        = jul,
	journal      = {Classical Quantum Gravity},
	volume       = 38,
	number       = 13,
	pages        = 135014,
	doi          = {10.1088/1361-6382/abfd85},
	eid          = 135014,
	archiveprefix = {arXiv},
	eprint       = {2101.11673},
	primaryclass = {astro-ph.IM},
	adsurl       = {https://ui.adsabs.harvard.edu/abs/2021CQGra..38m5014D}
}

@article{2025PhRvD.111l2005D,
	title        = {{Guiding interferometer improvements with the frequency-dependent inspiral range}},
	author       = {{Davis}, Derek and {Capote}, Elenna},
	year         = 2025,
	month        = jun,
	journal      = {Phys. Rev. D},
	volume       = 111,
	number       = 12,
	pages        = 122005,
	doi          = {10.1103/bfs5-dq5k},
	eid          = 122005,
	archiveprefix = {arXiv},
	eprint       = {2502.07253},
	primaryclass = {astro-ph.IM},
	adsurl       = {https://ui.adsabs.harvard.edu/abs/2025PhRvD.111l2005D}
}

@article{2022PhRvD.106j2006D,
	title        = {{Incorporating information from LIGO data quality streams into the PyCBC search for gravitational waves}},
	author       = {{Davis}, Derek and {Trevor}, Max and {Mozzon}, Simone and {Nuttall}, Laura K.},
	year         = 2022,
	month        = nov,
	journal      = {Phys. Rev. D},
	volume       = 106,
	number       = 10,
	pages        = 102006,
	doi          = {10.1103/PhysRevD.106.102006},
	eid          = 102006,
	archiveprefix = {arXiv},
	eprint       = {2204.03091},
	primaryclass = {gr-qc},
	adsurl       = {https://ui.adsabs.harvard.edu/abs/2022PhRvD.106j2006D}
}

@article{2024arXiv240115392D,
	title        = {{BRiSTOL -- a Band-limited RMS Stationarity Test Tool for Gravitational Wave Data}},
	author       = {{Di Renzo}, F. and {Fidecaro}, F. and {Razzano}, M. and {Sorrentino}, N.},
	year         = 2024,
	month        = jan,
	doi          = {10.48550/arXiv.2401.15392},
	archiveprefix = {arXiv},
	eprint       = {2401.15392},
	primaryclass = {gr-qc},
	adsurl       = {https://ui.adsabs.harvard.edu/abs/2024arXiv240115392D}
}

@article{2021SoftX..1400678D,
	title        = {{coherent WaveBurst, a pipeline for unmodeled gravitational-wave data analysis}},
	author       = {{Drago}, Marco and others},
	year         = 2021,
	month        = jun,
	journal      = {SoftwareX},
	volume       = 14,
	pages        = 100678,
	doi          = {10.1016/j.softx.2021.100678},
	eid          = 100678,
	archiveprefix = {arXiv},
	eprint       = {2006.12604},
	primaryclass = {gr-qc},
	adsurl       = {https://ui.adsabs.harvard.edu/abs/2021SoftX..1400678D}
}

@article{2015CQGra..32c5017E,
	title        = {{Environmental influences on the LIGO gravitational wave detectors during the 6th science run}},
	author       = {{Effler}, A. and {Schofield}, R.~M.~S. and {Frolov}, V.~V. and {Gonz{\'a}lez}, G. and {Kawabe}, K. and {Smith}, J.~R. and {Birch}, J. and {McCarthy}, R.},
	year         = 2015,
	month        = feb,
	journal      = {Classical Quantum Gravity},
	volume       = 32,
	number       = 3,
	pages        = {035017},
	doi          = {10.1088/0264-9381/32/3/035017},
	eid          = {035017},
	archiveprefix = {arXiv},
	eprint       = {1409.5160},
	primaryclass = {astro-ph.IM},
	adsurl       = {https://ui.adsabs.harvard.edu/abs/2015CQGra..32c5017E}
}

@article{2013CQGra..30o5010E,
	title        = {{Optimizing vetoes for gravitational-wave transient searches}},
	author       = {{Essick}, R. and {Blackburn}, L. and {Katsavounidis}, E.},
	year         = 2013,
	month        = aug,
	journal      = {Classical Quantum Gravity},
	volume       = 30,
	number       = 15,
	pages        = 155010,
	doi          = {10.1088/0264-9381/30/15/155010},
	eid          = 155010,
	archiveprefix = {arXiv},
	eprint       = {1303.7159},
	primaryclass = {astro-ph.IM},
	adsurl       = {https://ui.adsabs.harvard.edu/abs/2013CQGra..30o5010E}
}

@article{2020arXiv200512761E,
	title        = {{iDQ: Statistical Inference of Non-Gaussian Noise with Auxiliary Degrees of Freedom in Gravitational-Wave Detectors}},
	author       = {{Essick}, Reed and {Godwin}, Patrick and {Hanna}, Chad and {Blackburn}, Lindy and {Katsavounidis}, Erik},
	year         = 2020,
	month        = may,
	doi          = {10.48550/arXiv.2005.12761},
	archiveprefix = {arXiv},
	eprint       = {2005.12761},
	primaryclass = {astro-ph.IM},
	adsurl       = {https://ui.adsabs.harvard.edu/abs/2020arXiv200512761E}
}

@article{2021PhRvD.103d2003E,
	title        = {{A coincidence null test for Poisson-distributed events}},
	author       = {{Essick}, Reed and {Mo}, Geoffrey and {Katsavounidis}, Erik},
	year         = 2021,
	month        = feb,
	journal      = {Phys. Rev. D},
	volume       = 103,
	number       = 4,
	pages        = {042003},
	doi          = {10.1103/PhysRevD.103.042003},
	eid          = {042003},
	archiveprefix = {arXiv},
	eprint       = {2011.13787},
	primaryclass = {gr-qc},
	adsurl       = {https://ui.adsabs.harvard.edu/abs/2021PhRvD.103d2003E}
}

@article{2023PhRvX..13d1021G,
	title        = {{Broadband Quantum Enhancement of the LIGO Detectors with Frequency-Dependent Squeezing}},
	author       = {{Ganapathy}, D. and others},
	year         = 2023,
	month        = oct,
	journal      = {Phys. Rev. X},
	volume       = 13,
	number       = 4,
	pages        = {041021},
	doi          = {10.1103/PhysRevX.13.041021},
	eid          = {041021},
	adsurl       = {https://ui.adsabs.harvard.edu/abs/2023PhRvX..13d1021G}
}

@article{2023CQGra..40f5004G,
	title        = {{Data quality up to the third observing run of advanced LIGO: Gravity Spy glitch classifications}},
	author       = {{Glanzer}, J. and others},
	year         = 2023,
	month        = mar,
	journal      = {Classical Quantum Gravity},
	volume       = 40,
	number       = 6,
	pages        = {065004},
	doi          = {10.1088/1361-6382/acb633},
	eid          = {065004},
	archiveprefix = {arXiv},
	eprint       = {2208.12849},
	primaryclass = {gr-qc},
	adsurl       = {https://ui.adsabs.harvard.edu/abs/2023CQGra..40f5004G}
}

@article{2026arXiv260605959G,
	title        = {{Narrow spectral artifact investigation and mitigation in LIGO data from the fourth LIGO-Virgo-KAGRA observing run}},
	author       = {{Goetz}, E. and others},
	year         = 2026,
	month        = jun,
	doi          = {10.48550/arXiv.2606.05959},
	archiveprefix = {arXiv},
	eprint       = {2606.05959},
	primaryclass = {astro-ph.IM},
	adsurl       = {https://ui.adsabs.harvard.edu/abs/2026arXiv260605959G}
}

@article{2024CQGra..41n5003H,
	title        = {{Automated evaluation of environmental coupling for Advanced LIGO gravitational wave detections}},
	author       = {{Helmling-Cornell}, A.~F. and {Nguyen}, P. and {Schofield}, R.~M.~S. and {Frey}, R.},
	year         = 2024,
	month        = jul,
	journal      = {Classical Quantum Gravity},
	volume       = 41,
	number       = 14,
	pages        = 145003,
	doi          = {10.1088/1361-6382/ad5139},
	eid          = 145003,
	archiveprefix = {arXiv},
	eprint       = {2312.00735},
	primaryclass = {gr-qc},
	adsurl       = {https://ui.adsabs.harvard.edu/abs/2024CQGra..41n5003H}
}

@article{hooper2012summed,
       author = {{Hooper}, Shaun and {Chung}, Shin Kee and {Luan}, Jing and {Blair}, David and {Chen}, Yanbei and {Wen}, Linqing},
        title = "{Summed parallel infinite impulse response filters for low-latency detection of chirping gravitational waves}",
      journal = {Phys. Rev. D},
         year = 2012,
        month = jul,
       volume = {86},
       number = {2},
          eid = {024012},
        pages = {024012},
          doi = {10.1103/PhysRevD.86.024012},
archivePrefix = {arXiv},
       eprint = {1108.3186},
 primaryClass = {gr-qc},
       adsurl = {https://ui.adsabs.harvard.edu/abs/2012PhRvD..86b4012H}
}

@article{2025PhRvD.112h4006H,
	title        = {{Glitches far from transient gravitational-wave events do not bias inference}},
	author       = {{Hourihane}, Sophie and {Chatziioannou}, Katerina},
	year         = 2025,
	month        = oct,
	journal      = {Phys. Rev. D},
	volume       = 112,
	number       = 8,
	pages        = {084006},
	doi          = {10.1103/kxbb-dpwp},
	eid          = {084006},
	archiveprefix = {arXiv},
	eprint       = {2506.21869},
	primaryclass = {gr-qc},
	adsurl       = {https://ui.adsabs.harvard.edu/abs/2025PhRvD.112h4006H}
}

@article{2025PhRvD.112j2003J,
	title        = {{Coherent injection of magnetic noise and its impact on gravitational-wave searches}},
	author       = {{Janssens}, Kamiel and {Lawrence}, Jessica and {Effler}, Anamaria and {Schofield}, Robert M.~S. and {Lalleman}, Max and {Betzwieser}, Joseph and {Christensen}, Nelson and {Coughlin}, Michael W. and {Driggers}, Jennifer C. and {Helmling-Cornell}, Adrian F. and others},
	year         = 2025,
	month        = nov,
	journal      = {{Phys. Rev. D}},
	volume       = 112,
	number       = 10,
	pages        = 102003,
	doi          = {10.1103/hll9-qmtk},
	eid          = 102003,
	archiveprefix = {arXiv},
	eprint       = {2505.11903},
	primaryclass = {gr-qc},
	adsurl       = {https://ui.adsabs.harvard.edu/abs/2025PhRvD.112j2003J}
}

@article{2016RScI...87k4503K,
	title        = {{The Advanced LIGO photon calibrators}},
	author       = {{Karki}, S. and others},
	year         = 2016,
	month        = nov,
	journal      = {Rev. Sci. Instrum.},
	volume       = 87,
	number       = 11,
	pages        = 114503,
	doi          = {10.1063/1.4967303},
	eid          = 114503,
	archiveprefix = {arXiv},
	eprint       = {1608.05055},
	primaryclass = {astro-ph.IM},
	adsurl       = {https://ui.adsabs.harvard.edu/abs/2016RScI...87k4503K}
}

@article{2011CQGra..28x5001L,
	title        = {{First results from the {\textquoteleft}Violin-Mode{\textquoteright} tests on an advanced LIGO suspension at MIT}},
	author       = {{Lockerbie}, N.~A. and {Carbone}, L. and {Shapiro}, B. and {Tokmakov}, K.~V. and {Bell}, A. and {Strain}, K.~A.},
	year         = 2011,
	month        = dec,
	journal      = {Classical Quantum Gravity},
	volume       = 28,
	number       = 24,
	pages        = 245001,
	doi          = {10.1088/0264-9381/28/24/245001},
	eid          = 245001,
	adsurl       = {https://ui.adsabs.harvard.edu/abs/2011CQGra..28x5001L}
}

@article{luan2012towards,
       author = {{Luan}, Jing and {Hooper}, Shaun and {Wen}, Linqing and {Chen}, Yanbei},
        title = "{Towards low-latency real-time detection of gravitational waves from compact binary coalescences in the era of advanced detectors}",
      journal = {Phys. Rev. D},
         year = 2012,
        month = may,
       volume = {85},
       number = {10},
          eid = {102002},
        pages = {102002},
          doi = {10.1103/PhysRevD.85.102002},
archivePrefix = {arXiv},
       eprint = {1108.3174},
 primaryClass = {gr-qc},
       adsurl = {https://ui.adsabs.harvard.edu/abs/2012PhRvD..85j2002L}
}

@article{2024CQGra..41h5007A,
	title        = {{GSpyNetTree: a signal-vs-glitch classifier for gravitational-wave event candidates}},
	author       = {{{\'A}lvarez-L{\'o}pez}, Sof{\'\i}a and {Liyanage}, Annudesh and {Ding}, Julian and {Ng}, Raymond and {McIver}, Jess},
	year         = 2024,
	month        = apr,
	journal      = {Classical Quantum Gravity},
	volume       = 41,
	number       = 8,
	pages        = {085007},
	doi          = {10.1088/1361-6382/ad2194},
	eid          = {085007},
	archiveprefix = {arXiv},
	eprint       = {2304.09977},
	primaryclass = {gr-qc},
	adsurl       = {https://ui.adsabs.harvard.edu/abs/2024CQGra..41h5007A}
}

@article{2023PhRvD.108f3016M,
	title        = {{Sensitive test of non-Gaussianity in gravitational-wave detector data}},
	author       = {{Macas}, Ronaldas and {Lundgren}, Andrew},
	year         = 2023,
	month        = sep,
	journal      = {Phys. Rev. D},
	volume       = 108,
	number       = 6,
	pages        = {063016},
	doi          = {10.1103/PhysRevD.108.063016},
	eid          = {063016},
	archiveprefix = {arXiv},
	eprint       = {2306.09019},
	primaryclass = {gr-qc},
	adsurl       = {https://ui.adsabs.harvard.edu/abs/2023PhRvD.108f3016M}
}

@article{2021PhRvD.104j2005M,
	title        = {{Long-duration transient gravitational-wave search pipeline}},
	author       = {{Macquet}, A. and {Bizouard}, M.~A. and {Christensen}, N. and {Coughlin}, M.},
	year         = 2021,
	month        = nov,
	journal      = {Phys. Rev. D},
	volume       = 104,
	number       = 10,
	pages        = 102005,
	doi          = {10.1103/PhysRevD.104.102005},
	eid          = 102005,
	archiveprefix = {arXiv},
	eprint       = {2108.10588},
	primaryclass = {astro-ph.IM},
	adsurl       = {https://ui.adsabs.harvard.edu/abs/2021PhRvD.104j2005M}
}

@article{marx2025machine,
	title        = {Machine-learning pipeline for real-time detection of gravitational waves from compact binary coalescences},
	author       = {Marx, Ethan and Benoit, William and Gunny, Alec and Omer, Rafia and Chatterjee, Deep and Venterea, Ricco C and Wills, Lauren and Saleem, Muhammed and Moreno, Eric and Raikman, Ryan and others},
	year         = 2025,
	journal      = {Phys. Rev. D},
	publisher    = {APS},
	volume       = 111,
	number       = 4,
	pages        = {042010}
}

@article{2015CQGra..32r5003M,
	title        = {{Seismic isolation of Advanced LIGO: Review of strategy, instrumentation and performance}},
	author       = {{Matichard}, F. and others},
	year         = 2015,
	month        = sep,
	journal      = {Classical Quantum Gravity},
	volume       = 32,
	number       = 18,
	pages        = 185003,
	doi          = {10.1088/0264-9381/32/18/185003},
	eid          = 185003,
	archiveprefix = {arXiv},
	eprint       = {1502.06300},
	primaryclass = {physics.ins-det},
	adsurl       = {https://ui.adsabs.harvard.edu/abs/2015CQGra..32r5003M}
}

@article{messick2017analysis,
	title        = {Analysis framework for the prompt discovery of compact binary mergers in gravitational-wave data},
	author       = {Messick, Cody and Blackburn, Kent and Brady, Patrick and Brockill, Patrick and Cannon, Kipp and Cariou, Romain and Caudill, Sarah and Chamberlin, Sydney J and Creighton, Jolien DE and Everett, Ryan and others},
	year         = 2017,
	journal      = {Phys. Rev. D},
	publisher    = {APS},
	volume       = 95,
	number       = 4,
	pages        = {042001}
}

@article{mishra2022search,
	title        = {Search for binary black hole mergers in the third observing run of Advanced LIGO-Virgo using coherent WaveBurst enhanced with machine learning},
	author       = {Mishra, Tanmaya and O’Brien, Brendan and Szczepa{\'n}czyk, M and Vedovato, Gabriele and Bhaumik, Shubhagata and Gayathri, V and Prodi, Giovanni and Salemi, Francesco and Milotti, Edoardo and Bartos, Imre and others},
	year         = 2022,
	journal      = {Phys. Rev. D},
	publisher    = {APS},
	volume       = 105,
	number       = 8,
	pages        = {083018}
}

@article{2020CQGra..37u5014M,
	title        = {{Dynamic normalization for compact binary coalescence searches in non-stationary noise}},
	author       = {{Mozzon}, S. and {Nuttall}, L.~K. and {Lundgren}, A. and {Dent}, T. and {Kumar}, S. and {Nitz}, A.~H.},
	year         = 2020,
	month        = nov,
	journal      = {Classical Quantum Gravity},
	volume       = 37,
	number       = 21,
	pages        = 215014,
	doi          = {10.1088/1361-6382/abac6c},
	eid          = 215014,
	archiveprefix = {arXiv},
	eprint       = {2002.09407},
	primaryclass = {astro-ph.IM},
	adsurl       = {https://ui.adsabs.harvard.edu/abs/2020CQGra..37u5014M}
}

@article{2026arXiv260514143N,
	title        = {{Scattered light noise at LIGO Livingston Observatory during O4}},
	author       = {{Nandi}, Debasmita and {Effler}, Anamaria and {Soni}, Siddharth and {Aira Ferreira}, Tabata and {Schofield}, Robert and {Pham}, Huyen and {O'Hanlon}, Timothy and {Frolov}, V.~V. and {Gonz{\'a}lez}, Gabriela},
	year         = 2026,
	month        = may,
	doi          = {10.48550/arXiv.2605.14143},
	archiveprefix = {arXiv},
	eprint       = {2605.14143},
	primaryclass = {astro-ph.IM},
	adsurl       = {https://ui.adsabs.harvard.edu/abs/2026arXiv260514143N}
}

@article{2021CQGra..38n5001N,
	title        = {{Environmental noise in advanced LIGO detectors}},
	author       = {{Nguyen}, P. and {Schofield}, R.~M.~S. and {Effler}, A. and others},
	year         = 2021,
	month        = jul,
	journal      = {Classical Quantum Gravity},
	volume       = 38,
	number       = 14,
	pages        = 145001,
	doi          = {10.1088/1361-6382/ac011a},
	eid          = 145001,
	archiveprefix = {arXiv},
	eprint       = {2101.09935},
	primaryclass = {astro-ph.IM},
	adsurl       = {https://ui.adsabs.harvard.edu/abs/2021CQGra..38n5001N}
}

@article{2023ApJ...952...25R,
	title        = {{pygwb: A Python-based Library for Gravitational-wave Background Searches}},
	author       = {{Renzini}, Arianna I. and others},
	year         = 2023,
	month        = jul,
	journal      = {Astrophys. J.},
	volume       = 952,
	number       = 1,
	pages        = 25,
	doi          = {10.3847/1538-4357/acd775},
	eid          = 25,
	archiveprefix = {arXiv},
	eprint       = {2303.15696},
	primaryclass = {gr-qc},
	adsurl       = {https://ui.adsabs.harvard.edu/abs/2023ApJ...952...25R}
}

@article{2020SoftX..1200620R,
	title        = {{Omicron: A tool to characterize transient noise in gravitational-wave detectors}},
	author       = {{Robinet}, Florent and {Arnaud}, Nicolas and {Leroy}, Nicolas and {Lundgren}, Andrew and {Macleod}, Duncan and {McIver}, Jessica},
	year         = 2020,
	month        = jul,
	journal      = {SoftwareX},
	volume       = 12,
	pages        = 100620,
	doi          = {10.1016/j.softx.2020.100620},
	eid          = 100620,
	archiveprefix = {arXiv},
	eprint       = {2007.11374},
	primaryclass = {astro-ph.IM},
	adsurl       = {https://ui.adsabs.harvard.edu/abs/2020SoftX..1200620R}
}

@article{2019CQGra..36e5011D,
	title        = {{Improving the sensitivity of Advanced LIGO using noise subtraction}},
	author       = {{Davis}, Derek and {Massinger}, Thomas and {Lundgren}, Andrew and {Driggers}, Jennifer C. and {Urban}, Alex L. and {Nuttall}, Laura},
	year         = 2019,
	month        = mar,
	journal      = {Classical Quantum Gravity},
	volume       = 36,
	number       = 5,
	pages        = {055011},
	doi          = {10.1088/1361-6382/ab01c5},
	eid          = {055011},
	archiveprefix = {arXiv},
	eprint       = {1809.05348},
	primaryclass = {astro-ph.IM},
	adsurl       = {https://ui.adsabs.harvard.edu/abs/2019CQGra..36e5011D}
}

@article{2020CQGra..37w5007S,
	title        = {{Improving the robustness of the advanced LIGO detectors to earthquakes}},
	author       = {{Schwartz}, E. and others},
	year         = 2020,
	month        = dec,
	journal      = {Classical Quantum Gravity},
	volume       = 37,
	number       = 23,
	pages        = 235007,
	doi          = {10.1088/1361-6382/abbc8c},
	eid          = 235007,
	archiveprefix = {arXiv},
	eprint       = {2007.12847},
	primaryclass = {physics.ins-det},
	adsurl       = {https://ui.adsabs.harvard.edu/abs/2020CQGra..37w5007S}
}

@article{2021CQGra..38b5016S,
	title        = {{Reducing scattered light in LIGO's third observing run}},
	author       = {{Soni}, S. and others},
	year         = 2021,
	month        = jan,
	journal      = {Classical Quantum Gravity},
	volume       = 38,
	number       = 2,
	pages        = {025016},
	doi          = {10.1088/1361-6382/abc906},
	eid          = {025016},
	archiveprefix = {arXiv},
	eprint       = {2007.14876},
	primaryclass = {astro-ph.IM},
	adsurl       = {https://ui.adsabs.harvard.edu/abs/2021CQGra..38b5016S}
}

@article{2015PhRvL.114p1102E,
	title        = {{Observation of Parametric Instability in Advanced LIGO}},
	author       = {{Evans}, M. and others},
	year         = 2015,
	month        = apr,
	journal      = {Phys. Rev. Lett.},
	volume       = 114,
	number       = 16,
	pages        = 161102,
	doi          = {10.1103/PhysRevLett.114.161102},
	eid          = 161102,
	archiveprefix = {arXiv},
	eprint       = {1502.06058},
	primaryclass = {astro-ph.IM}
}

@article{2025CQGra..42h5016S,
	title        = {{LIGO Detector Characterization in the first half of the fourth Observing run}},
	author       = {{Soni}, S. and others},
	year         = 2025,
	month        = apr,
	journal      = {Classical Quantum Gravity},
	volume       = 42,
	number       = 8,
	pages        = {085016},
	doi          = {10.1088/1361-6382/adc4b6},
	eid          = {085016},
	archiveprefix = {arXiv},
	eprint       = {2409.02831},
	primaryclass = {astro-ph.IM},
	adsurl       = {https://ui.adsabs.harvard.edu/abs/2025CQGra..42h5016S}
}

@article{usman2016pycbc,
	title        = {The PyCBC search for gravitational waves from compact binary coalescence},
	author       = {Usman, Samantha A and Nitz, Alexander H and Harry, Ian W and Biwer, Christopher M and Brown, Duncan A and Cabero, Miriam and Capano, Collin D and Canton, Tito Dal and Dent, Thomas and Fairhurst, Stephen and others},
	year         = 2016,
	journal      = {Classical Quantum Gravity},
	publisher    = {IOP Publishing},
	volume       = 33,
	number       = 21,
	pages        = 215004
}

@article{2023CQGra..40c5008V,
	title        = {{Identifying glitches near gravitational-wave signals from compact binary coalescences using the Q-transform}},
	author       = {{Vazsonyi}, Leah and {Davis}, Derek},
	year         = 2023,
	month        = feb,
	journal      = {Classical Quantum Gravity},
	volume       = 40,
	number       = 3,
	pages        = {035008},
	doi          = {10.1088/1361-6382/acafd2},
	eid          = {035008},
	archiveprefix = {arXiv},
	eprint       = {2208.12338},
	primaryclass = {astro-ph.IM},
	adsurl       = {https://ui.adsabs.harvard.edu/abs/2023CQGra..40c5008V}
}

@article{2025PhRvD.111h2005V,
	title        = {{Impact of correlated magnetic noise on directional stochastic gravitational-wave background searches}},
	author       = {{Venikoudis}, Stavros and {De Lillo}, Federico and {Janssens}, Kamiel and {Suresh}, Jishnu and {Bruno}, Giacomo},
	year         = 2025,
	month        = apr,
	journal      = {{Phys. Rev. D}},
	volume       = 111,
	number       = 8,
	pages        = {082005},
	doi          = {10.1103/PhysRevD.111.082005},
	eid          = {082005},
	archiveprefix = {arXiv},
	eprint       = {2411.11746},
	primaryclass = {gr-qc},
	adsurl       = {https://ui.adsabs.harvard.edu/abs/2025PhRvD.111h2005V}
}

@article{2018CQGra..35v5002W,
	title        = {{Identifying correlations between LIGO{\textquoteright}s astronomical range and auxiliary sensors using lasso regression}},
	author       = {{Walker}, Marissa and {Agnew}, Alfonso F. and {Bidler}, Jeffrey and {Lundgren}, Andrew and {Macedo}, Alexandra and {Macleod}, Duncan and {Massinger}, T.~J. and {Patane}, Oliver and {Smith}, Joshua R.},
	year         = 2018,
	month        = nov,
	journal      = {Classical Quantum Gravity},
	volume       = 35,
	number       = 22,
	pages        = 225002,
	doi          = {10.1088/1361-6382/aae593},
	eid          = 225002,
	archiveprefix = {arXiv},
	eprint       = {1807.02592},
	primaryclass = {astro-ph.IM},
	adsurl       = {https://ui.adsabs.harvard.edu/abs/2018CQGra..35v5002W}
}

@article{2017CQGra..34f4003Z,
	title        = {{Gravity Spy: integrating advanced LIGO detector characterization, machine learning, and citizen science}},
	author       = {{Zevin}, M. and others},
	year         = 2017,
	month        = mar,
	journal      = {Classical Quantum Gravity},
	volume       = 34,
	number       = 6,
	pages        = {064003},
	doi          = {10.1088/1361-6382/aa5cea},
	eid          = {064003},
	archiveprefix = {arXiv},
	eprint       = {1611.04596},
	primaryclass = {gr-qc},
	adsurl       = {https://ui.adsabs.harvard.edu/abs/2017CQGra..34f4003Z}
}

@article{davis2026rapid,
	author = {{Davis}, Derek and others},
        title = "{Rapid data quality investigations of gravitational-wave events with the Data Quality Report Builder toolkit}",
      journal = {arXiv e-prints},
         year = 2026,
        month = may,
          eid = {arXiv:2605.16183},
        pages = {arXiv:2605.16183},
          doi = {10.48550/arXiv.2605.16183},
archivePrefix = {arXiv},
       eprint = {2605.16183},
 primaryClass = {astro-ph.IM},
       adsurl = {https://ui.adsabs.harvard.edu/abs/2026arXiv260516183D}
}

@article{acernese2023virgo,
	title        = {Virgo detector characterization and data quality: tools},
	author       = {Acernese, F and Agathos, M and Ain, A and Albanesi, S and Allocca, A and Amato, A and Andrade, T and Andres, N and Andr{\'e}s-Carcasona, M and Andri{\'c}, T and others},
	year         = 2023,
	journal      = {Classical Quantum Gravity},
	publisher    = {IOP Publishing},
	volume       = 40,
	number       = 18,
	pages        = 185005
}

@article{skliris2024toward,
	title        = {Toward real-time detection of unmodeled gravitational wave transients using convolutional neural networks},
	author       = {Skliris, Vasileios and Norman, Michael RK and Sutton, Patrick J},
	year         = 2024,
	journal      = {Phys. Rev. D},
	publisher    = {APS},
	volume       = 110,
	number       = 10,
	pages        = 104034
}

@article{2025GCN.38856....1L,
	title        = {{LIGO/Virgo/KAGRA S250108ha: Retraction of GW compact binary merger candidate}},
	author       = {{Ligo Scientific Collaboration} and {VIRGO Collaboration} and {Kagra Collaboration}},
	year         = 2025,
	month        = jan,
	journal      = {GRB Coordinates Network},
	volume       = 38856,
	pages        = 1,
	adsurl       = {https://ui.adsabs.harvard.edu/abs/2025GCN.38856....1L}
}

@article{2024GCN.38070....1L,
	title        = {{LIGO/Virgo/KAGRA S241104a: Retraction of GW compact binary merger candidate}},
	author       = {{Ligo Scientific Collaboration} and {VIRGO Collaboration} and {Kagra Collaboration}},
	year         = 2024,
	month        = nov,
	journal      = {GRB Coordinates Network},
	volume       = 38070,
	pages        = 1,
	adsurl       = {https://ui.adsabs.harvard.edu/abs/2024GCN.38070....1L}
}

@article{2024GCN.36747....1L,
	title        = {{LIGO/Virgo/KAGRA S240624cd: Retraction of GW unmodeled transient candidate}},
	author       = {{Ligo Scientific Collaboration} and {VIRGO Collaboration} and {Kagra Collaboration}},
	year         = 2024,
	month        = jun,
	journal      = {GRB Coordinates Network},
	volume       = 36747,
	pages        = 1,
	adsurl       = {https://ui.adsabs.harvard.edu/abs/2024GCN.36747....1L}
}

@article{2024GCN.36190....1L,
	title        = {{LIGO/Virgo/KAGRA S240420aw: Retraction of GW compact binary merger candidate}},
	author       = {{Ligo Scientific Collaboration} and {VIRGO Collaboration} and {Kagra Collaboration}},
	year         = 2024,
	month        = apr,
	journal      = {GRB Coordinates Network},
	volume       = 36190,
	pages        = 1,
	adsurl       = {https://ui.adsabs.harvard.edu/abs/2024GCN.36190....1L}
}

@misc{alog:78693,
	title        = {{BS HEPI injections imply large amount of scattered light in the vertex, from ITM elliptical baffles}},
	author       = {{Pham}, H. and {O'Hanlon}, T. and {Effler}, A.},
	howpublished = {\url{https://alog.ligo-la.caltech.edu/aLOG/index.php?callRep=78693}}
}

@misc{64337llo,
	title        = {{Electrical glitches remain at EY after ESD transitioned to EX}},
	author       = {Lundgren, A.},
	howpublished = {\url{https://alog.ligo-la.caltech.edu/aLOG/index.php?callRep=64377}}
}

@misc{Ashton2024CBCFlow,
	title        = {{CBCFlow}},
	author       = {{Ashton, G. et. al.}},
	year         = 2024,
	howpublished = {\url{https://cbc.docs.ligo.org/projects/cbcflow/index.html}}
}

@misc{alog:78903,
	title        = {{Greater angular offload to L1}},
	author       = {{Bossilkov}, V.},
	howpublished = {\url{https://alog.ligo-la.caltech.edu/aLOG/index.php?callRep=78903}}
}

@misc{alog:78942,
	title        = {{Summary of Commissioning Changes}},
	author       = {{Bossilkov}, V.},
	howpublished = {\url{https://alog.ligo-la.caltech.edu/aLOG/index.php?callRep=78942}}
}

@misc{alog:74834,
	title        = {{Environmental coupling to the ITMs due to temperature excursions even with stabilized lab temperature}},
	author       = {{Brooks}, A.},
	howpublished = {\url{https://alog.ligo-la.caltech.edu/aLOG/index.php?callRep=74834}}
}

@misc{alog:64609,
	title        = {{Comment to: Another failed attmept to switch OMC ASC to dither alignment}},
	author       = {{Cahillane}, C.},
	howpublished = {\url{https://alog.ligo-wa.caltech.edu/aLOG/index.php?callRep=64609}}
}

@misc{alog:48161,
	title        = {Long-duration non-linearities between violin modes and calibration lines may explain some lines in the ER14 spectra},
	author       = {{Covas}, P.},
	howpublished = {\url{https://alog.ligo-wa.caltech.edu/aLOG/index.php?callRep=48161}}
}

@misc{alog:53439,
	title        = {Summary of O3a H1 lines and combs},
	author       = {{Goetz}, E. and others},
	howpublished = {\url{https://alog.ligo-wa.caltech.edu/aLOG/index.php?callRep=53439}}
}

@misc{alog:87923,
	title        = {Intermodulation lines and $\beta$ estimation in {H1:OMC-DCPD\_SUM\_OUT\_DQ}},
	author       = {{Calafat}, A. and {M\'erou}, J. and {Dwyer}, S. and {Schofield}, R. and {Driggers}, J.},
	howpublished = {\url{https://alog.ligo-wa.caltech.edu/aLOG/index.php?callRep=87923}}
}

@misc{alog:80232,
	title        = {{Wandering lines in PEM VMON channels at EX line up with BNS range oscillations}},
	author       = {{Berger}, B.},
	howpublished = {\url{https://alog.ligo-la.caltech.edu/aLOG/index.php?callRep=80232}}
}

@misc{alog:80556,
	title        = {{Time variation of wandering lines in PEM EX VMON ESDPOWER channels}},
	author       = {{Berger}, B.},
	howpublished = {\url{https://alog.ligo-la.caltech.edu/aLOG/index.php?callRep=80556}}
}

@misc{alog:89929,
	title        = {{PEM November 2025 Injection Coupling Analysis Complete}},
	author       = {{Callos}, S.R. and {Connolly}, G. and {Schofield}, R. and {Short}, R. and {Campos}, C.},
	howpublished = {\url{https://alog.ligo-wa.caltech.edu/aLOG/index.php?callRep=89929}}
}

@misc{alog:80182,
	title        = {{24-30 Hz Range Noise in DARM Identification}},
	author       = {{Callos}, S.R. and {Schofield}, R.},
	howpublished = {\url{https://alog.ligo-wa.caltech.edu/aLOG/index.php?callRep=80182}}
}

@misc{alog:86257,
	title        = {{20-40 Hz HVAC Noise}},
	author       = {{Callos}, S.R. and {Schofield}, R.},
	howpublished = {\url{https://alog.ligo-wa.caltech.edu/aLOG/index.php?callRep=86257}}
}

@misc{alog:85740,
	title        = {{Trying High Bandwidth control for earthquake}},
	author       = {{Capote}, E. and {Warner}, J.},
	howpublished = {\url{https://alog.ligo-wa.caltech.edu/aLOG/index.php?callRep=85740}}
}

@misc{alog:81386,
	title        = {{Comparison of PSL glitches before locklosses tagged: glitches dramatically changed after NPRO swap}},
	author       = {{Compton}, C.},
	howpublished = {\url{https://alog.ligo-wa.caltech.edu/aLOG/index.php?callRep=81386}}
}

@misc{alog:79546,
	title        = {{Hunting 30-40 Hz peaks with portable accelerometer}},
	author       = {{Connolly}, G.},
	howpublished = {\url{https://alog.ligo-wa.caltech.edu/aLOG/index.php?callRep=79546}}
}

@misc{alog:80655,
	title        = {{Identification of 30-40 Hz peaks in DARM}},
	author       = {{Connolly}, G.},
	howpublished = {\url{https://alog.ligo-wa.caltech.edu/aLOG/index.php?callRep=80655}}
}

@misc{alog:75579,
	title        = {{aLIGO LLO Logbook}},
	author       = {D. Nandi},
	year         = 2025,
	howpublished = {\url{https://alog.ligo-la.caltech.edu/aLOG/index.php?callRep=75579}}
}

@misc{alog:74939,
	title        = {{New DARM offloading reduces non-stationarity by reducing ESD drive RMS.}},
	author       = {{Dartez}, L. and {Dwyer}, S. and {Vajente, G.}},
	howpublished = {\url{https://alog.ligo-wa.caltech.edu/aLOG/index.php?callRep=74939}}
}

@misc{selfgating,
	title        = {{Self-gating of O4a h(t) for use in continuous-wave searches}},
	author       = {{Davis}, D. and {Neunzert}, A. and {Goetz}, E. and {Riles}, K. and {Wette}, K. and {Lalleman}, M.},
	howpublished = {\url{https://dcc.ligo.org/LIGO-T2400003/public}}
}

@misc{alog:80837,
	title        = {{Comment to: Test setup of spare NPRO does not seem to see glitches; beginning process to swap to spare NPRO}},
	author       = {{Driggers}, J.},
	howpublished = {\url{https://alog.ligo-wa.caltech.edu/aLOG/index.php?callRep=80837}}
}

@misc{alog:77368,
	title        = {{IFO very unhappy, something likely wrong with alignment, AS\_C needs strange offsets}},
	author       = {{Driggers}, J. and others},
	howpublished = {\url{https://alog.ligo-wa.caltech.edu/aLOG/index.php?callRep=77368}}
}

@misc{alog:79363,
	title        = {{FS Assmembly Reinstalled in OFI}},
	author       = {{Dwyer}, S and {Kawabe}, K. and {Driggers}, J. and {Compton}, C.},
	howpublished = {\url{https://alog.ligo-wa.caltech.edu/aLOG/index.php?callRep=79363}}
}

@misc{alog:77427,
	title        = {{Summary of alogs related to changes in our output arm this week}},
	author       = {{Dwyer}, S.},
	howpublished = {\url{https://alog.ligo-wa.caltech.edu/aLOG/index.php?callRep=77427}}
}

@misc{alog:79101,
	title        = {{some symptoms that seem similar to April 22nd/23rd output change}},
	author       = {{Dwyer}, S.},
	howpublished = {\url{https://alog.ligo-wa.caltech.edu/aLOG/index.php?callRep=79101}}
}

@misc{alog:77392,
	title        = {{we suspect that something in OFI has changed}},
	author       = {{Dwyer}, S. and others},
	howpublished = {\url{https://alog.ligo-wa.caltech.edu/aLOG/index.php?callRep=77392}}
}

@misc{alog:64609LLO,
	title        = {{PEM Injection Status}},
	author       = {{Effler}, A.},
	howpublished = {\url{https://alog.ligo-la.caltech.edu/aLOG/index.php?callRep=64609}}
}

@misc{alog:67149,
	title        = {{Commissioning tests today}},
	author       = {{Effler}, A.},
	howpublished = {\url{https://alog.ligo-la.caltech.edu/aLOG/index.php?callRep=67149}}
}

@misc{alog:79137,
	title        = {{Vault Tests for LEMI magnetometers calibration}},
	author       = {{Effler}, A. and {Pham}, K. and {Guidry}, T. and {MacDonald}, T. and {Connolly}, G.},
	howpublished = {\url{https://alog.ligo-la.caltech.edu/aLOG/index.php?callRep=79137}}
}

@misc{alog:69626,
	title        = {{Significant discrepancy between the spectrograms from GDS\_CALIB\_STRAIN and GDS\_CALIB\_STRAIN\_CLEAN/NOLINES}},
	author       = {{Ferreira}, T. and {Gonzalez}, G.},
	howpublished = {\url{https://alog.ligo-la.caltech.edu/aLOG/index.php?callRep=69626}}
}

@misc{alog:69850,
	title        = {{Omicron at GDS\_CALIB\_STRAIN and GDS\_CALIB\_STRAIN\_NOLINES}},
	author       = {{Ferreira}, T. and {Gonzalez}, G.},
	howpublished = {\url{https://alog.ligo-la.caltech.edu/aLOG/index.php?callRep=69850}}
}

@misc{alog:73288,
	title        = {{EX tapping tests and microseism drive}},
	author       = {{Frolov}, V. and {Effler}, E.},
	howpublished = {\url{https://alog.ligo-la.caltech.edu/aLOG/index.php?callRep=73288}}
}

@misc{alog:72819,
	title        = {{Comment to: Excess noise during range oscillation}},
	author       = {{Glanzer}, J.},
	howpublished = {\url{https://alog.ligo-la.caltech.edu/aLOG/index.php?callRep=72819}}
}

@misc{alog:73009,
	title        = {{Potential correlation between long duration range drops and CS temp}},
	author       = {{Glanzer}, J.},
	howpublished = {\url{https://alog.ligo-la.caltech.edu/aLOG/index.php?callRep=73009}}
}

@misc{l1:alog:78194,
	title        = {{Comment to: BRUCO comparision for range oscillations}},
	author       = {{Glanzer}, J.},
	howpublished = {\url{https://alog.ligo-la.caltech.edu/aLOG/index.php?callRep=78194}}
}

@misc{alog:79579,
	title        = {{O4a lines statistics and most problematic artifacts}},
	author       = {{Goetz}, E. and {Knee}, A. and {Neunzert}, A.},
	howpublished = {\url{https://alog.ligo-wa.caltech.edu/aLOG/index.php?callRep=79579}}
}

@misc{goetz_2025_18776397,
	title        = {Fscan},
	author       = {{Goetz}, E. and {Neunzert}, A. and {Suyamprakasam}, S.},
	year         = 2025,
	month        = dec,
	publisher    = {Zenodo},
	doi          = {10.5281/zenodo.18776397},
	url          = {https://doi.org/10.5281/zenodo.18776397},
	version      = {0.6.4}
}

@misc{LIGO-T2100200,
	title        = {{O3 lines and combs in found in self-gated C01 data}},
	author       = {{Goetz}, E. and others},
	howpublished = {\url{https://dcc.ligo.org/LIGO-T2100200/public}}
}

@misc{LIGO-T2400204,
	title        = {{O4a lines and combs in found in self-gated C00 cleaned data}},
	author       = {{Goetz}, E. and others},
	howpublished = {\url{https://dcc.ligo.org/LIGO-T2400204/public}}
}

@misc{jane_surf2025,
	title        = {\textit{Investigations of Binary Neutron Star Range Oscillations at LIGO Livingston}},
	author       = {Grant, G. and Glanzer, J.},
	howpublished = {\url{https://dcc.ligo.org/LIGO-T2500220/public}}
}

@misc{alog:79690,
	title        = {{H1 Back To Observing!!!}},
	author       = {{Gray}, C. and {Patane}, O. and {Short}, R. and {Dwyer}, S.},
	howpublished = {\url{https://alog.ligo-wa.caltech.edu/aLOG/index.php?callRep=79690}}
}

@misc{alog:72674,
	title        = {{aLIGO LLO Logbook}},
	author       = {H. Pham},
	year         = 2024,
	howpublished = {\url{https://alog.ligo-la.caltech.edu/aLOG/index.php?callRep=72674}}
}

@misc{alog:72583,
	title        = {{Low range, sqz on off test: It's not the squeezer}},
	author       = {{Kabagoz}, B. and {Effler}, A.},
	howpublished = {\url{https://alog.ligo-la.caltech.edu/aLOG/index.php?callRep=72583}}
}

@misc{alog:70869,
	title        = {{Wandering lines in h(t) around 580 Hz and other related lines}},
	author       = {{Kandhasamy}, S.},
	year         = 2024,
	howpublished = {\url{https://alog.ligo-la.caltech.edu/aLOG/index.php?callRep=70869}}
}

@misc{alog:79326,
	title        = {{KTP wedge chipped, back up spare installed and ready to go to chamber}},
	author       = {{Kawabe}, K. and {Dwyer}, S},
	howpublished = {\url{https://alog.ligo-wa.caltech.edu/aLOG/index.php?callRep=79326}}
}

@misc{alog:79082,
	title        = {{DARM offset causing locklosses again}},
	author       = {{Kawabe}, K. and {Dwyer}, S. and {Crouch}, R. and {Abouelfettouh}, I.},
	howpublished = {\url{https://alog.ligo-wa.caltech.edu/aLOG/index.php?callRep=79082}}
}

@misc{alog:87414,
	title        = {{Witness channels of the set of near-30Hz and near-100Hz combs at LHO}},
	author       = {{M\'erou}, J. and {Calafat}, A. and {Effler}, A. and {Dwyer}, S. and {Schofield}, R. and {Driggers}, J.},
	howpublished = {\url{https://alog.ligo-wa.caltech.edu/aLOG/index.php?callRep=87414}}
}

@misc{alog:75125,
	title        = {{Unlocking ALS end-station PLLs}},
	author       = {{Mullavey}, A.},
	howpublished = {\url{https://alog.ligo-la.caltech.edu/aLOG/index.php?callRep=75125}}
}

@misc{alog:78226,
	title        = {{Comparing the rate of scattered light glitches pre and post vent}},
	author       = {{Nandi}, D. and {Ferreira}, T.},
	howpublished = {\url{https://alog.ligo-la.caltech.edu/aLOG/index.php?callRep=78226}}
}

@misc{alog:71964,
	title        = {{CAL\_AWG\_LINES extra calibration lines create many narrow artifacts below 100 Hz}},
	author       = {{Neunzert}, A.},
	howpublished = {\url{https://alog.ligo-wa.caltech.edu/aLOG/index.php?callRep=71964}}
}

@misc{alog:82320,
	title        = {{Relationship between violin mode height and narrow spectral artifact contamination, revisited}},
	author       = {{Neunzert}, A.},
	howpublished = {\url{https://alog.ligo-wa.caltech.edu/aLOG/index.php?callRep=82320}}
}

@misc{alog:79825,
	title        = {{Deeper study of a 2023 test shows calibration lines + violin modes = more line contamination than previously understood}},
	author       = {{Neunzert}, A. and {Goetz}, E. and {Knee}, A. and {Collier}, T. and {Marceau}, A.},
	howpublished = {\url{https://alog.ligo-wa.caltech.edu/aLOG/index.php?callRep=79825}}
}

@misc{alog:73481,
	title        = {{EX Manifold Wrapped to try and affect 30 min oscillations}},
	author       = {{O'Hanlon}, T. and {Effler}, E.},
	howpublished = {\url{https://alog.ligo-la.caltech.edu/aLOG/index.php?callRep=73481}}
}

@misc{alog:81409,
	title        = {{PSL NPRO Swap: Here We Go Again (WP 12210)}},
	author       = {{Oberling}, J. and {Short}, R.},
	howpublished = {\url{https://alog.ligo-wa.caltech.edu/aLOG/index.php?callRep=81409}}
}

@misc{alog:81193,
	title        = {{Overview of PSL Story so far}},
	author       = {{Oberling}, J. and {Short}, R. and {Dwyer}, S. and {Sanchez}, A. and {Xu}, V. and {Capote}, C. and {Compton}, C.},
	howpublished = {\url{https://alog.ligo-wa.caltech.edu/aLOG/index.php?callRep=81193}}
}

@misc{alog:78104,
	title        = {{Comparison before and after changing AHU settings}},
	author       = {{Iwaguchi}, S.},
	howpublished = {\url{https://alog.ligo-la.caltech.edu/aLOG/index.php?callRep=78104}}
}

@misc{alog:78563,
	title        = {{AHU behavior and LVEA temperature after change of AHU1 setting }},
	author       = {{Iwaguchi}, S.},
	howpublished = {\url{https://alog.ligo-la.caltech.edu/aLOG/index.php?callRep=78563}}
}

@misc{alog:78617,
	title        = {{LVEA HVAC Controls Change}},
	author       = {{O'Hanlon}, T.},
	howpublished = {\url{https://alog.ligo-la.caltech.edu/aLOG/index.php?callRep=78617}}
}

@misc{alog:78473,
	title        = {{Start testing EQ control with ASC cut-off}},
	author       = {{Pham}, H.},
	howpublished = {\url{https://alog.ligo-la.caltech.edu/aLOG/index.php?callRep=78473}}
}

@misc{alog:84924,
	title        = {{May 15 LEMI calibration}},
	author       = {{Pham}, K. and {Schofield}, R. and {Campos}, C.},
	howpublished = {\url{https://alog.ligo-wa.caltech.edu/aLOG/index.php?callRep=84924}}
}

@misc{alog:77400,
	title        = {{Re-aligning SQZ to current AS alignment ~4.5dB, required big ZM moves}},
	author       = {{Sanchez}, A. and {Kijbunchoo}, N. and {Xu}, V.},
	howpublished = {\url{https://alog.ligo-wa.caltech.edu/aLOG/index.php?callRep=77400}}
}

@misc{alog:72118,
	title        = {{ ETMX bias that minimizes grounding noise coupling hasnt changed much since January, and PEM injections at 60W}},
	author       = {{Schofield}, R.},
	howpublished = {\url{https://alog.ligo-wa.caltech.edu/aLOG/index.php?callRep=72118}}
}

@misc{alog:74175,
	title        = {{CER ACs and main HVAC likely couple to DARM in input arm, possibly through MCA2 baffle}},
	author       = {{Schofield}, R.},
	howpublished = {\url{https://alog.ligo-wa.caltech.edu/aLOG/index.php?callRep=74175}}
}

@misc{alog:74772,
	title        = {{Confirmation from laser vibrometry and beating shakers: the MC baffle by HAM3 has been producing noise in DARM and the one by HAM2 produces noise when 18 Hz vibration is increased by about 5; also, suggestions for mitigation during the January break}},
	author       = {{Schofield}, R.},
	howpublished = {\url{https://alog.ligo-wa.caltech.edu/aLOG/index.php?callRep=74772}}
}

@misc{alog:76969,
	title        = {{Evaluation of stray light work done during the break: desired reductions in vibration coupling attained for EX cryobaffle, and for input arm - after adjustment of ITMY CP yaw}},
	author       = {{Schofield}, R.},
	howpublished = {\url{https://alog.ligo-wa.caltech.edu/aLOG/index.php?callRep=76969}}
}

@misc{alog:78734,
	title        = {{ Slow changes of ITMY ESD bias during observation}},
	author       = {{Schofield}, R.},
	howpublished = {\url{https://alog.ligo-wa.caltech.edu/aLOG/index.php?callRep=78734}}
}

@misc{alog:78925,
	title        = {{Slow bias changes during observation completed; New ESD bias settings: ITMX 0V, ITMY -40V (-4.0 in offset window)}},
	author       = {{Schofield}, R.},
	howpublished = {\url{https://alog.ligo-wa.caltech.edu/aLOG/index.php?callRep=78925}}
}

@misc{alog:67075,
	title        = {{Update on grounding noise studies}},
	author       = {{Schofield}, R. and {Dwyer}, S},
	howpublished = {\url{https://alog.ligo-wa.caltech.edu/aLOG/index.php?callRep=67075}}
}

@misc{alog:78194,
	title        = {{ ETMX bias sweep gives significantly different minimum in electronics ground noise coupling after charge change: from 150 V to new value of 58 V (1.5 bias offset)}},
	author       = {{Schofield}, R. and {Dwyer}, S. and {Compton}, C.},
	howpublished = {\url{https://alog.ligo-wa.caltech.edu/aLOG/index.php?callRep=78194}}
}

@misc{alog:82986,
	title        = {{35 Hz peak in DARM greatly reduced by changing the frequency of the OSB HVAC drive.}},
	author       = {{Schofield}, R. and {McCarthy}, R. and {Guidry}, T. and {Connolly}, G. and {Mannix}, B. and {Olson}, J.},
	howpublished = {\url{https://alog.ligo-wa.caltech.edu/aLOG/index.php?callRep=82986}}
}

@misc{alog:69745,
	title        = {{PEM injections successfully completed - will inform HVAC experiments}},
	author       = {{Short}, R. and {Glanzer}, J. and {Helmling-Cornell}, A. and {Effler}, A. and {Schofield}, R},
	howpublished = {\url{https://alog.ligo-wa.caltech.edu/aLOG/index.php?callRep=69745}}
}

@misc{alog:80908,
	title        = {{PSL NPRO Swap: It Lives! (WP 12155)}},
	author       = {{Short}, R. and {Oberling}, J.},
	howpublished = {\url{https://alog.ligo-wa.caltech.edu/aLOG/index.php?callRep=80908}}
}

@misc{alog:71501,
	title        = {{Effect of violin mode amplitude on narrow lines near the violin modes}},
	author       = {{Starkman}, T.},
	howpublished = {\url{https://alog.ligo-wa.caltech.edu/aLOG/index.php?callRep=71501}}
}

@misc{alog:71800,
	title        = {{Asymmetry in narrow line contamination in bands around violin mode frequencies}},
	author       = {{Starkman}, T.},
	howpublished = {\url{https://alog.ligo-wa.caltech.edu/aLOG/index.php?callRep=71800}}
}

@misc{alog:81107,
	title        = {{PSL NPRO Crystal Temperature Tuning}},
	author       = {{Xu}, V. and {Oberling}, J.},
	howpublished = {\url{https://alog.ligo-wa.caltech.edu/aLOG/index.php?callRep=81107}}
}

@phdthesis{helmlingcornellthesis,
	title        = {Noise Mitigation in Searches for Gravitational Waves from Compact Binaries and Cosmic Strings},
	author       = {{Helmling-Cornell}, A.},
	year         = 2025,
	school       = {University of Oregon}
}

@software{duncan_macleod_2024_10999143,
	title        = {gwpy/gwsumm: 2.2.6},
	author       = {Macleod, D. and Goetz, E. and Ota, I. and Isi, M. and Massinger, T. and Davis, D. and Pitkin, M. and Altin, P. and Nitz, A.},
	year         = 2024,
	month        = apr,
	publisher    = {Zenodo},
	doi          = {10.5281/zenodo.10999143},
	url          = {https://doi.org/10.5281/zenodo.10999143},
	version      = {2.2.6}
}

\end{document}